%% file: RPM3.tex
\documentclass[onecolumn, twoside]{IEEEtran}
\usepackage[table]{xcolor}
\usepackage{amssymb,epsfig,graphics, graphicx, url,times,mathrsfs,amsmath,array,latexsym,fancyhdr,xspace,wrapfig, xcolor,multirow,amsthm,caption,cite,helvet,sidecap,bm,hyperref,url,dsfont,nicefrac, subcaption}

\usepackage[normalem]{ulem}

\usepackage[inline]{enumitem}
\usepackage{arydshln}
\usepackage[makeroom]{cancel}
\usepackage{varwidth}
\usepackage{pgfplots}
\usepackage[ruled,vlined]{algorithm2e}

\usepackage{tikz}
\usetikzlibrary{shapes.geometric}
\usetikzlibrary{shapes,snakes,patterns}
\usetikzlibrary{arrows,positioning,automata,calc}
\usetikzlibrary{plotmarks}

\definecolor{green1}{rgb}{0,0.5,0}
\definecolor{magenta}{rgb}{1.0, 0.11, 0.81}
\definecolor{mulberry}{rgb}{0.77, 0.29, 0.55}
\definecolor{xgray}{rgb}{0.9, 0.9, 0.9}

\newcommand{\temp}[1]{\textcolor{red}{#1}}
\newcommand{\fig}[1]{Figure~{#1}}

\def \bes{\begin{equation*}}
\def \ees{\end{equation*}}
\def \bas{\begin{align*}}
\def \eas{\end{align*}}
\def \be{\begin{equation}}
\def \ee{\end{equation}}
\def \bbm{\begin{bmatrix}}
\def \ebm{\end{bmatrix}}

\newcommand{\taulb}[1]{\tau^{(\text{LB})}_{#1}}
\newcommand{\tauideal}[1]{\tau^{(\text{ideal})}_{#1}}
\newcommand{\taunc}[1]{\tau^{(\text{no col})}_{#1}}
\newcommand{\taugasp}[1]{\tau^{(\text{S-GASP})}_{#1}}
\newcommand{\taukk}[1]{\tau^{(\text{L-GASP})}_{#1}}
\newcommand{\taulag}[1]{\tau^{(\text{Lag})}_{#1}}
\newcommand{\fr}{KES}

\def \rvA{\texttt{A}}
\def \rvB{\texttt{B}}
\def \rvC{\texttt{C}}

\def \rvT{\texttt{T}}
\def \rvY{\texttt{Y}}
\def \rvX{\texttt{X}}

\def \F{\mathbb{F}}
\newcommand{\Fdata}[1]{\textcolor{blue}{\mathbf{#1}}}
\newcommand{\ftr}[2]{f_{#1}^{(#2)}}
\newcommand{\gtr}[2]{g_{#1}^{(#2)}}
\newcommand{\htr}[2]{h_{#1}^{(#2)}}

\newcommand{\cA}{{\cal A}}

\newcommand{\cF}{{\cal F}}

\newcommand{\cK}{{\cal K}}

\newcommand{\cR}{{\cal R}}

\newcommand{\cW}{{\cal W}}

\newcommand{\cZ}{{\cal Z}}

\newcommand{\mA}{\mathbf{A}}
\newcommand{\mB}{\mathbf{B}}
\newcommand{\mC}{\mathbf{C}}
\newcommand{\mR}{\mathbf{R}}
\newcommand{\mS}{\mathbf{S}}

\newcommand{\mtA}[2]{\widetilde{\mA}_{#1}^{(#2)}}
\newcommand{\mtB}[2]{\widetilde{\mB}_{#1}^{(#2)}}

\newtheorem*{theorem*}{Theorem}
\newtheorem{theorem}{Theorem}
\newtheorem{corollary}{Corollary}
\newtheorem{lemma}{Lemma}

\newtheorem{definition}{Definition}

\newtheorem{example}{Example}
\newtheorem{remark}{Remark}

\usetikzlibrary{decorations.pathreplacing,calc}
\tikzset{brace/.style={decorate, decoration={brace}},
 brace mirrored/.style={decorate, decoration={brace,mirror}},
}

\newcounter{brace}
\newcommand{\PreserveBackslash}[1]{\let\temp=\\#1\let\\=\temp}
\newcolumntype{C}[1]{>{\PreserveBackslash\centering}p{#1}}

\newcommand{\fcA}[3]{\widetilde{\mA}_{#1,#2}^{(#3)}}
\newcommand{\fcB}[3]{\widetilde{\mB}_{#1,#2}^{(#3)}}
\def \interval{\Delta}

\IEEEoverridecommandlockouts

\SetArgSty{textnormal}

\SetCommentSty{mycommfont}
\begin{document}
\newlength\figureheight
\newlength\figurewidth

\title{Adaptive Private Distributed Matrix Multiplication}

\author{\IEEEauthorblockN{
Rawad Bitar, Marvin Xhemrishi and Antonia Wachter-Zeh }\\
\IEEEauthorblockA{Institute for Communications Engineering, Technical University of Munich, Munich, Germany\\
\{\texttt{rawad.bitar, marvin.xhemrishi, antonia.wachter-zeh}\}\texttt{@tum.de}}
\thanks{Parts of these results of this work were presented at the IEEE International Symposium on Information Theory and its Applications (ISITA)~\cite{BXWZ20}.}
\thanks{
This work was partly supported by the Technical University of Munich - Institute for Advanced Studies, funded by the German Excellence Initiative
and European Union Seventh Framework Programme under Grant Agreement
No. 291763.
}
}


\maketitle
\begin{abstract}
We consider the problem of designing codes with flexible rate (referred to as 
\emph{rateless} codes), for private distributed matrix-matrix multiplication. A master server owns two private matrices $\mA$ and $\mB$ and hires worker nodes to help computing their multiplication. The matrices should remain information-theoretically private from the workers.
Codes with fixed rate require the master to assign tasks to the workers and then wait for a predetermined number of workers to finish their assigned tasks. The size of the tasks, hence the rate of the scheme, depends on the number of workers that the master waits for.

We design a rateless private matrix-matrix multiplication scheme, called RPM3. 
In contrast to fixed-rate schemes, our scheme fixes the size of the tasks and allows the master to send multiple tasks to the workers. The master keeps sending tasks and receiving results until it can decode the multiplication; rendering the scheme flexible and adaptive to heterogeneous environments. 

Despite resulting in a smaller rate than known straggler-tolerant schemes, RPM3 provides a smaller mean waiting time of the master by leveraging the heterogeneity of the workers. The waiting time is studied under two different models for the workers' service time. We provide upper bounds for the mean waiting time under both models. In addition, we provide lower bounds on the mean waiting time under the worker-dependent fixed service time model.
\end{abstract}
\begin{IEEEkeywords} Private rateless codes, double-sided private matrix multiplication, partial stragglers, information-theoretic privacy   \end{IEEEkeywords}

\section{Introduction}

\input{intro.tex}

\section{Preliminaries}\label{sec:Preliminaries}
\input{problem.tex}

\section{Main Results}
\input{main_results.tex}


\section{RPM3 Scheme}\label{sec:rpm3}

\input{sys_model}

\section{Rate Analysis}\label{sec:rate}
\input{rate_analysis.tex}

\section{Time Analysis}\label{sec:time}
\input{time_analysis.tex}

\section{Perfect Load Balancing}\label{sec:load_balancing}
\input{perfect_load_balancing.tex}


\section{Conclusion}\label{sec:conc}
\input{conclusion.tex}

\bibliographystyle{ieeetr}
\bibliography{IEEEabrv,RPM3}

\appendices
\section{Proof of privacy}\label{app:proof_privacy}
\input{proof_privacy.tex}

\section{Proofs for the mean waiting time}\label{app:proof_waiting}
\input{app_proof_waiting_time.tex}

\section{Algorithms}
\input{algorithms}

\end{document}

%% file: intro.tex
{Matrix-matrix multiplication is at the core of several machine learning algorithms. In such applications the multiplied matrices are large and distributing the tasks to several machines is required.} We consider the problem in which a master server owns two private matrices $\mA$ and $\mB$ and wants to compute $\mC=\mA \mB$. The master splits the computation into smaller tasks and distributes them to several worker nodes that can run those computations in parallel. However, waiting for all workers to finish their tasks suffers from the presence of slow processing nodes \cite{DB13,dean2012large}, referred to as \emph{stragglers}. The presence of stragglers can outweigh the benefit of parallelism, see e.g., \cite{AGSS13,FASTC,speeding} and references therein. 

Moreover, the master's data must remain private from the workers. We are interested in information-theoretic privacy which does not impose any constraints on the computational power of the compromised workers. On the other hand, information-theoretic privacy assumes that the number of compromised workers is limited by a certain threshold.

We consider applications where the resources of the workers are different, limited and time-varying. Examples of this setting include edge computing in which the devices collecting the data (e.g., sensors, tablets, etc.) cooperate to run the intensive computations. In such applications, the workers have different computation power, battery life and network latency which can change in time. We refer to this setting as heterogeneous and time-varying setting.

We develop a coding scheme that allows the master to offload the computational tasks to the workers while satisfying the following requirements: \begin{enumerate*}[label=\textit{\roman*)}] \item leverage the heterogeneity of the workers, i.e., assign a number of tasks to the workers that is proportional to their resources; \item adapt to the time-varying nature of the workers; and \item maintain the privacy of the master's data.
\end{enumerate*}

We focus on matrix-matrix multiplication since this is a building block of several machine learning algorithms \cite{suykens1999least,seber2012linear}. We use coding-theoretic techniques to encode the tasks sent to the workers. We illustrate the use of codes to distribute the tasks in the following example.

\begin{table*}[t]
    \centering
    \renewcommand{\arraystretch}{1.2}
    \resizebox{\textwidth}{!}{
    \begin{tabular}{c|c|c|c|c|c}
    ~ & Worker~$1$ & Worker~$2$ & Worker~$3$ & Worker~$4$ & Worker~$5$ \\ \hline%
      \multirow{2}{*}{Round $1$} & $\mR_1(1-a_1)+\Fdata{A_1} a_1$ &  $\mR_1(1-a_2)+\Fdata{A_1} a_2$ &    $\mR_1(1-a_3)+\Fdata{A_1} a_3$&  $\mR_1(1-a_4)+\Fdata{A_2} a_4$&  $\mR_1(1-a_5)+\Fdata{A_2} a_5$ \\ %
       ~ & $\mS_1(1-a_1)+\Fdata{B} a_1$ &  $\mS_1(1-a_2)+\Fdata{B} a_2$ &  $\mS_1(1-a_3)+\Fdata{B} a_3$&  $\mS_1(1-a_4)+\Fdata{B} a_4$&   $\mS_1(1-a_5)+\Fdata{B} a_5$ \\[0.5em] \hdashline%
      \multirow{2}{*}{Round $2$} & $\mR_2(1-a_1)+(\Fdata{A_1+A_2}) a_1$ &  $\mR_2(1-a_2)+(\Fdata{A_1+A_2}) a_2$ &    $\mR_2(1-a_3)+(\Fdata{A_1+A_2}) a_3$& ~& ~ \\ %
       ~ & $\mS_2(1-a_1)+\Fdata{B} a_1$ &  $\mS_2(1-a_2)+\Fdata{B} a_2$ &  $\mS_2(1-a_3)+\Fdata{B} a_3$& ~ &   ~\\
    \end{tabular}
    }
    
    \caption{A depiction of the tasks sent to the workers in Example~\ref{ex:intro2}.}
    \label{tab:intro-ex2}
    \vspace{-0.5cm}
\end{table*}

\begin{example}\label{ex:intro}
Let $\mA \in \F_q^{r\times s}$ and $\mB \in \F_q^{s\times \ell}$ be two private matrices owned by the master who wants to compute $\mC=\mA \mB$. The master has access to $5$ workers. At most $2$ workers can be stragglers. The workers do not collude, i.e., the workers do not share with each other the tasks sent to them by the master. 
To encode the tasks, the master generates two random matrices $\mR \in \F_q^{r\times s}$ and $\mS \in \F_q^{s\times \ell}$ uniformly at random and independently from $\mA$ and $\mB$. The master creates two polynomials\footnote{The multiplication and addition within the polynomials is element-wise, e.g., each element of $\mA$ is multiplied by $x$.} $f(x) = \mR(1-x)+ \mA x$ and $g(x) = \mS(1-x)+\mB x$. The data sent to worker $i$ is $f(a_i)$ and $g(a_i)$,  $i=1,\dots,5$, where $a_i\in \F_q\setminus\{1\}$. Each worker computes $h(a_i) \triangleq f(a_i) g (a_i) = \mR\mS(1-a_i)^2 + \mR\mB a_i(1-a_i) + \mA\mS a_i(1-a_i) + \mA\mB a_i^2$ and sends the result to the master. When the master receives at least three evaluations of $h(x) \triangleq f(x)g(x)$, it can interpolate the polynomial of degree $2$. In particular, the master can compute $\mA\mB=h(1)$. On a high level, the privacy of $\mA$ and $\mB$ is maintained because each matrix is masked by a random matrix before being sent to a worker.
\end{example}

In Example~\ref{ex:intro}, even if there are no stragglers, the master ignores the responses of two workers. In addition, all the workers obtain computational tasks of the same complexity\footnote{Each evaluation of the polynomial $f(x)$ (or $g(x)$) is a matrix of the same dimension as $\mA$ (or $\mB$). The computational complexity of the task is therefore proportional to the dimension of the created polynomial.}.

We highlight in Example~\ref{ex:intro2} the main ideas of our scheme that allows the master to assign tasks of different complexity to the workers and use all the responses of the non stragglers.

\begin{example}\label{ex:intro2}
Consider the same setting as in Example~\ref{ex:intro}. Assume that workers $1,2$ and $3$ are more powerful than the others. The master splits $\mA$ into $\mA= \begin{bmatrix}\mA_1^T & \mA_2^T
\end{bmatrix}^T$ and 
wants $\mC=\begin{bmatrix}(\mA_1\mB)^T & (\mA_2\mB)^T
\end{bmatrix}^T$. 
The master divides the computations into two rounds. In the first round, the master generates two random matrices $\mR_1 \in \F_q^{r/2\times s}$ and $\mS_1 \in \F_q^{s\times \ell}$ uniformly at random and independently from $\mA$ and $\mB$. The master creates four polynomials: 
\begin{align*}
  f_1^{(1)}(x) & = \mR_1(1-x)+ \mA_1x,\\
  \ftr{1}{2}(x) & = \mR_1(1-x)+\mA_2 x,\\
  \gtr{1}{1}(x) & =\gtr{1}{2}(x) = \mS_1(1-x)+\mB x.   
\end{align*}
The master sends $\ftr{1}{1}(a_i)$ and $\gtr{1}{1}(a_i)$ to workers $1,2,3$, and sends $\ftr{1}{2}(a_i)$ and $\gtr{1}{2}(a_i)$ to workers $4,5$, where $a_i\in \F_q\setminus\{0,1\}$. Workers $1$, $2$, $3$ compute $\htr{1}{1}(a_i) \triangleq \ftr{1}{1}(a_i) \gtr{1}{1} (a_i)$ and workers $4$, $5$ compute $\htr{1}{2}(a_i)\triangleq \ftr{1}{2}(a_i) \gtr{1}{2} (a_i)$.

The master starts round $2$ when workers $1,2,3$ finish their tasks. It generates two random matrices $\mR_2 \in \F_q^{r/2\times s}$ and $\mS_2 \in \F_q^{s\times \ell}$ and creates 
\begin{align*}
f_2^{(1)}(x) &= \mR_2 (1-x) + (\mA_1+\mA_2)x,\\ 
\gtr{2}{1}(x) &= \mS_2(1-x)+\mB x,
\end{align*}
and sends evaluations to the first three workers which compute $\htr{2}{1}(a_i)\triangleq \ftr{2}{1}(a_i)\gtr{2}{1}(a_i)$. One main component of our scheme is to generate $\widetilde{\mA}_1\triangleq \mA_1$, $\widetilde{\mA}_2\triangleq \mA_2$ and $\widetilde{\mA}_3\triangleq \mA_1+\mA_2$ as Fountain-coded~\cite{Fountain,LT,Raptor,factored_lt} codewords of $\mA_1$ and $\mA_2$.
The tasks sent to the workers are depicted in Table~\ref{tab:intro-ex2}.

\textbf{Decoding $\mC$:} The master has two options:
\begin{enumerate}[label=\textit{\arabic*)}] \item workers $4$ and $5$ finish their first task before workers $1,2,3$ finish their second tasks, i.e., no stragglers. The master interpolates $\htr{1}{1}(x)$ and obtains $\htr{1}{1}(1) = \mA_1\mB$ and $\htr{1}{1}(0) = \mR_1\mS_1$. Notice that  $\htr{1}{2}(0)=\htr{1}{1}(0) = \mR_1\mS_1$. Thus, the master has three evaluations of $\htr{1}{2}(x)$ and can interpolate it to obtain $\mA_2\mB=\htr{1}{2}(1)$.
\item workers $4$ and $5$ are stragglers and do not finish their first task before workers $1,2,3$ finish their second tasks. The master interpolates (decodes) both $\htr{1}{1}(x)$ and $\htr{2}{1}(x)$. In particular, the master obtains $\mA_1\mB = \htr{1}{1}(1)$ and $\mA_2\mB = (\mA_1+\mA_2)\mB - \mA_1\mB= \htr{2}{1}(1) - \mA_1\mB.$ \end{enumerate}
On a high level, the privacy of $\mA$ and $\mB$ is maintained because each matrix is masked by a different random matrix before being sent to a worker.
\end{example}
In this paper, we generalize the setting of Example~\ref{ex:intro2} and show how to encode the data and generate the tasks sent to the workers in such a heterogeneous server setting. {We compare the proposed scheme to schemes with fixed straggler tolerance and to perfect load balancing where we assume the master knows how the resources of the workers would change in time.}

{\em Related work:} The use of codes to mitigate stragglers in distributed linear computations was first proposed in \cite{speeding} without privacy constraints. Several works such as \cite{mallick2018rateless,baharav2018straggler,wang2018coded,yu2017polynomial,li2016fundamental,yu2018straggler,fahim2017optimal,KS18,factored_lt,nodehi2019secure,behrouzi2020efficient,9174030,coded_sparse_mm,Alex_paper} propose different techniques improving on \cite{speeding} and provide fundamental limits on distributed computing. Of particular importance is the work of~\cite{factored_lt} where the authors show how to construct rateless codes for non-private distributed matrix-matrix multiplication.
Straggler mitigation with privacy constraints is considered in \cite{BPR17,bitar2019private,yang2018secure,d2018gasp,chang2018capacity,kakar2018rate,yu2018lagrange,yu2020entangled,kim2019private,kakar2019capacity,aliasgari2020private,jia2019cross,mital2020secure,kakar2020uplink}. The majority of the literature assumes a threshold of fixed number of stragglers. 
In \cite{BPR17,bitar2019private} the authors consider the setting in which the number of stragglers is not known a priori and design schemes that can cope with this setting. However, \cite{BPR17,bitar2019private} consider the matrix-vector multiplication setting in which only the input matrix must remain private. Our proposed scheme can be seen as a generalization of the coding scheme in \cite{bitar2019private} to handle matrix-matrix multiplication. 
{In~\cite{d2020notes}, the authors characterize the regimes in which distributing the private matrix-matrix multiplication is faster than computing the product locally.}

{\em Contributions:} We present RPM3 a coding scheme for private matrix-matrix multiplication that have flexible straggler tolerance. Our scheme is based on dividing the input matrices into smaller parts and encode the small parts using rateless Fountain codes. The Fountain-coded matrices are then encoded into small computational tasks (using several Lagrange polynomials) and sent to the workers. The master adaptively sends tasks to the workers. In other words, the master first sends a small task to each worker and then starts sending new small tasks to workers who finished their previous task. We show that our scheme satisfies the following properties: \begin{enumerate*}[label={\textit{\roman*)}}]\item it maintains the privacy of the input matrices against a given number of colluding workers; \item it leverages the heterogeneity of the resources at the workers; and \item it adapts to the time-varying resources of the workers\end{enumerate*}. 

{We study the rate, i.e., the number of assigned tasks versus number of useful computations, and the mean waiting time of the master when using RPM3. We give an upper bound on the mean waiting time of the master under two different delay models and provide a probabilistic guarantee of succeeding when given a deadline. We show that despite having a lower rate, RPM3 outperforms schemes assuming a fixed number of stragglers in heterogeneous environments. In addition, we provide lower bounds on the mean waiting time of the master. The bounds are derived by assuming perfect load balancing, i.e., assuming the master has full knowledge of the variation of the compute power at the workers during the matrix-matrix multiplication process. We shed light on the properties that the encoding functions of any flexible-rate codes should satisfy in order to have mean waiting time that matches with our lower bounds. We leave the problem of finding lower bounds on the rate, and thus tighter lower bounds on the mean waiting time, as an interesting open problem.}

{{\em Organization:} In Section~\ref{sec:Preliminaries} we set the notation and define the model. We provide the details of our RPM3 scheme in Section~\ref{sec:rpm3}. In Section~\ref{sec:rate}, we analyze the efficiency of RPM3 and compare it to the straggler tolerant fixed rate scheme with the best known rate. We introduce the delay models and analyze the waiting of the master in Section~\ref{sec:time}. In Section~\ref{sec:load_balancing} we explain the perfect load balancing scheme, provide lower bounds on the mean waiting time and compare them to the mean waiting time of to RPM3. We conclude the paper in Section~\ref{sec:conc}.}

%% file: problem.tex
\subsection{Notation} For any positive integer $n$ we define $[n]\triangleq \{1,\dots,n\}$. We denote by $n$ the total number of workers. For $i\in[n]$ we denote worker $i$ by $w_i$. 
For a prime power $q$, we denote by $\F_q$ the finite field of size $q$. We denote by $H(\rvA)$ the entropy of the random variable $\rvA$ and by $I(\rvA;\rvB)$ the mutual information between two random variables $\rvA$ and $\rvB$. All logarithms are to the base $q$.

\subsection{Problem setting} The master possesses two private matrices $\mA\in\F_q^{r\times s}$ and $\mB\in \F_q^{s\times \ell}$ uniformly distributed over their respective fields and wants to compute $\mC=\mA\mB \in \F_q^{r \times \ell}$. The master has access to $n$ workers that satisfy the following properties: 

\begin{enumerate*}[label=\textit{\arabic*)}] \item The workers have different resources. They can be grouped into $c>1$ clusters with $n_u$ workers, $u=1,\dots,c,$ with similar resources such that $\sum_{u\in[c]} n_u = n$.\\ 
\item The resources available at the workers can change with time. Therefore, the size of the clusters and their number can change throughout the multiplication of $\mA$ and $\mB$.\\
\item The workers have limited computational capacity.\\
\item Up to $z$, $\displaystyle 1\leq z<\min_{u\in[c]}{n_u}$, workers collude to obtain information about $\mA$ and/or $\mB$. If $z = 1$, we say the workers do not collude. 
\end{enumerate*}

The master splits $\mA$ row-wise and $\mB$ column-wise into $m$ and $k$ smaller sub-matrices, respectively, i.e., $\mA=\bbm \mA_1^T, \dots, \mA_m^T\ebm^T$, and $\mB=\bbm \mB_1, \dots, \mB_k\ebm$. The master sends several computational tasks to each of the workers such that each task has the same computational complexity as $\mA_i \mB_j$, $i\in [m], j\in [k]$. After receiving $km(1+\varepsilon)$ responses from the workers, where $0\leq \epsilon\leq 1$ is a parameter of the scheme, the master should be able to compute $\mC=\mA\mB$. The value of $\varepsilon$ depends on the encoding strategy\footnote{For regular Fountain codes, $\varepsilon$ is of the order of $0.05$~\cite{LT,Fountain}.} and decreases with $km$.

A matrix-matrix multiplication scheme is double-sided $z$-private if any collection of $z$ colluding workers learns nothing about the input matrices involved in the multiplication.

\begin{definition}[Double-sided $z$-private matrix-matrix multiplication scheme]\label{def:doubly_private}
Let $\rvA$ and $\rvB$ be the random variables representing the input matrices. We denote by $\cW_i$ the set of random variables representing all the tasks assigned to worker $w_i$, $i=1,\dots,n$. For a set $\mathcal{A}\subseteq [n]$ we define $\cW_\mathcal{A}$ as the set of random variables representing all tasks sent to workers indexed by $\mathcal{A}$, i.e., $\cW_\mathcal{A}=\{\cW_i| i\in \cA\}$. Then the privacy constraint can be expressed as
\begin{equation}\label{eq:privacy}
    {I}\left(\rvA,\rvB;\cW_\mathcal{Z}\right) = 0, \forall \cZ \subset [n], \text{ s.t. } |\cZ| = z.
\end{equation}

Let $\cR_i$ be the set of random variable representing all the computational results of $w_i$ received at the master. Let $\rvC$ be the random variable representing the matrix $\mC$. The decodability constraint can be expressed as
\begin{equation}\label{eq:decoding}
    {H}\left(\rvC|\cR_1,\ldots,\cR_n\right) = 0.
\end{equation}
\end{definition}

Note that the sets $\cR_i$ can be of different cardinality, and some may be empty, reflecting the heterogeneity of the system and the straggler tolerance.

Let the \emph{download rate}, $\rho$, of the scheme be defined as the ratio between the number of needed tasks to compute $\mC$ and the number of responses sent by the workers to the master, %

\begin{equation*}
\rho = \frac{mk}{\text{number of received responses}}.
\end{equation*}
We are interested in designing \emph{rateless} double-sided $z$-private codes for this setting. By \emph{rateless}, we mean that the download rate, or simply rate, of the scheme is not fixed a priori, but it changes depending on the resources available at the workers.
For instance, the rate of the scheme in Example~\ref{ex:intro} is fixed to $1/3$, whereas the rate of the scheme in Example~\ref{ex:intro2} is either $2/5$ or $1/3$ depending on the behavior of the workers.

\begin{figure*}[t!]
\hspace*{-.5cm}
\begin{subfigure}[b]{0.33\textwidth}
\centering
 \setlength\figureheight{0.65\textwidth}
 \setlength\figurewidth{0.75\textwidth}
 \resizebox{.99\textwidth}{!}{
  \input{Figures/Intro_model1} }
  \captionsetup{width = 0.8\textwidth}
  \caption{Empirical average waiting time under model~1.}
  \label{subfig:intro_model1}
\end{subfigure}%
\begin{subfigure}[b]{0.33\textwidth}
\centering
 \setlength\figureheight{0.65\textwidth}
 \setlength\figurewidth{0.75\textwidth}
\resizebox{\textwidth}{!}{
 \input{Figures/Intro_model2} }
 \captionsetup{width = 0.8\textwidth}
 \caption{Empirical average waiting time under model~2.}
 \label{subfig:intro_model2}
\end{subfigure}%
\begin{subfigure}[b]{0.33\textwidth}
\captionsetup{width = 0.8\textwidth}
\centering
 \setlength\figureheight{0.65\textwidth}
 \setlength\figurewidth{0.75\textwidth}
 \resizebox{.99\textwidth}{!}{
  \input{Figures/Intro_bound} }
  \caption{Theoretical bound and empirical average waiting time under model~1.}
  \label{subfig:intro_bound}
\end{subfigure}%
\caption{Average waiting time of RPM3 and the fixed rate scheme from~\cite{kakar2018rate}. We model the response time of the individual workers as a shifted exponential random variable. We consider $n=1000$ workers grouped in $5$ different clusters, referred to as setting 1, and shown in Table~\ref{tab:parameters}. We consider two scenarios where workers of different clusters have similar service time (homogeneous environment) and different service times (heterogeneous environment). In \fig{\ref{subfig:intro_model1}} we plot the average waiting time over $100$ experiments when considering Model~1. %
In \fig{\ref{subfig:intro_model2}} we plot the average waiting time over $100$ experiments when considering Model~2. %
We observe that for both models RPM3 outperforms the scheme of~\cite{kakar2018rate} in a heterogeneous environment. However, when the workers of different clusters have similar service time, RPM3 outperforms the scheme of~\cite{kakar2018rate} only in the more accurate model~2.
In \fig{\ref{subfig:intro_bound}} we plot the theoretical bounds on the mean waiting time~\eqref{eq:avg_waiting_model1} of RPM3 under model~1 derived in Theorem~\ref{thm:waiting_exp} and the empirical average waiting time of RPM3. We observe that the bound is a good representation of the empirical average waiting time.} 
\label{fig:intro}
\end{figure*}
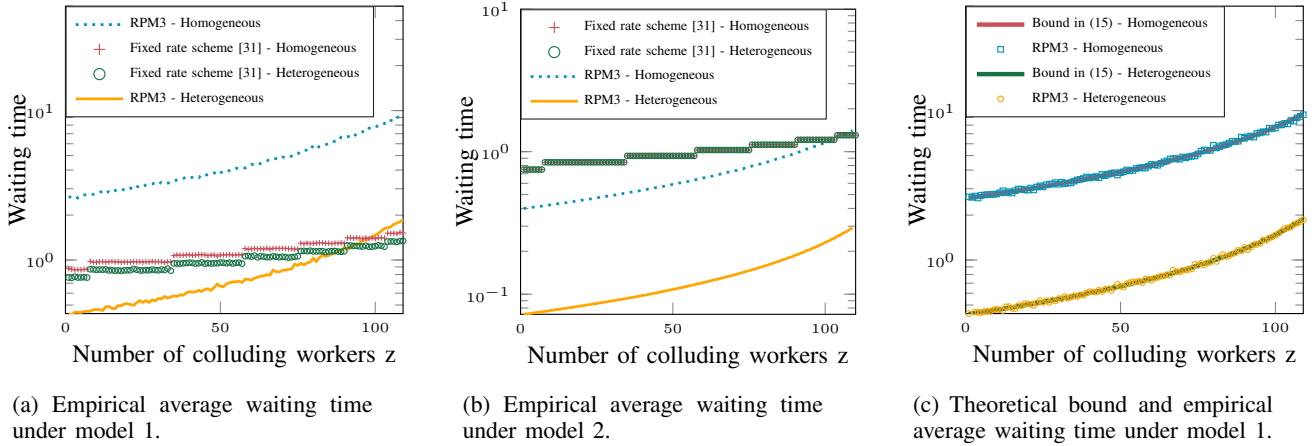

\subsection{Model of the service time} 
\emph{Workers model:} The time spent by a worker to compute a task is a shifted exponential random variable with shift $s$ and rate $\lambda$, i.e., the probability density function of the service (upload, compute and download) time is given by \cite{TOFEC,speeding,9174030,BPR17}
\begin{equation*}
    \mathrm{Pr}_{\rvT}(x) = \lambda \exp {(-\lambda(x - s))}, \text{ if } x > s, \text{ and } 0 \text{ otherwise.}
\end{equation*}
The service time includes the time spent to send the task from the master to the worker, the computation time at the worker and the time spent to send the result back from the worker to the master.

Following the literature, we make the following assumption. Let $\Pr_{\rvT}(x)$ be the probability density function of the time spent to compute the whole product $\mA\mB$ at one worker, we assume that the time spent to compute a fraction $1/mk$ of the whole computation follows a scaled distribution of $\Pr_{\rvT}(x)$ where the shift $s$ becomes $s/mk$ and the rate becomes $\lambda m k$.

In our delay analysis of RPM3, we consider two different models to account for sending several tasks from the master to the workers. Model~1 is rather simplistic and treats all the tasks sent to a worker as one large task. However, it provides significant engineering insights. Model~2 is more accurate and treats all tasks separately.

{\noindent{\em \textbf{Model 1: Worker-dependent fixed service time.}} In this model, we assume that the time spent to compute a task at worker~$w_i$ in cluster $u$, $i=1,\dots,n_u$, is fixed to $\rvY_i^u$ throughout the whole algorithm. We take $\rvY_i^u$ to be a random variable that is exponentially distributed with mean $\lambda_u km$, for a given $\lambda_u\in \mathbb{R}$. In addition, for each task there is an initial handshake time $s_u/km$, for a given $s_u\in \mathbb{R}$, after which the worker computes the task. Therefore, for each worker $w_i$ in cluster $u$, the time spent to compute $\tau_u$ tasks under this model is $\rvT_u^i = \tau_u s_u /km + \tau_u \rvY_i^u$ for all $i=1,\dots,n_u$. That is, the random variable $\rvT_u^i$ is a shifted exponential random variable with shift $\tau_u s_u /km$ and rate $\lambda_u km/\tau_u$. The random variable $\rvY_i^u$ is different for every worker; hence the name of the model.}

{\noindent{\em \textbf{Model 2: Cluster-dependent additive service time.}} In this model, we assume that the time spent to compute a task at worker $w_i$, $i=1,\dots,n_u$, in cluster $u$ at a given round is a shifted exponential random variable with shift $s_u/km$ and rate $\lambda_u km$, $s_u, \lambda_u \in \mathbb{R}$. Namely, at a given round, the service time of a worker depends only on the cluster $u$ to which he belongs. Therefore, for each worker $w_i$ in cluster $u$, the time spent to compute $\tau_u$ tasks under this model is $\rvT_u^i = \sum_{j=1}^{\tau_i} \rvX_j$ where $\rvX_1,\dots,\rvX_{\tau_u}$ are \emph{iid} random variables following the shifted exponential distribution with shift $s_u/km$ and rate $\lambda_u km$. That is, the random variable $\rvT_u^i$ follows a shifted Erlang distribution with shape $\tau_u$ and rate $\lambda_u km$. The shape $\tau_u$ of the Erlang random variable is cluster-dependent; hence the name of the model.}


%% file: Figures/Intro_model1.tex
\definecolor{mycolor1}{rgb}{0.0, 0.58, 0.71}
\definecolor{mycolor2}{rgb}{1.0, 0.65, 0.0}
\definecolor{mycolor3}{rgb}{0.8, 0.31, 0.36}
\definecolor{mycolor4}{rgb}{0.09, 0.45, 0.27}

\begin{tikzpicture}

\begin{axis}[%
width=0.951\figurewidth,
height=\figureheight,
at={(0\figurewidth,0\figureheight)},
scale only axis,
xmin=0,
xmax=109,
xlabel style={font=\color{white!15!black}, font =\small, yshift = 0.25cm},
xlabel={Number of colluding workers z},
ymode=log,
ymin=0.438362667605437,
ymax=50,
yminorticks=true,
ylabel style={font=\color{white!15!black}, font =\small, yshift=-0.65cm},
tick label style={font=\tiny},
ylabel={Waiting time}, 
axis background/.style={fill=white},
legend style={legend cell align=left, align=left, draw=white!15!black, font =\tiny, at ={(0,1)}, anchor = north west}
]

\addlegendimage{color=mycolor1, dotted, line width = 1 pt};
\addlegendentry{RPM3 - Homogeneous}

\addlegendimage{color=mycolor3, only marks, mark=+, mark size = 2pt};
\addlegendentry{Fixed rate scheme~\cite{kakar2018rate} - Homogeneous};

\addlegendimage{color=mycolor4, only marks, mark=o, mark size = 2pt};
\addlegendentry{Fixed rate scheme~\cite{kakar2018rate} - Heterogeneous};

\addlegendimage{color=mycolor2, solid, line width = 1 pt};
\addlegendentry{RPM3 - Heterogeneous};

\addplot [color=mycolor1, mark=none, mark options={solid, mycolor1}, mark size = 1pt, line width = 1pt, dotted ]
   table[row sep=crcr]{%
1	2.64981021526051\\
2	2.62880187917758\\
3	2.68175285590732\\
4	2.60199416388582\\
5	2.70631668735611\\
6	2.72820421269209\\
7	2.73595745139773\\
8	2.74780556468997\\
9	2.75175591809391\\
10	2.76550267201894\\
11	2.82064461339855\\
12	2.90731007625124\\
13	2.81653297545948\\
14	2.87028285747134\\
15	2.80969349851864\\
16	2.91006456888509\\
17	2.98798061827175\\
18	2.97746881443346\\
19	2.94953071519945\\
20	2.93906401572888\\
21	3.03377673148333\\
22	2.94856702857895\\
23	3.07790710873341\\
24	3.1365899121219\\
25	3.08744859761023\\
26	3.18417492059472\\
27	3.21851938557707\\
28	3.22581920783744\\
29	3.2992242565346\\
30	3.21671816056814\\
31	3.31322871360057\\
32	3.24849458645311\\
33	3.27648162554987\\
34	3.29020321500755\\
35	3.37321744480175\\
36	3.43674680994762\\
37	3.46594552800184\\
38	3.51044070270294\\
39	3.59897281128164\\
40	3.53877073287988\\
41	3.57778947657727\\
42	3.61738516239577\\
43	3.55429250804843\\
44	3.73575769998943\\
45	3.66990227066825\\
46	3.74825940578157\\
47	3.79737116373415\\
48	3.84083188703877\\
49	3.70815562217224\\
50	3.84256721127915\\
51	3.90227152355229\\
52	4.01013703334107\\
53	3.95340140324411\\
54	3.97889886040311\\
55	4.09734688697814\\
56	4.13256328912186\\
57	4.16175463136977\\
58	4.12597537184908\\
59	4.24812523054642\\
60	4.36322649468651\\
61	4.33515129805725\\
62	4.30797048000216\\
63	4.43565765389892\\
64	4.54226653957916\\
65	4.60089138224536\\
66	4.74699046016749\\
67	4.75801237745882\\
68	4.77285270835812\\
69	4.72844489507639\\
70	4.88943321824746\\
71	4.83055561217075\\
72	4.90632023775044\\
73	4.95814489678511\\
74	4.96924797551685\\
75	5.12606955123641\\
76	5.35569106761249\\
77	5.44812109906841\\
78	5.42965023012641\\
79	5.41641476252975\\
80	5.85130962106556\\
81	5.62531299147585\\
82	5.7472757134105\\
83	5.85792303534535\\
84	5.9155890500043\\
85	6.03550504038336\\
86	6.22198487478911\\
87	6.21881497716803\\
88	6.24417991186473\\
89	6.64369568094623\\
90	6.53092929070135\\
91	6.67191377736543\\
92	6.76185624108436\\
93	6.64478331909053\\
94	7.02651050489631\\
95	7.09703479795951\\
96	7.2264327538551\\
97	7.18417193115282\\
98	7.62561576583249\\
99	7.60262686501089\\
100	7.65631536206243\\
101	8.01841673949865\\
102	8.02878489331865\\
103	8.2074975892786\\
104	8.14316955555644\\
105	8.58106172560717\\
106	8.50679475619414\\
107	8.95189102619333\\
108	9.14306793953416\\
109	9.39065963854684\\
};

\addplot [color=mycolor2, mark=none, mark options={solid, mycolor2}, mark size = 1pt, line width = 1pt]
  table[row sep=crcr]{%
1	0.438362667605437\\
2	0.429542125907438\\
3	0.4485066970975\\
4	0.446872319837296\\
5	0.44577371114046\\
6	0.452081482166505\\
7	0.454058688677409\\
8	0.454571054890661\\
9	0.465365065076253\\
10	0.467489397178146\\
11	0.465226475235491\\
12	0.459768677588848\\
13	0.486371338519949\\
14	0.488312016931621\\
15	0.475147658015721\\
16	0.473327853789495\\
17	0.496012009586637\\
18	0.493442642189989\\
19	0.509234953121497\\
20	0.509675979181333\\
21	0.504411320616168\\
22	0.497662507571395\\
23	0.518389605697941\\
24	0.534854941370773\\
25	0.520439406932147\\
26	0.534373510576465\\
27	0.520559172077219\\
28	0.541425293193153\\
29	0.533027134988294\\
30	0.552861914513089\\
31	0.550269490221019\\
32	0.570071608046356\\
33	0.556380242034015\\
34	0.565406320542248\\
35	0.576429099121181\\
36	0.571132956678732\\
37	0.585643180921689\\
38	0.581974326786339\\
39	0.60189553798915\\
40	0.61385116348438\\
41	0.612545512275636\\
42	0.610923896052482\\
43	0.628112105733589\\
44	0.616098788298847\\
45	0.637088344644633\\
46	0.642160045104582\\
47	0.626460207473605\\
48	0.655824165011707\\
49	0.689333970511497\\
50	0.668459180754596\\
51	0.662446334429014\\
52	0.693779444031541\\
53	0.705876028814092\\
54	0.715238443273996\\
55	0.713948543785817\\
56	0.723407627060655\\
57	0.706773833496394\\
58	0.74459942731525\\
59	0.744779299077716\\
60	0.732133554829519\\
61	0.756866216348889\\
62	0.766175682630512\\
63	0.77959531914591\\
64	0.782760791703589\\
65	0.798228979754516\\
66	0.792348889876669\\
67	0.855662125186119\\
68	0.835805810506846\\
69	0.832430744272724\\
70	0.852225835240183\\
71	0.857984477184297\\
72	0.865444317480912\\
73	0.871869183731368\\
74	0.937554869165771\\
75	0.920315023836095\\
76	0.951960711689018\\
77	0.92668286931046\\
78	0.956424066119576\\
79	0.968829671420866\\
80	1.02771654050934\\
81	0.978416283278185\\
82	1.03788103416805\\
83	1.08581786558246\\
84	1.05717666888008\\
85	1.07328203104535\\
86	1.10001987820158\\
87	1.1161938738178\\
88	1.16311211466122\\
89	1.15111368320011\\
90	1.2172934986428\\
91	1.1736685050299\\
92	1.21969297032856\\
93	1.26800794787698\\
94	1.27313391595645\\
95	1.29157291904929\\
96	1.36451388390494\\
97	1.35380171721583\\
98	1.40001254403882\\
99	1.42817588033128\\
100	1.46744267079365\\
101	1.50867926865824\\
102	1.56379938667266\\
103	1.62872139294354\\
104	1.64103606466724\\
105	1.66722068334357\\
106	1.73210522392929\\
107	1.76826176500397\\
108	1.80329245008804\\
109	1.85833783285227\\
};

\addplot [color=mycolor3, only marks, mark=+, mark options={solid, mycolor3}, mark size = 1pt]
  table[row sep=crcr]{%
1	0.879310799492365\\
2	0.865950385076752\\
3	0.859563154650205\\
4	0.862585421407764\\
5	0.861084301484378\\
6	0.865197351494956\\
7	0.868997648115174\\
8	0.980053492508223\\
9	0.966498359398763\\
10	0.967830443879773\\
11	0.975112023951606\\
12	0.974367890227269\\
13	0.970891118198766\\
14	0.971973662635536\\
15	0.979667186671699\\
16	0.972117823421057\\
17	0.974724047856215\\
18	0.965692596651316\\
19	0.97212112334427\\
20	0.968987314068088\\
21	0.968645293532332\\
22	0.975760643570873\\
23	0.971735241436995\\
24	0.974831525929003\\
25	0.96914130364626\\
26	0.976944438162248\\
27	0.970186959885842\\
28	0.980026593864967\\
29	0.972198052882721\\
30	0.973138478529255\\
31	0.973275949834295\\
32	0.962927875629078\\
33	0.973657774692276\\
34	0.971164027479141\\
35	1.07013379280142\\
36	1.08234728097931\\
37	1.0751837345991\\
38	1.07677291881841\\
39	1.08195527817028\\
40	1.0844036957972\\
41	1.07776574680297\\
42	1.07870794327193\\
43	1.08554846146426\\
44	1.08451080896833\\
45	1.08463598972413\\
46	1.07037900956961\\
47	1.08045401834197\\
48	1.08193357195218\\
49	1.06389241899029\\
50	1.08204969658426\\
51	1.08247935733099\\
52	1.08141343704405\\
53	1.08958331846101\\
54	1.07976335486538\\
55	1.08169369253879\\
56	1.0879446553268\\
57	1.08145308492864\\
58	1.19401381241364\\
59	1.18282719081211\\
60	1.18691766571208\\
61	1.19590602352342\\
62	1.18864039440011\\
63	1.19083354541058\\
64	1.1858021707088\\
65	1.1977968114747\\
66	1.18251310419583\\
67	1.19661402806221\\
68	1.1972401057646\\
69	1.19670219097261\\
70	1.19697012939462\\
71	1.19562700157455\\
72	1.19169910289445\\
73	1.18923459923175\\
74	1.18227252847568\\
75	1.18540933034962\\
76	1.29563560049909\\
77	1.28598212911549\\
78	1.29753724164464\\
79	1.28783434475323\\
80	1.29428789607114\\
81	1.29738489655096\\
82	1.30343943329321\\
83	1.30284893514175\\
84	1.29396486421914\\
85	1.29591701240944\\
86	1.28864172542306\\
87	1.29605447355336\\
88	1.29380815466875\\
89	1.30377004477274\\
90	1.29498799717056\\
91	1.4075642665353\\
92	1.40893686430264\\
93	1.39750319497015\\
94	1.4101232030834\\
95	1.40135413803702\\
96	1.38831255460668\\
97	1.4093278158397\\
98	1.39149907805992\\
99	1.40506526604644\\
100	1.39119261981918\\
101	1.41150923867863\\
102	1.40789762363677\\
103	1.40323255683446\\
104	1.51277850527911\\
105	1.50631628739772\\
106	1.51192523359519\\
107	1.50001116432069\\
108	1.53072697013123\\
109	1.51995149540916\\
110	1.51934138131265\\
111	1.50031911949527\\
112	1.52110210608712\\
113	1.51668418178114\\
114	1.50544244062426\\
115	1.60727091192059\\
116	1.63327477121505\\
117	1.60799339317126\\
118	1.62952643542755\\
119	1.62387166082317\\
120	1.62171445696596\\
121	1.62483250199515\\
122	1.62693592315339\\
123	1.60970968898583\\
124	1.73264466233216\\
125	1.73924979150141\\
126	1.74319284167785\\
127	1.72973003132712\\
128	1.73519587247569\\
129	1.71849586800672\\
130	1.72823352823808\\
131	1.72892854696342\\
132	1.71135793226496\\
133	1.83426473380515\\
134	1.84081060164517\\
135	1.8319683362415\\
136	1.82972602671237\\
137	1.83957497749535\\
138	1.85155012584904\\
139	1.84637094238066\\
140	1.94965138107208\\
141	1.9468649004554\\
142	1.93063378824433\\
143	1.94486703959567\\
144	1.94565844290292\\
145	1.94266200039272\\
146	2.07643507044523\\
147	2.06230131258937\\
148	2.06880217609037\\
149	2.0458135959573\\
150	2.06715098790112\\
151	2.06690695602669\\
152	2.16292118449178\\
153	2.15084668519641\\
154	2.15362471956585\\
155	2.15912341227348\\
156	2.1564367563536\\
157	2.16946726011422\\
158	2.2676587206627\\
159	2.29162268884064\\
160	2.25583721391151\\
161	2.26206513753742\\
162	2.35940006382086\\
163	2.39058751314201\\
164	2.37931147769333\\
165	2.36889826288184\\
166	2.38607955952732\\
167	2.47580399567975\\
168	2.49457600395347\\
169	2.47458428284139\\
170	2.46745433358111\\
171	2.59927552769384\\
172	2.60742203506519\\
173	2.57994432514923\\
174	2.70799139297643\\
175	2.7147245071626\\
176	2.72723378958895\\
177	2.69970007427533\\
178	2.83049230113194\\
179	2.80945076298366\\
180	2.82409654555684\\
181	2.90321878323189\\
182	2.90639812462445\\
183	3.028722123821\\
184	3.01929253174602\\
185	3.01865667755717\\
186	3.13633071877338\\
187	3.12175474353605\\
188	3.14506769516198\\
189	3.22582336506195\\
190	3.23928546278525\\
191	3.37220293197054\\
192	3.35454924147435\\
193	3.46341538972957\\
194	3.44892364604957\\
195	3.56148973030689\\
196	3.57363177920106\\
197	3.68179081926872\\
198	3.66887688459925\\
199	3.80740503615691\\
200	3.91731663376451\\
201	3.88541834156997\\
202	3.97594771765232\\
203	4.09277054300325\\
204	4.10637108892888\\
205	4.19438575199192\\
206	4.30399047773743\\
207	4.34300867016463\\
208	4.43264476995002\\
209	4.54872774697073\\
210	4.63787515721101\\
211	4.76555574094998\\
212	4.86572104974515\\
213	4.96318991197106\\
214	5.05952130298364\\
215	5.15071310536721\\
216	5.29520831469565\\
217	5.42057077188134\\
218	5.6805329466969\\
219	5.72496576066828\\
220	5.9818848691356\\
221	6.03176487646831\\
222	6.28353673393007\\
223	6.36221102624498\\
224	6.54840758784422\\
225	6.79630318691604\\
226	7.07440154374865\\
227	7.17090801918308\\
228	7.52321225014604\\
229	7.80969041421808\\
230	8.06345401735089\\
231	8.30454730470528\\
232	8.62457733977346\\
233	9.08110957454905\\
234	9.34076812480192\\
235	9.9217859133018\\
236	10.4491107207634\\
237	10.806835947303\\
238	11.560614095069\\
239	12.0891242791514\\
240	12.8270522881517\\
241	13.6521128914738\\
242	14.4677871445783\\
243	15.5356971017637\\
244	16.8266202070907\\
245	18.2278102498266\\
246	19.5132127962011\\
247	21.5178706285991\\
248	23.8809879813617\\
249	27.2579345318869\\
250	31.0248423429328\\
251	35.5060826290653\\
252	43.3941005630055\\
253	53.8887035986303\\
254	71.8173913788893\\
255	109.48071270997\\
256	215.07164904932\\
};

\addplot [color=mycolor3, only marks, mark=o, mark options={solid, mycolor4}, mark size = 1pt]
  table[row sep=crcr]{%
1	0.763375912728432\\
2	0.759710514423163\\
3	0.772836775258338\\
4	0.764002022748855\\
5	0.765266540374356\\
6	0.767638086833723\\
7	0.763536077599703\\
8	0.867845697723851\\
9	0.866616042563247\\
10	0.860637823297394\\
11	0.863925481914477\\
12	0.862739315628761\\
13	0.865538157790476\\
14	0.852528490046849\\
15	0.854522111857067\\
16	0.849823777259183\\
17	0.852241900796817\\
18	0.848340973654875\\
19	0.853193044932312\\
20	0.854508002642754\\
21	0.854707332651197\\
22	0.871879076976538\\
23	0.861842986264457\\
24	0.852438843956144\\
25	0.852591082804575\\
26	0.855106532104441\\
27	0.854577384629191\\
28	0.863003450746603\\
29	0.857294955435274\\
30	0.854260136176278\\
31	0.864343133850721\\
32	0.857258653011654\\
33	0.875279951119076\\
34	0.848813748254946\\
35	0.951019509200043\\
36	0.947436062867624\\
37	0.946461025711445\\
38	0.951308217847093\\
39	0.955307009392025\\
40	0.959875284487428\\
41	0.957053245195442\\
42	0.946515108113832\\
43	0.95115936810187\\
44	0.965133069104929\\
45	0.946374214173125\\
46	0.952607632815452\\
47	0.947980227475836\\
48	0.951502220478408\\
49	0.952222006509388\\
50	0.962115864377146\\
51	0.959146096634362\\
52	0.951915912794546\\
53	0.960792711765281\\
54	0.947801037587487\\
55	0.957003175603927\\
56	0.961277673986182\\
57	0.945943877282371\\
58	1.06050974841385\\
59	1.06377692192996\\
60	1.05347930815732\\
61	1.07035573683182\\
62	1.04322485194029\\
63	1.05991907828321\\
64	1.05420700411856\\
65	1.04461797908527\\
66	1.04576599309763\\
67	1.05921190683229\\
68	1.05325193980057\\
69	1.04402746766981\\
70	1.05317835927116\\
71	1.04777029904069\\
72	1.0379792611226\\
73	1.0443884978413\\
74	1.05051299986778\\
75	1.06020121765895\\
76	1.15012069362737\\
77	1.15120602579702\\
78	1.14589747343256\\
79	1.15473518327684\\
80	1.14294536698395\\
81	1.1425647668089\\
82	1.14698247647102\\
83	1.14682469641177\\
84	1.13154187382727\\
85	1.14204181654017\\
86	1.14564531085159\\
87	1.14890154182329\\
88	1.13994829655719\\
89	1.13657113601503\\
90	1.14349724893679\\
91	1.25301442806729\\
92	1.25039938026637\\
93	1.24389696671633\\
94	1.24301312662152\\
95	1.23473473769425\\
96	1.23129861740062\\
97	1.24619587823634\\
98	1.22793948230619\\
99	1.22903267220411\\
100	1.23926058937003\\
101	1.25511520427971\\
102	1.24417262849722\\
103	1.22839079700477\\
104	1.32986388727522\\
105	1.33156416736873\\
106	1.33195442286975\\
107	1.32094524611208\\
108	1.3392402215088\\
109	1.34359622637916\\
110	1.34335666036199\\
111	1.33433463278195\\
112	1.33554158392638\\
113	1.32735907659997\\
114	1.33860940518125\\
115	1.41990536951073\\
116	1.44262042991234\\
117	1.4276403758472\\
118	1.42588551285798\\
119	1.42990270126787\\
120	1.42701199609788\\
121	1.42148802158891\\
122	1.43326596729534\\
123	1.43718693501454\\
124	1.52567610799041\\
125	1.52426069521736\\
126	1.51538788345122\\
127	1.51771805479426\\
128	1.53020280870045\\
129	1.52364005746599\\
130	1.52331575306666\\
131	1.53455289148738\\
132	1.51937555345502\\
133	1.6379355965443\\
134	1.60109145521983\\
135	1.62442848786636\\
136	1.61783805043913\\
137	1.62005100936013\\
138	1.62035492715111\\
139	1.61472605906497\\
140	1.71089432814304\\
141	1.71859974084756\\
142	1.71299122970713\\
143	1.72654531230403\\
144	1.70754304855809\\
145	1.71441173620863\\
146	1.83272210347763\\
147	1.81798453567062\\
148	1.81205651945255\\
149	1.82398142387919\\
150	1.82477043040602\\
151	1.82753960759941\\
152	1.89986956362326\\
153	1.91425944094854\\
154	1.92194694652004\\
155	1.90613196414739\\
156	1.90889549501867\\
157	1.92624985341788\\
158	1.98205188930666\\
159	2.00422814617294\\
160	1.99683812905607\\
161	2.00322314852879\\
162	2.09102367338731\\
163	2.13114051588005\\
164	2.10643480694831\\
165	2.08298617389203\\
166	2.10464649141565\\
167	2.21289327971634\\
168	2.19412240043835\\
169	2.18208292812091\\
170	2.17182083130255\\
171	2.27601702209054\\
172	2.27490744101496\\
173	2.29284035127136\\
174	2.38626793325672\\
175	2.40443947373077\\
176	2.37587358893907\\
177	2.39691275333682\\
178	2.50529592826001\\
179	2.48597490726347\\
180	2.50118414304857\\
181	2.57053980298048\\
182	2.5714045797779\\
183	2.65844555186785\\
184	2.68000476940808\\
185	2.67672567608001\\
186	2.77345179861314\\
187	2.79683445652376\\
188	2.75300388933102\\
189	2.88019528928524\\
190	2.79863149009639\\
191	2.96655309702827\\
192	2.9708095983042\\
193	3.07569718621403\\
194	3.06143113123328\\
195	3.15738627080871\\
196	3.16926769966176\\
197	3.27658653459868\\
198	3.25449236063714\\
199	3.3514172262007\\
200	3.42859802696068\\
201	3.42507406664027\\
202	3.53817811709883\\
203	3.59518370390477\\
204	3.64845482188182\\
205	3.70001930303728\\
206	3.81863918684923\\
207	3.8003459245813\\
208	3.8908236072925\\
209	3.99724035289179\\
210	4.10611230093566\\
211	4.14778336539574\\
212	4.32903217381811\\
213	4.42498075583549\\
214	4.51115622630925\\
215	4.54301094483467\\
216	4.69830357198442\\
217	4.78128979370449\\
218	4.94395170883727\\
219	5.00082425641603\\
220	5.22371526279043\\
221	5.37182074468053\\
222	5.53247564726795\\
223	5.59952260750684\\
224	5.85999713761211\\
225	6.00359647138191\\
226	6.22775892064466\\
227	6.38918220360292\\
228	6.61326781802256\\
229	6.8764424499927\\
230	7.12851607499402\\
231	7.33004668792487\\
232	7.55016777671517\\
233	7.99698000522026\\
234	8.35655830086399\\
235	8.63364838321096\\
236	9.11782288255568\\
237	9.50427901955212\\
238	10.0772448803329\\
239	10.6856830805068\\
240	11.2505830970629\\
241	11.9606013170954\\
242	12.8173433634176\\
243	13.736731291261\\
244	14.7141494509354\\
245	15.8373542316498\\
246	17.4351441812248\\
247	19.1261637157905\\
248	21.2745427337596\\
249	23.7618580629396\\
250	27.1980506477164\\
251	31.5843758345921\\
252	37.7972350646249\\
253	47.4842665613613\\
254	63.0181905186068\\
255	95.6025054840482\\
256	191.510553686552\\
};

\end{axis}
\end{tikzpicture}%

%% file: Figures/Intro_model2.tex
\definecolor{mycolor1}{rgb}{0.0, 0.58, 0.71}
\definecolor{mycolor2}{rgb}{1.0, 0.65, 0.0}
\definecolor{mycolor3}{rgb}{0.8, 0.31, 0.36}
\definecolor{mycolor4}{rgb}{0.09, 0.45, 0.27}

\begin{tikzpicture}

\begin{axis}[%
width=0.951\figurewidth,
height=\figureheight,
at={(0\figurewidth,0\figureheight)},
scale only axis,
xmin=0,
xmax=110,
xlabel style={font=\color{white!15!black}, font =\small, yshift = 0.25cm},
xlabel={Number of colluding workers z},
ymode=log,
ymin=0.0722242876699268,
ymax=10,
yminorticks=true,
tick label style={font=\tiny},
ylabel style={font=\color{white!15!black},font =\small, yshift=-0.55cm  
},
ylabel={Waiting time}, 
axis background/.style={fill=white},
legend style={legend cell align=left, align=left, draw=white!15!black, font =\tiny, at ={(0,1)}, anchor = north west}
]

\addlegendimage{color=mycolor3, only marks, mark=+, mark size = 2pt};
\addlegendentry{Fixed rate scheme~\cite{kakar2018rate} - Homogeneous};

\addlegendimage{color=mycolor4, only marks, mark=o, mark size = 2pt};
\addlegendentry{Fixed rate scheme~\cite{kakar2018rate} - Heterogeneous};

\addlegendimage{color=mycolor1, dotted, line width = 1 pt};
\addlegendentry{RPM3 - Homogeneous}

\addlegendimage{color=mycolor2, solid, line width = 1 pt};
\addlegendentry{RPM3 - Heterogeneous};

\addplot [color=mycolor1, dotted, mark=none, thin, mark size=1pt, mark options={solid, mycolor1}, line width = 1pt]
 table[row sep=crcr]{%
1	0.398797505877992\\
2	0.402555576149379\\
3	0.404096888679591\\
4	0.407953668023388\\
5	0.409666224340223\\
6	0.413634058802602\\
7	0.415112886931128\\
8	0.419189788120906\\
9	0.420690177914351\\
10	0.425044082510061\\
11	0.426669296389991\\
12	0.43098464853228\\
13	0.432712902775327\\
14	0.4371483040391\\
15	0.438838273026781\\
16	0.443340602442795\\
17	0.445190933327267\\
18	0.450013351353087\\
19	0.451922011694236\\
20	0.456941271908463\\
21	0.458865548546028\\
22	0.463868345089385\\
23	0.465933337654212\\
24	0.470948646498484\\
25	0.472943069554714\\
26	0.478386308735325\\
27	0.48047137654007\\
28	0.486184500399961\\
29	0.488334459336137\\
30	0.494160781327209\\
31	0.496441851770145\\
32	0.502222817264471\\
33	0.50464972194514\\
34	0.510756062759285\\
35	0.513130015873959\\
36	0.519375965887215\\
37	0.521883694722551\\
38	0.528532808915061\\
39	0.531089696572224\\
40	0.537941668223232\\
41	0.540504105946518\\
42	0.547656547646788\\
43	0.550532670636309\\
44	0.557798999023488\\
45	0.560571416258833\\
46	0.568370281494652\\
47	0.571371371030846\\
48	0.579202358686338\\
49	0.582418781033966\\
50	0.590499838477102\\
51	0.593865144027299\\
52	0.602320460055623\\
53	0.605682485444778\\
54	0.614783382381929\\
55	0.618358113907632\\
56	0.627473509081678\\
57	0.631167119926951\\
58	0.641070243156672\\
59	0.644814617724145\\
60	0.655026631426341\\
61	0.659154369451873\\
62	0.669712406752134\\
63	0.673900140737516\\
64	0.684944846540484\\
65	0.689167016452883\\
66	0.700803605170963\\
67	0.705475019912782\\
68	0.717640127572539\\
69	0.722691866715767\\
70	0.735711346281806\\
71	0.740710912824958\\
72	0.75399753984974\\
73	0.759150618043039\\
74	0.773354082484491\\
75	0.778925323484388\\
76	0.793901522336433\\
77	0.79983953963735\\
78	0.815604294598659\\
79	0.821700248841216\\
80	0.838642970280036\\
81	0.845267121238903\\
82	0.862854581188277\\
83	0.869877063286925\\
84	0.8883866314647\\
85	0.895942232549266\\
86	0.916045702936263\\
87	0.923860108287978\\
88	0.944679938981879\\
89	0.953348157847422\\
90	0.976191712548126\\
91	0.985134187119625\\
92	1.00940157482201\\
93	1.01893485601343\\
94	1.04467906862935\\
95	1.0549596859627\\
96	1.08285855508889\\
97	1.09422832790629\\
98	1.12396367552123\\
99	1.13615131926717\\
100	1.16846625124633\\
101	1.18140418799808\\
102	1.21676597710386\\
103	1.23080071235025\\
104	1.26920454547392\\
105	1.28435471989672\\
106	1.32566565779357\\
107	1.34279497573684\\
108	1.38852122354887\\
109	1.40711207353609\\
};

\addplot [color=mycolor2, mark=none, mark options={solid, mycolor2}, mark size = 1pt, line width = 1pt]
  table[row sep=crcr]{%
1	0.0722242876699268\\
2	0.0728455770383496\\
3	0.0730254972545038\\
4	0.0738194281871396\\
5	0.0742055896543366\\
6	0.0748277496668231\\
7	0.0752228158225371\\
8	0.0757974273778697\\
9	0.076170087206631\\
10	0.0769640445867082\\
11	0.0773722494264231\\
12	0.0779710700544566\\
13	0.0785308224107284\\
14	0.0791053557296572\\
15	0.0796794853436661\\
16	0.0802331482801014\\
17	0.0808348889577849\\
18	0.0816279403672203\\
19	0.0820583601976449\\
20	0.0828622008140717\\
21	0.0834569640028979\\
22	0.0842057908105111\\
23	0.084600295367134\\
24	0.0855348496413465\\
25	0.0859301967851714\\
26	0.0869639058809019\\
27	0.0873773387008448\\
28	0.0882937422815934\\
29	0.0888474119375729\\
30	0.0898098455922475\\
31	0.0903498803679849\\
32	0.0912870312565677\\
33	0.0918542923781244\\
34	0.092775428431474\\
35	0.0933448342116971\\
36	0.0945466294691716\\
37	0.095083813760873\\
38	0.0960166239345186\\
39	0.0968173304961924\\
40	0.0979477992800906\\
41	0.0985089355247137\\
42	0.0997015448602476\\
43	0.100466339686425\\
44	0.101616859546032\\
45	0.102404603289399\\
46	0.103800229025286\\
47	0.104612660415736\\
48	0.105995302299001\\
49	0.106794681277755\\
50	0.108176574840051\\
51	0.108891625912196\\
52	0.110442332878395\\
53	0.111194756440637\\
54	0.11278169184568\\
55	0.11377024677877\\
56	0.11528661354545\\
57	0.116313868526745\\
58	0.118104900497612\\
59	0.119068870670654\\
60	0.120744268016815\\
61	0.121676602769711\\
62	0.123676276169548\\
63	0.124795507545306\\
64	0.126683661020161\\
65	0.127803900999703\\
66	0.129783665405981\\
67	0.131072016622578\\
68	0.1332036581747\\
69	0.134526333918887\\
70	0.13663250480384\\
71	0.137945751525793\\
72	0.140473027539685\\
73	0.141835047434473\\
74	0.144446291639559\\
75	0.145935347713409\\
76	0.148546918936699\\
77	0.150045739316082\\
78	0.153039303073977\\
79	0.154703999973067\\
80	0.157700539287367\\
81	0.159687152231696\\
82	0.162941969263496\\
83	0.164864542621741\\
84	0.16836775080219\\
85	0.170477301428916\\
86	0.174361611381579\\
87	0.176531710748142\\
88	0.180579272069072\\
89	0.182749802012817\\
90	0.187071136475288\\
91	0.189700568967051\\
92	0.194369209445302\\
93	0.197185790347861\\
94	0.202421615458041\\
95	0.20544843385189\\
96	0.210977570747765\\
97	0.214351844650093\\
98	0.220350095882562\\
99	0.224176786316515\\
100	0.230892935798299\\
101	0.235013294176888\\
102	0.242220418144878\\
103	0.246706956557019\\
104	0.255117675724155\\
105	0.260060900609735\\
106	0.269060387257996\\
107	0.2744747873944\\
108	0.28451936300092\\
109	0.290809156417885\\
};

\addplot [color=mycolor3, only marks, mark=+, mark options={solid, mycolor3}, mark size = 1pt]
  table[row sep=crcr]{%
1	0.749731321362557\\
2	0.749731321362557\\
3	0.749731321362557\\
4	0.749731321362557\\
5	0.749731321362557\\
6	0.749731321362557\\
7	0.749731321362557\\
8	0.842405484128658\\
9	0.842405484128658\\
10	0.842405484128658\\
11	0.842405484128658\\
12	0.842405484128658\\
13	0.842405484128658\\
14	0.842405484128658\\
15	0.842405484128658\\
16	0.842405484128658\\
17	0.842405484128658\\
18	0.842405484128658\\
19	0.842405484128658\\
20	0.842405484128658\\
21	0.842405484128658\\
22	0.842405484128658\\
23	0.842405484128658\\
24	0.842405484128658\\
25	0.842405484128658\\
26	0.842405484128658\\
27	0.842405484128658\\
28	0.842405484128658\\
29	0.842405484128658\\
30	0.842405484128658\\
31	0.842405484128658\\
32	0.842405484128658\\
33	0.842405484128658\\
34	0.842405484128658\\
35	0.934890325127308\\
36	0.934890325127308\\
37	0.934890325127308\\
38	0.934890325127308\\
39	0.934890325127308\\
40	0.934890325127308\\
41	0.934890325127308\\
42	0.934890325127308\\
43	0.934890325127308\\
44	0.934890325127308\\
45	0.934890325127308\\
46	0.934890325127308\\
47	0.934890325127308\\
48	0.934890325127308\\
49	0.934890325127308\\
50	0.934890325127308\\
51	0.934890325127308\\
52	0.934890325127308\\
53	0.934890325127308\\
54	0.934890325127308\\
55	0.934890325127308\\
56	0.934890325127308\\
57	0.934890325127308\\
58	1.02742027727898\\
59	1.02742027727898\\
60	1.02742027727898\\
61	1.02742027727898\\
62	1.02742027727898\\
63	1.02742027727898\\
64	1.02742027727898\\
65	1.02742027727898\\
66	1.02742027727898\\
67	1.02742027727898\\
68	1.02742027727898\\
69	1.02742027727898\\
70	1.02742027727898\\
71	1.02742027727898\\
72	1.02742027727898\\
73	1.02742027727898\\
74	1.02742027727898\\
75	1.02742027727898\\
76	1.11985318588893\\
77	1.11985318588893\\
78	1.11985318588893\\
79	1.11985318588893\\
80	1.11985318588893\\
81	1.11985318588893\\
82	1.11985318588893\\
83	1.11985318588893\\
84	1.11985318588893\\
85	1.11985318588893\\
86	1.11985318588893\\
87	1.11985318588893\\
88	1.11985318588893\\
89	1.11985318588893\\
90	1.11985318588893\\
91	1.2127648161775\\
92	1.2127648161775\\
93	1.2127648161775\\
94	1.2127648161775\\
95	1.2127648161775\\
96	1.2127648161775\\
97	1.2127648161775\\
98	1.2127648161775\\
99	1.2127648161775\\
100	1.2127648161775\\
101	1.2127648161775\\
102	1.2127648161775\\
103	1.2127648161775\\
104	1.30514618967008\\
105	1.30514618967008\\
106	1.30514618967008\\
107	1.30514618967008\\
108	1.30514618967008\\
109	1.30514618967008\\
110	1.30514618967008\\
111	1.30514618967008\\
112	1.30514618967008\\
113	1.30514618967008\\
114	1.30514618967008\\
115	1.3972467872374\\
116	1.3972467872374\\
117	1.3972467872374\\
118	1.3972467872374\\
119	1.3972467872374\\
120	1.3972467872374\\
121	1.3972467872374\\
122	1.3972467872374\\
123	1.3972467872374\\
124	1.48976383543461\\
125	1.48976383543461\\
126	1.48976383543461\\
127	1.48976383543461\\
128	1.48976383543461\\
129	1.48976383543461\\
130	1.48976383543461\\
131	1.48976383543461\\
132	1.48976383543461\\
133	1.58269179185286\\
134	1.58269179185286\\
135	1.58269179185286\\
136	1.58269179185286\\
137	1.58269179185286\\
138	1.58269179185286\\
139	1.58269179185286\\
140	1.67485336121274\\
141	1.67485336121274\\
142	1.67485336121274\\
143	1.67485336121274\\
144	1.67485336121274\\
145	1.67485336121274\\
146	1.76703255341535\\
147	1.76703255341535\\
148	1.76703255341535\\
149	1.76703255341535\\
150	1.76703255341535\\
151	1.76703255341535\\
152	1.85955404519911\\
153	1.85955404519911\\
154	1.85955404519911\\
155	1.85955404519911\\
156	1.85955404519911\\
157	1.85955404519911\\
158	1.95205001996333\\
159	1.95205001996333\\
160	1.95205001996333\\
161	1.95205001996333\\
162	2.04462306371799\\
163	2.04462306371799\\
164	2.04462306371799\\
165	2.04462306371799\\
166	2.04462306371799\\
167	2.13715541049757\\
168	2.13715541049757\\
169	2.13715541049757\\
170	2.13715541049757\\
171	2.22920598329332\\
172	2.22920598329332\\
173	2.22920598329332\\
174	2.32126184793004\\
175	2.32126184793004\\
176	2.32126184793004\\
177	2.32126184793004\\
178	2.41337356991074\\
179	2.41337356991074\\
180	2.41337356991074\\
181	2.50562464426544\\
182	2.50562464426544\\
183	2.59769347227863\\
184	2.59769347227863\\
185	2.59769347227863\\
186	2.68959683076906\\
187	2.68959683076906\\
188	2.68959683076906\\
189	2.78219280031205\\
190	2.78219280031205\\
191	2.87438524705382\\
192	2.87438524705382\\
193	2.96640376554957\\
194	2.96640376554957\\
195	3.05888283530292\\
196	3.05888283530292\\
197	3.15066806779468\\
198	3.15066806779468\\
199	3.24284291540182\\
200	3.33504620899404\\
201	3.33504620899404\\
202	3.42736200152417\\
203	3.51935327743723\\
204	3.51935327743723\\
205	3.61165602326887\\
206	3.70393402096591\\
207	3.70393402096591\\
208	3.79616505775071\\
209	3.88835696046899\\
210	3.98072902464795\\
211	4.07276268206215\\
212	4.16474034947654\\
213	4.25687662193995\\
214	4.34889781007742\\
215	4.44075110547329\\
216	4.53280302991874\\
217	4.62489940131541\\
218	4.80949027936095\\
219	4.90155098372066\\
220	5.08583141619017\\
221	5.17777134100427\\
222	5.36227387118886\\
223	5.45423326226269\\
224	5.63845526591634\\
225	5.82264119866887\\
226	6.00602394583151\\
227	6.19016360554938\\
228	6.37411138267844\\
229	6.64973743566913\\
230	6.92548827552279\\
231	7.1094475316889\\
232	7.38503789243435\\
233	7.75277456844769\\
234	8.02919878182696\\
235	8.39711987597035\\
236	8.85635599013144\\
237	9.22423685732075\\
238	9.77565963532219\\
239	10.3272000238849\\
240	10.8794263675221\\
241	11.5229564621127\\
242	12.3500644261108\\
243	13.1772364427073\\
244	14.1885034636865\\
245	15.383151575779\\
246	16.7600013224273\\
247	18.4139242575986\\
248	20.5273362771407\\
249	23.005815823797\\
250	26.3141899250284\\
251	30.720149002085\\
252	36.7808941016256\\
253	45.9610631653932\\
254	61.2897987211074\\
255	91.8533074333356\\
256	183.587577846564\\
};

\addplot [color=mycolor4, only marks, mark=o, mark options={solid, mycolor4}, mark size = 1pt]
 table[row sep=crcr]{%
1	0.749256730113756\\
2	0.749256730113756\\
3	0.749256730113756\\
4	0.749256730113756\\
5	0.749256730113756\\
6	0.749256730113756\\
7	0.749256730113756\\
8	0.842129023080965\\
9	0.842129023080965\\
10	0.842129023080965\\
11	0.842129023080965\\
12	0.842129023080965\\
13	0.842129023080965\\
14	0.842129023080965\\
15	0.842129023080965\\
16	0.842129023080965\\
17	0.842129023080965\\
18	0.842129023080965\\
19	0.842129023080965\\
20	0.842129023080965\\
21	0.842129023080965\\
22	0.842129023080965\\
23	0.842129023080965\\
24	0.842129023080965\\
25	0.842129023080965\\
26	0.842129023080965\\
27	0.842129023080965\\
28	0.842129023080965\\
29	0.842129023080965\\
30	0.842129023080965\\
31	0.842129023080965\\
32	0.842129023080965\\
33	0.842129023080965\\
34	0.842129023080965\\
35	0.934857110416997\\
36	0.934857110416997\\
37	0.934857110416997\\
38	0.934857110416997\\
39	0.934857110416997\\
40	0.934857110416997\\
41	0.934857110416997\\
42	0.934857110416997\\
43	0.934857110416997\\
44	0.934857110416997\\
45	0.934857110416997\\
46	0.934857110416997\\
47	0.934857110416997\\
48	0.934857110416997\\
49	0.934857110416997\\
50	0.934857110416997\\
51	0.934857110416997\\
52	0.934857110416997\\
53	0.934857110416997\\
54	0.934857110416997\\
55	0.934857110416997\\
56	0.934857110416997\\
57	0.934857110416997\\
58	1.02719995224394\\
59	1.02719995224394\\
60	1.02719995224394\\
61	1.02719995224394\\
62	1.02719995224394\\
63	1.02719995224394\\
64	1.02719995224394\\
65	1.02719995224394\\
66	1.02719995224394\\
67	1.02719995224394\\
68	1.02719995224394\\
69	1.02719995224394\\
70	1.02719995224394\\
71	1.02719995224394\\
72	1.02719995224394\\
73	1.02719995224394\\
74	1.02719995224394\\
75	1.02719995224394\\
76	1.11979343956886\\
77	1.11979343956886\\
78	1.11979343956886\\
79	1.11979343956886\\
80	1.11979343956886\\
81	1.11979343956886\\
82	1.11979343956886\\
83	1.11979343956886\\
84	1.11979343956886\\
85	1.11979343956886\\
86	1.11979343956886\\
87	1.11979343956886\\
88	1.11979343956886\\
89	1.11979343956886\\
90	1.11979343956886\\
91	1.21208719228578\\
92	1.21208719228578\\
93	1.21208719228578\\
94	1.21208719228578\\
95	1.21208719228578\\
96	1.21208719228578\\
97	1.21208719228578\\
98	1.21208719228578\\
99	1.21208719228578\\
100	1.21208719228578\\
101	1.21208719228578\\
102	1.21208719228578\\
103	1.21208719228578\\
104	1.30440146938838\\
105	1.30440146938838\\
106	1.30440146938838\\
107	1.30440146938838\\
108	1.30440146938838\\
109	1.30440146938838\\
110	1.30440146938838\\
111	1.30440146938838\\
112	1.30440146938838\\
113	1.30440146938838\\
114	1.30440146938838\\
115	1.39682122379278\\
116	1.39682122379278\\
117	1.39682122379278\\
118	1.39682122379278\\
119	1.39682122379278\\
120	1.39682122379278\\
121	1.39682122379278\\
122	1.39682122379278\\
123	1.39682122379278\\
124	1.48880870364757\\
125	1.48880870364757\\
126	1.48880870364757\\
127	1.48880870364757\\
128	1.48880870364757\\
129	1.48880870364757\\
130	1.48880870364757\\
131	1.48880870364757\\
132	1.48880870364757\\
133	1.58111953106286\\
134	1.58111953106286\\
135	1.58111953106286\\
136	1.58111953106286\\
137	1.58111953106286\\
138	1.58111953106286\\
139	1.58111953106286\\
140	1.67368725328853\\
141	1.67368725328853\\
142	1.67368725328853\\
143	1.67368725328853\\
144	1.67368725328853\\
145	1.67368725328853\\
146	1.76613173058666\\
147	1.76613173058666\\
148	1.76613173058666\\
149	1.76613173058666\\
150	1.76613173058666\\
151	1.76613173058666\\
152	1.85853386037734\\
153	1.85853386037734\\
154	1.85853386037734\\
155	1.85853386037734\\
156	1.85853386037734\\
157	1.85853386037734\\
158	1.95104526910268\\
159	1.95104526910268\\
160	1.95104526910268\\
161	1.95104526910268\\
162	2.04334837414389\\
163	2.04334837414389\\
164	2.04334837414389\\
165	2.04334837414389\\
166	2.04334837414389\\
167	2.13572850521339\\
168	2.13572850521339\\
169	2.13572850521339\\
170	2.13572850521339\\
171	2.2280501169605\\
172	2.2280501169605\\
173	2.2280501169605\\
174	2.32027761568997\\
175	2.32027761568997\\
176	2.32027761568997\\
177	2.32027761568997\\
178	2.412708732359\\
179	2.412708732359\\
180	2.412708732359\\
181	2.50472168452462\\
182	2.50472168452462\\
183	2.596589883748\\
184	2.596589883748\\
185	2.596589883748\\
186	2.68874015934004\\
187	2.68874015934004\\
188	2.68874015934004\\
189	2.78123871937428\\
190	2.78123871937428\\
191	2.8735576934362\\
192	2.8735576934362\\
193	2.96578127011771\\
194	2.96578127011771\\
195	3.0578341696657\\
196	3.0578341696657\\
197	3.15019621316359\\
198	3.15019621316359\\
199	3.24236302449583\\
200	3.33417197413602\\
201	3.33417197413602\\
202	3.42630026189207\\
203	3.51855774645629\\
204	3.51855774645629\\
205	3.6107083059369\\
206	3.70273769324052\\
207	3.70273769324052\\
208	3.79491972535114\\
209	3.88703360868599\\
210	3.9793310924373\\
211	4.07139112059702\\
212	4.16347065475632\\
213	4.25565383897586\\
214	4.34751896916156\\
215	4.43919787290773\\
216	4.53145760379904\\
217	4.62351949681395\\
218	4.80773044599771\\
219	4.90008702097114\\
220	5.08411934873971\\
221	5.17599431195675\\
222	5.36013773136129\\
223	5.45249489272888\\
224	5.63675901702296\\
225	5.82076394556694\\
226	6.00487594246378\\
227	6.189221550735\\
228	6.37296269587981\\
229	6.64910181054127\\
230	6.92548025242477\\
231	7.10961813245816\\
232	7.38536434278325\\
233	7.75359465564018\\
234	8.02943870560922\\
235	8.39838259106334\\
236	8.85783156075978\\
237	9.22563472559387\\
238	9.77717344931324\\
239	10.3290059410463\\
240	10.879764350512\\
241	11.5229003937838\\
242	12.3505328584333\\
243	13.177410208462\\
244	14.1872748413636\\
245	15.3816271871347\\
246	16.7595824255571\\
247	18.4131190564683\\
248	20.5270449127638\\
249	23.0073713679926\\
250	26.3135427104456\\
251	30.7204297926149\\
252	36.7794789519169\\
253	45.9574144429915\\
254	61.2874924593564\\
255	91.8455710070936\\
256	183.593401209663\\
};

\end{axis}
\end{tikzpicture}%

%% file: Figures/Intro_bound.tex
\definecolor{mycolor1}{rgb}{0.0, 0.58, 0.71}
\definecolor{mycolor2}{rgb}{1.0, 0.65, 0.0}
\definecolor{mycolor3}{rgb}{0.8, 0.31, 0.36}
\definecolor{mycolor4}{rgb}{0.09, 0.45, 0.27}

\begin{tikzpicture}

\begin{axis}[%
width=0.951\figurewidth,
height=\figureheight,
at={(0\figurewidth,0\figureheight)},
scale only axis,
xmin=0,
xmax=109,
xlabel style={font=\color{white!15!black}, font =\small, yshift = 0.25cm},
xlabel={Number of colluding workers z},
ymode=log,
ymin=0.438362667605437,
ymax=50,
yminorticks=true,
tick label style={font=\tiny},
ylabel style={font=\color{white!15!black}, font =\small, yshift=-0.65cm},
ylabel={Waiting time}, 
axis background/.style={fill=white},
legend style={legend cell align=left, align=left, draw=white!15!black, font =\tiny, at = {(0,1)}, anchor = north west}
]

\addplot [color=mycolor3, ultra thick] 
  table[row sep=crcr]{%
1	2.60181650516877\\
2	2.62710140803727\\
3	2.63721536918467\\
4	2.66250027205317\\
5	2.673878478344\\
6	2.70042762635592\\
7	2.71054158750332\\
8	2.73835498065867\\
9	2.74846894180607\\
10	2.77754658010485\\
11	2.78892478639568\\
12	2.81800242469445\\
13	2.82938063098528\\
14	2.85972251442748\\
15	2.87110072071831\\
16	2.90144260416051\\
17	2.91408505559476\\
18	2.94569118418038\\
19	2.95833363561463\\
20	2.99120400934369\\
21	3.00384646077794\\
22	3.03798107965041\\
23	3.05188777622809\\
24	3.08602239510057\\
25	3.09992909167824\\
26	3.13659220083757\\
27	3.15049889741524\\
28	3.188426251718\\
29	3.20233294829567\\
30	3.24152454774185\\
31	3.25669548946295\\
32	3.29588708890912\\
33	3.31232227577365\\
34	3.35277812036325\\
35	3.36921330722778\\
36	3.41219764210423\\
37	3.42863282896876\\
38	3.47288140898863\\
39	3.49058084099658\\
40	3.53609366615989\\
41	3.55379309816784\\
42	3.60183441361799\\
43	3.62079809076936\\
44	3.67010365136294\\
45	3.68906732851432\\
46	3.74090137939474\\
47	3.76112930168955\\
48	3.8142275977134\\
49	3.83571976515162\\
50	3.8900823063189\\
51	3.91283871890055\\
52	3.96972975035468\\
53	3.99248616293633\\
54	4.05316992982073\\
55	4.07719058754581\\
56	4.13913859957364\\
57	4.16442350244214\\
58	4.23016424990024\\
59	4.25544915276874\\
60	4.3237183905137\\
61	4.35153178366905\\
62	4.42232951170085\\
63	4.45140714999963\\
64	4.52599761346171\\
65	4.55507525176048\\
66	4.63345845065284\\
67	4.66506457923846\\
68	4.74724051356109\\
69	4.78011088729014\\
70	4.86607955704305\\
71	4.90021417591552\\
72	4.99123982624213\\
73	5.02663869025803\\
74	5.12272132115833\\
75	5.16064867546108\\
76	5.26178828693509\\
77	5.30224413152469\\
78	5.4084407235724\\
79	5.45016081330542\\
80	5.56394287621368\\
81	5.60819145623355\\
82	5.72703049971551\\
83	5.77507181516566\\
84	5.90149632950817\\
85	5.95206613524517\\
86	6.08734036559165\\
87	6.1404386616155\\
88	6.28329836282254\\
89	6.34145363942009\\
90	6.49442730177452\\
91	6.55511106865892\\
92	6.71946293730418\\
93	6.78393943961886\\
94	6.95966951455494\\
95	7.03046724258674\\
96	7.21883976895708\\
97	7.29469447756258\\
98	7.49697370051059\\
99	7.57914963483322\\
100	7.79912828978918\\
101	7.88762544982893\\
102	8.12530353679284\\
103	8.22138616769315\\
104	8.48055642209529\\
105	8.58548876899957\\
106	8.86741543598335\\
107	8.98246174403503\\
108	9.29220180417417\\
109	9.41862631851668\\
};
\addlegendentry{Bound in~\eqref{eq:avg_waiting_model1} - Homogeneous}

\addplot [color=mycolor1, mark=square, only marks, mark options={solid, mycolor1, thin}, mark size = 1]
   table[row sep=crcr]{%
1	2.64981021526051\\
2	2.62880187917758\\
3	2.68175285590732\\
4	2.60199416388582\\
5	2.70631668735611\\
6	2.72820421269209\\
7	2.73595745139773\\
8	2.74780556468997\\
9	2.75175591809391\\
10	2.76550267201894\\
11	2.82064461339855\\
12	2.90731007625124\\
13	2.81653297545948\\
14	2.87028285747134\\
15	2.80969349851864\\
16	2.91006456888509\\
17	2.98798061827175\\
18	2.97746881443346\\
19	2.94953071519945\\
20	2.93906401572888\\
21	3.03377673148333\\
22	2.94856702857895\\
23	3.07790710873341\\
24	3.1365899121219\\
25	3.08744859761023\\
26	3.18417492059472\\
27	3.21851938557707\\
28	3.22581920783744\\
29	3.2992242565346\\
30	3.21671816056814\\
31	3.31322871360057\\
32	3.24849458645311\\
33	3.27648162554987\\
34	3.29020321500755\\
35	3.37321744480175\\
36	3.43674680994762\\
37	3.46594552800184\\
38	3.51044070270294\\
39	3.59897281128164\\
40	3.53877073287988\\
41	3.57778947657727\\
42	3.61738516239577\\
43	3.55429250804843\\
44	3.73575769998943\\
45	3.66990227066825\\
46	3.74825940578157\\
47	3.79737116373415\\
48	3.84083188703877\\
49	3.70815562217224\\
50	3.84256721127915\\
51	3.90227152355229\\
52	4.01013703334107\\
53	3.95340140324411\\
54	3.97889886040311\\
55	4.09734688697814\\
56	4.13256328912186\\
57	4.16175463136977\\
58	4.12597537184908\\
59	4.24812523054642\\
60	4.36322649468651\\
61	4.33515129805725\\
62	4.30797048000216\\
63	4.43565765389892\\
64	4.54226653957916\\
65	4.60089138224536\\
66	4.74699046016749\\
67	4.75801237745882\\
68	4.77285270835812\\
69	4.72844489507639\\
70	4.88943321824746\\
71	4.83055561217075\\
72	4.90632023775044\\
73	4.95814489678511\\
74	4.96924797551685\\
75	5.12606955123641\\
76	5.35569106761249\\
77	5.44812109906841\\
78	5.42965023012641\\
79	5.41641476252975\\
80	5.85130962106556\\
81	5.62531299147585\\
82	5.7472757134105\\
83	5.85792303534535\\
84	5.9155890500043\\
85	6.03550504038336\\
86	6.22198487478911\\
87	6.21881497716803\\
88	6.24417991186473\\
89	6.64369568094623\\
90	6.53092929070135\\
91	6.67191377736543\\
92	6.76185624108436\\
93	6.64478331909053\\
94	7.02651050489631\\
95	7.09703479795951\\
96	7.2264327538551\\
97	7.18417193115282\\
98	7.62561576583249\\
99	7.60262686501089\\
100	7.65631536206243\\
101	8.01841673949865\\
102	8.02878489331865\\
103	8.2074975892786\\
104	8.14316955555644\\
105	8.58106172560717\\
106	8.50679475619414\\
107	8.95189102619333\\
108	9.14306793953416\\
109	9.39065963854684\\
};
\addlegendentry{RPM3 - Homogeneous}

\addplot [color=mycolor4, ultra thick]
  table[row sep=crcr]{%
1	0.43490032933822\\
2	0.438693064768495\\
3	0.43995730991192\\
4	0.44501429048562\\
5	0.44754278077247\\
6	0.451335516202746\\
7	0.453864006489596\\
8	0.457656741919871\\
9	0.460185232206721\\
10	0.465242212780421\\
11	0.467770703067271\\
12	0.471563438497546\\
13	0.475356173927822\\
14	0.479148909358097\\
15	0.482941644788372\\
16	0.486734380218647\\
17	0.490527115648922\\
18	0.495584096222623\\
19	0.498112586509473\\
20	0.503169567083173\\
21	0.506962302513448\\
22	0.512019283087148\\
23	0.514547773373998\\
24	0.520868999091124\\
25	0.523397489377974\\
26	0.529718715095099\\
27	0.532247205381949\\
28	0.538568431099074\\
29	0.54236116652935\\
30	0.548682392246475\\
31	0.55247512767675\\
32	0.558796353393875\\
33	0.562589088824151\\
34	0.568910314541276\\
35	0.572703049971551\\
36	0.580288520832101\\
37	0.584081256262377\\
38	0.590402481979502\\
39	0.595459462553202\\
40	0.603044933413752\\
41	0.606837668844028\\
42	0.614423139704578\\
43	0.619480120278278\\
44	0.627065591138829\\
45	0.632122571712529\\
46	0.640972287716504\\
47	0.646029268290204\\
48	0.65487898429418\\
49	0.65993596486788\\
50	0.668785680871856\\
51	0.673842661445556\\
52	0.683956622592956\\
53	0.689013603166656\\
54	0.699127564314057\\
55	0.705448790031182\\
56	0.715562751178583\\
57	0.721883976895708\\
58	0.733262183186533\\
59	0.739583408903659\\
60	0.750961615194484\\
61	0.757282840911609\\
62	0.76992529234586\\
63	0.77751076320641\\
64	0.790153214640661\\
65	0.797738685501211\\
66	0.810381136935462\\
67	0.819230852939437\\
68	0.833137549517113\\
69	0.841987265521088\\
70	0.855893962098764\\
71	0.864743678102739\\
72	0.881178864967265\\
73	0.89002858097124\\
74	0.907728012979191\\
75	0.917841974126592\\
76	0.935541406134543\\
77	0.945655367281943\\
78	0.964619044433319\\
79	0.975997250724145\\
80	0.996225173018945\\
81	1.0088676244532\\
82	1.03035979189142\\
83	1.04300224332567\\
84	1.06575865590732\\
85	1.079665352485\\
86	1.1049502553535\\
87	1.11885695193118\\
88	1.14667034508653\\
89	1.16184128680763\\
90	1.1909189251064\\
91	1.20861835711435\\
92	1.23896024055656\\
93	1.25792391770793\\
94	1.29205853658041\\
95	1.31228645887521\\
96	1.34894956803454\\
97	1.37170598061619\\
98	1.41089758006236\\
99	1.43618248293087\\
100	1.48043106295074\\
101	1.50698021096267\\
102	1.55502152641282\\
103	1.58536340985502\\
104	1.6397259510223\\
105	1.67259632475135\\
106	1.73201584649233\\
107	1.76867895565166\\
108	1.83694819339661\\
109	1.87866828312964\\
};
\addlegendentry{Bound in~\eqref{eq:avg_waiting_model1} - Heterogeneous}

\addplot [color=mycolor2, mark=o, only marks, mark options={solid, mycolor2}, mark size = 1pt]
  table[row sep=crcr]{%
1	0.438362667605437\\
2	0.429542125907438\\
3	0.4485066970975\\
4	0.446872319837296\\
5	0.44577371114046\\
6	0.452081482166505\\
7	0.454058688677409\\
8	0.454571054890661\\
9	0.465365065076253\\
10	0.467489397178146\\
11	0.465226475235491\\
12	0.459768677588848\\
13	0.486371338519949\\
14	0.488312016931621\\
15	0.475147658015721\\
16	0.473327853789495\\
17	0.496012009586637\\
18	0.493442642189989\\
19	0.509234953121497\\
20	0.509675979181333\\
21	0.504411320616168\\
22	0.497662507571395\\
23	0.518389605697941\\
24	0.534854941370773\\
25	0.520439406932147\\
26	0.534373510576465\\
27	0.520559172077219\\
28	0.541425293193153\\
29	0.533027134988294\\
30	0.552861914513089\\
31	0.550269490221019\\
32	0.570071608046356\\
33	0.556380242034015\\
34	0.565406320542248\\
35	0.576429099121181\\
36	0.571132956678732\\
37	0.585643180921689\\
38	0.581974326786339\\
39	0.60189553798915\\
40	0.61385116348438\\
41	0.612545512275636\\
42	0.610923896052482\\
43	0.628112105733589\\
44	0.616098788298847\\
45	0.637088344644633\\
46	0.642160045104582\\
47	0.626460207473605\\
48	0.655824165011707\\
49	0.689333970511497\\
50	0.668459180754596\\
51	0.662446334429014\\
52	0.693779444031541\\
53	0.705876028814092\\
54	0.715238443273996\\
55	0.713948543785817\\
56	0.723407627060655\\
57	0.706773833496394\\
58	0.74459942731525\\
59	0.744779299077716\\
60	0.732133554829519\\
61	0.756866216348889\\
62	0.766175682630512\\
63	0.77959531914591\\
64	0.782760791703589\\
65	0.798228979754516\\
66	0.792348889876669\\
67	0.855662125186119\\
68	0.835805810506846\\
69	0.832430744272724\\
70	0.852225835240183\\
71	0.857984477184297\\
72	0.865444317480912\\
73	0.871869183731368\\
74	0.937554869165771\\
75	0.920315023836095\\
76	0.951960711689018\\
77	0.92668286931046\\
78	0.956424066119576\\
79	0.968829671420866\\
80	1.02771654050934\\
81	0.978416283278185\\
82	1.03788103416805\\
83	1.08581786558246\\
84	1.05717666888008\\
85	1.07328203104535\\
86	1.10001987820158\\
87	1.1161938738178\\
88	1.16311211466122\\
89	1.15111368320011\\
90	1.2172934986428\\
91	1.1736685050299\\
92	1.21969297032856\\
93	1.26800794787698\\
94	1.27313391595645\\
95	1.29157291904929\\
96	1.36451388390494\\
97	1.35380171721583\\
98	1.40001254403882\\
99	1.42817588033128\\
100	1.46744267079365\\
101	1.50867926865824\\
102	1.56379938667266\\
103	1.62872139294354\\
104	1.64103606466724\\
105	1.66722068334357\\
106	1.73210522392929\\
107	1.76826176500397\\
108	1.80329245008804\\
109	1.85833783285227\\
};
\addlegendentry{RPM3 - Heterogeneous}

\end{axis}
\end{tikzpicture}%

%% file: main_results.tex
We introduce a new scheme for \textbf{r}ateless \textbf{p}rivate \textbf{m}atrix-\textbf{m}atrix \textbf{m}ultiplication that we call RPM3. This scheme allows the master to cluster the workers into groups of workers of similar speed. Then, RPM3 takes any fixed-rate private scheme, couples it with the special Fountain coding technique~\cite{factored_lt} of the input matrices to create a rateless code. We prove, in Theorem~\ref{thm:RPM3}, those properties of RPM3 when the fixed-rate scheme is a Lagrange polynomial.




\subsection{Rate analysis}On a high level, under privacy constraints, clustering the workers comes at the expense of reducing the efficiency of the scheme by $2z-1$ for every new cluster. We reduce this loss to $z-1$ per cluster by using Lagrange polynomials that share $z$ evaluations with each other, c.f., Remark~\ref{rem:loss}. To remove the penalty of flexibility, the encoding polynomials of a rateless scheme must share $2z-1$ evaluations.

We analyze the rate of RPM3 when using Lagrange polynomials and show in Lemma~\ref{lemma:rate} that the rate of RPM3 is%
\begin{align*}
        \rho_{RPM3} 
        & = \dfrac{mk}{\displaystyle 2mk(1+\varepsilon)+ (z-1)\tau_c\sum_{u=1}^c\gamma_{u} + z\tau_c \gamma_{1}},
\end{align*} 
where $\tau_c$ is the number of tasks finished by cluster the slowest cluster $c$, $\gamma_u\tau_c$ is the number of tasks finished by cluster $u$ and $\epsilon$ is the overhead required by Fountain codes.

In contrast to other schemes, the rate of RPM3 does not depend on the number of workers $n$. Therefore, the master can design the size of the tasks to fit the computational power and storage constraints of the workers. The penalty of clustering the workers appears in the term $(z-1)\tau_c\sum_{u=2}^c\gamma_{u}$. Clustering the workers does not affect the rate of RPM3 for $z=1$. However, for $z>1$ the rate decreases linearly with the number of clusters and with $z$. Hence, the existence of a tradeoff between the rate and the flexibility (number of clusters) of the scheme.

We compare the rate of RPM3 to the rate of the fixed-rate scheme of~\cite{kakar2018rate} that tolerates a fixed number of stragglers. This scheme is the closest to our model and has the best known rate among schemes that tolerate stragglers. We observe that the rate of RPM3 is lower than the rate of the scheme in~\cite{kakar2018rate} except for a small subset of parameters. 

\subsection{Time analysis}We analyze the waiting time of the master for both considered models of the service time. The first model is a simplified model that provides both theoretical and engineering insights about the system and its design. The second model reflects our RPM3 scheme, but its analysis does not provide engineering insights about the system design. For both models, we give an upper bound on the probability distribution and on the mean of the waiting time of the master, Theorem~\ref{thm:waiting_exp} and Theorem~\ref{thm:waiting_erlang}, respectively. The bound on the probability distribution can be used to tune the parameters of the scheme so that the waiting time of the master is less than a fixed deadline with high probability. We analyze the waiting time of the master when using the scheme of~\cite{kakar2018rate} for both models of the service time, Corollary~\ref{cor:kakar_exp} and Corollary~\ref{cor:erlang_kakr}. Note that model~1 is the accurate model for fixed-rate schemes that send only one task to the workers. However, we allow the fixed-rate scheme to divide its tasks into smaller tasks to benefit from the advantages of the second model. For the first model, we give a theoretical guarantee on when the master has a smaller mean waiting time when using RPM3. However, for the second model the derived expressions are hard to analyze. We provide numerical evidence (e.g., Figure~\ref{subfig:intro_model1} and~\ref{subfig:intro_model2}), showing that when the workers have different service times (heterogeneous environment) RPM3 has a smaller waiting time for small values of $z$. However, when the workers have similar service times, RPM3 outperforms the fixed rate scheme only under the second time model. We also show numerically that the provided upper bound on the mean waiting time of PRM3 under model 1 is a good representative of the actual mean waiting time, e.g., Figure~\ref{subfig:intro_bound}.

\subsection{Comparison to perfect load balancing}
To check the effect of clustering on the mean waiting time of the master, we compare RPM3 to the setting where the master has previous knowledge of the computation power of the workers. In such a setting, the master can simply assign tasks that are proportional to the compute power of the workers without the need of rateless codes.
We show in Theorem~\ref{thm:load_bal} that, under model~1, the mean waiting time of the master when using RPM3 is far from the mean waiting time of the load balancing scheme by a factor of $\frac{\tau_{u^\star}}{\taulb{c}} \frac{\lambda_c}{\lambda_{u^\star}}$, where $\tau_{u^\star}$ is the number of tasks computed by the slowest cluster of RPM3 and $\taulb{c}$ is the number of tasks computed by the slowest cluster in the load balancing scheme.

Since $\taulb{c}$ depends on the rate of the scheme used for load balancing, we consider two settings. An ideal setting in which there exists a scheme with the best possible rate $(n-z)/n$ that the master uses. Schemes achieving this rate for matrix-matrix multiplication exist for $z=1$. Since this rate may not be always achievable, we present an achievable load balancing scheme based on GASP codes~\cite{d2018gasp}. We derive the value of $\taulb{c}$ under both settings and different parameter regimes, see Table~\ref{tab:LB_tauc}.

We make a key observation when comparing RPM3 to the ideal setting and to the achievable setting with large $z$. Those settings are the two extreme settings of the spectrum of achievable rates.
\begin{corollary}[Informal; see Corollary~\ref{cor:LB_ideal} for a formal version]
The mean waiting time of the master when using RPM3 is bounded away from that of the ideal setting by%
\begin{equation*}
    {\mathbb{E}[T_{RPM3}]} \leq \mathbb{E}[T_{\text{LB}}]\dfrac{2(1+\epsilon)}{R_1(z)}\frac{\lambda_c}{\lambda_{u^\star}}.
\end{equation*}
The mean waiting time of RPM3 is bounded away from the mean waiting time of the achievable GASP scheme with large values of $z$ by
\begin{align*}
     {\mathbb{E}[T_{RPM3}]} \leq \mathbb{E}[T_{\text{LB}}] \dfrac{1+\epsilon}{R_2(z)}\frac{\lambda_c}{\lambda_{u^\star}}.
\end{align*}
\end{corollary}

Recall that $0\leq \epsilon \leq 1$ is the overhead required by Fountain codes. $R_1(z), R_2(z) \in [0,1)$ are decreasing functions of $z$ that reach $0$ when $z$ reaches its maximal value $z=n_1+1/2 = n_u+1$ for all $u=2,\dots,c$. The functions $R_1(z)$ and $R_2(z)$ show the effect of the rate loss due to clustering on the waiting time. When alleviating the penalty of clustering, $R_1(z)=R_2(z)=1$ we would get back the expected bound on the waiting time, see Remark~\ref{rem:open}. Finding rateless codes with $R_1(z)$ or $R_2(z)=1$ remains an open problem, see Remark~\ref{rem:optimal}. In other words, if the clustering comes at no extra cost, the only increase in the waiting time would be due to the overhead of Fountain codes. Note that the rate of GASP codes for large $z$ coincide with the rate of Lagrange polynomials. However, when comparing to the ideal scheme an extra factor of $2$ is present. This is due to the fact that Lagrange polynomials have rate $(n-z)/2n$ which is half of the rate of the ideal scheme. 
The ratio ${\lambda_c}/{\lambda_{u^\star}}\leq 1$ shows the ratio of the expected service rates of the cluster that finishes the last task of RPM3 to the slowest cluster $c$.



%% file: sys_model.tex
We provide a detailed explanation of our RPM3 
scheme and prove the following theorem.
\begin{theorem}\label{thm:RPM3}
    Consider a matrix-matrix multiplication setting as described in Section~\ref{sec:Preliminaries}. The RPM3 scheme defined next is a rateless double-sided $z$-private matrix-matrix multiplication scheme that adapts to the heterogeneous behavior of the workers.
\end{theorem}

\begin{IEEEproof}
The proof is constructive. We give the details of the construction in Sections~\ref{sec:data_enc} and~\ref{subsec:clustering}. In Section~\ref{sec:decoding} we show that the master can obtain the desired the computation. We prove the privacy constraint in Section~\ref{sec:privacy}. 
\end{IEEEproof}

\subsection{Data encoding}\label{sec:data_enc}
The master divides the encoding into rounds. At a given round $t$, the workers are grouped into $c$ clusters each of $n_u$ workers, $u=1,\dots,c$ and $\sum_{u=1}^c n_u = n$. We defer the clustering technique to the next section. Dividing workers into several clusters adds flexibility in the decoding at the master. The results returned from a cluster of workers allow the master to decode new Fountain-coded computations as explained next. We define $d_1 \triangleq \lfloor \frac{n_1 - 2z + 1}{2}\rfloor$ and $d_u \triangleq \lfloor \frac{n_u - z + 1}{2}\rfloor$ for $u = 2,\dots, c$. The master generates $c$ Lagrange polynomial pairs $\ftr{t}{u}(x)$ and $\gtr{t}{u}(x)$. Each polynomial $\ftr{t}{u}(x)$ contains $d_u$ Fountain-coded matrices $\widetilde{\mA}_{t,\kappa}^{(u)}$, $\kappa = 1,\dots, d_u$, defined as $\widetilde{\mA}_{t,\kappa}^{(u)} \triangleq \sum_{i=1}^m b_{\kappa,i}^{(u)} \mA_i$, where\footnote{Note that $b^{(u)}_{\kappa,i}$ also depends on $t$, but we remove the subscript $t$ for the ease of notation.} $b_{\kappa,i}^{(u)} \in \{0,1\}$. Similarly, each polynomial $\gtr{t}{u}(x)$ contains $d_u$ Fountain-coded matrices $\widetilde{\mB}_{t,\kappa}^{(u)} \triangleq \sum_{j=1}^k b_{\kappa,j}^{(u)} \mB_j$ where  $b_{\kappa,j}^{(u)} \in \{0,1\}$ are chosen randomly \cite{factored_lt}. {The distributions from which the $b_{\kappa,i}^{(u)}$ and $b_{\kappa,j}^{(u)}$ are drawn must be designed jointly as in \cite{factored_lt} to guarantee that the master can decode $\mA\mB$ after receiving $(1+\epsilon)mk$ products of the form $\widetilde{\mA}\widetilde{\mB}$ with small values of $\epsilon$.} The master generates $2z$ uniformly random matrices $\mR_{t,1},\dots, \mR_{t,z} \in \F_q^{r/m\times s}$ and $\mS_{t,1},\dots, \mS_{t,z} \in \F_q^{s\times \ell/k}$.

\begin{remark}[Penalty on clustering the workers]\label{rem:loss}
{From the definition of $d_1$ and $d_u$, $u=2,\dots,c$, we can see that clustering the workers and assigning one polynomials to each cluster incurs an extra penalty of loosing $z-1$ computations per cluster. The number $d_u$ is the number of computations of the form $\widetilde{\mA}\widetilde{\mB}$ computed at every cluster. Assigning one Lagrange polynomial to all the workers results in $d=\lfloor \frac{n - 2z + 1}{2}\rfloor$, i.e., for $d_1=\left\lfloor \frac{n_1-2z+1}{2}\right\rfloor$, every $d_u$ should be equal to $\lfloor \frac{n_u}{2}\rfloor$. Initially, one would expect a loss of $2z-1$ computations per cluster. We reduce this loss by allowing the encoding polynomials to share $z$ evaluations as explained in the decoding part. To alleviate the penalty of clustering, the encoding polynomials of every round must share $2z-1$ evaluations. More details are given in Remark~\ref{rem:optimal} after formally defining the decoding process.}
\end{remark}
Let $d_{\text{max}} = \max_u d_u$ and $\alpha_1,\dots,\alpha_{d_{\text{max}}} \in \F_q$ be distinct elements of $\F_q$.  The polynomials are constructed as shown in~\eqref{eq:fx} and~\eqref{eq:gx}.
\begin{align}
\ftr{t}{u}(x) &= \sum_{\delta=1}^{z} \mR_{t,\delta} \prod_{\nu \in [{d_u}+z]\setminus \{\delta\}} \frac{x - \alpha_\nu}{\alpha_\delta-\alpha_\nu} \nonumber \\
 &+ \sum_{\delta=z+1}^{{d_u}+z} \widetilde{\mA}^{(u)}_{t,\delta-z} \prod_{\nu \in [d_u+z]\setminus \{\delta\}} \frac{x - \alpha_\nu}{\alpha_\delta-\alpha_\nu}, \label{eq:fx}\\
\gtr{t}{u}(x) &= \sum_{\delta=1}^{z} \mS_{t,\delta} \prod_{\nu \in [d_u+z]\setminus \{\delta\}} \frac{x - \alpha_\nu}{\alpha_\delta-\alpha_\nu} \nonumber \\
&+ \sum_{\delta=z+1}^{d_u+z} \widetilde{\mB}^{(u)}_{t,\delta-z} \prod_{\nu \in [d_u+z]\setminus \{\delta\}} \frac{x - \alpha_\nu}{\alpha_\delta-\alpha_\nu}. \label{eq:gx}
\end{align}

The master chooses $n$ distinct\footnote{Choosing the $\beta_i$'s carefully is needed to maintain the privacy constraints as explained in the sequel.} elements $\beta_i\in \F_q\setminus\{\alpha_1,\cdots,\alpha_{d_{\text{max}+z}}\}$, $i=1,\dots,n$. 
For each worker, $w_i$ the master checks the cluster $u$ to which this worker belongs, and sends $\ftr{t}{u}(\beta_i)$, $\gtr{t}{u}(\beta_i)$ to that worker. 



\subsection{Clustering of the workers and task distribution}\label{subsec:clustering}

\noindent{\em Clustering: }For the first round $t=1$, the master groups all the workers in one cluster of size $n_1 = n$. The master generates tasks as explained above and sends them to the workers. 

For $t>1$, the master wants to put workers that have similar response times in the same cluster. In other words, workers that send their results in round $t-1$ to the master within a pre-specified interval of time will be put in the same cluster. Let $\interval$ be the length of the time interval desired by the master.

In addition to the time constraint, the number of workers per cluster and the privacy parameter $z$ must satisfy
\begin{equation}\label{eq:clustering}
    n_u \geq \begin{cases}
        2z-1 & \text{if } u = 1,\\
        z+1 & \text{otherwise.}
    \end{cases}
\end{equation}
Those constraints ensure that the master can decode the respective polynomials $\htr{t}{u}(x)$ as explained in the next section.

Let $\eta_1$ be\footnote{In this section, all variables depend on $t$. However, we omit $t$ for the clarity of presentation.} the time spent until the result of $w_{i_1}$ is received by the master (at round $t-1$). All workers that send their results before time $\eta_1 + \interval$ are put in cluster $1$. If $n_1\geq 2z-1$, the master moves to cluster $2$. Otherwise, the master increases $\interval$ so that $n_1 \geq 2z-1$. The master repeats the same until putting all the workers in different clusters guaranteeing $n_u\geq z+1$, $u=2,\dots,c$.

{In the remaining of the paper we assume that the number of workers per cluster is fixed during the whole algorithm and is known as a system parameter.}

Over the course of the computation process, the master keeps measuring the empirical response time of the workers. The response time of a worker is the time spent by that worker to receive, compute and return the result of one task. Having those measurements, the master can update the clustering accordingly when needed using the same time intervals.

{\em Task distribution: }
At the beginning of the algorithm, the master generates tasks assuming all workers are in the same cluster and sends those tasks to the workers. For round $2$ the master arranges the workers in their respective clusters and sends tasks accordingly. Afterwards, when the master receives\footnote{To avoid idle time at the workers, the master can measure the expected computation time of each worker at round $t_i-1$. Using this information, the master can then send a task to a worker in a way that this worker will receive the task right after finishing its current computation. This will guarantee that the worker will not be idle during the transmission of tasks to and from the master. See \cite{KS18} for more details.} a task from worker $w_i$, it checks at which round $t_i$ this worker is (how many tasks did the worker finish so far) and to which cluster $u$ it belongs. The  master generates $\ftr{t_i+1}{u}(x), \htr{t_i+1}{u}(x)$ if $w_i$ is the first worker of cluster $u_i$ to finish round $t_i$ and sends $\ftr{t_i+1}{u}(\beta_i), \htr{t_i+1}{u}(\beta_i)$ to $w_i$.

\subsection{Decoding}\label{sec:decoding}
 
 
 
   
    

    
At a given round $t$, the master first waits for the $n_1$ fastest workers belonging to cluster $1$ to finish computing their tasks so that it can interpolate $\htr{t}{1}(x)$. This is possible because the master obtains $n_1 = 2d_1+2z-1$ evaluations of $\htr{t}{1}(x)$ equal to the degree of $\htr{t}{1}(x)$ plus one. By construction, for a given $t$, the polynomials $\ftr{t}{u}(x)$ and $\gtr{t}{u}(x)$ share the same random matrices as coefficients, see~\eqref{eq:fx} and~\eqref{eq:gx}. Thus, for $i=1,\dots,z$, the polynomials $\htr{t}{u}(x)$ share the following $z$ evaluations
\begin{equation}
    \htr{t}{1}(\alpha_{i})=\htr{t}{2}(\alpha_{i})=\dots=\htr{t}{c}(\alpha_{i})=\mR_{t,i}\mS_{t,i}.
\end{equation}

Therefore, the master can interpolate $\htr{t}{u}(x)$ when $n_u$ workers of cluster $u, u=2,\dots,c,$ return their results. This is possible because the master receives $n_u=2d_u+z-1$ evaluations of $\htr{t}{u}(x)$ and possesses the $z$ evaluations shared with $\htr{t}{1}(x)$. Allowing the polynomials to share the randomness enables us to reduce the number of workers from every cluster $u>1$ by $z$ workers.


After successfully interpolating a polynomial $\htr{t}{u}(x)$ for a given round $t$ and a cluster $u$, the master computes $d_u$ products of Fountain-coded matrices
\begin{equation}
\htr{t}{u}(\alpha_{\kappa+z}) = \mtA{t,\kappa}{u}\mtB{t,\kappa}{u}
\end{equation}
for $\kappa=1,\dots,d_u$. The master feeds those $d_u$ computations to a peeling decoder \cite{LT,factored_lt,Raptor,Fountain} and continues this process until the peeling decoder can successfully decode all the components of the matrix $\mC$. Thus, allowing a flexibility in the rate and leveraging the rateless property. {The peeling decoder works by searching for a received computation which is equal to any of the components of the desired matrix $\mC$, i.e., a Fountain coded matrix product $\widetilde{\mA}\widetilde{\mB} = \mA_i\mB_j$. If such a computation exists, the decoder extracts (decodes) its value and then subtracts it from all other Fountain coded packets that contain it as a summand. This procedure is done iteratively until the decoding succeeds, i.e., until decoding the values of all of the components of $\mC$.}

\subsection{Proof of double-sided privacy}\label{sec:privacy}
Since the master generates new random matrices at each round, it is sufficient to prove that the privacy constraint given in~\eqref{eq:privacy} holds at each round separately. The proof is rather standard and follows the same steps as~\cite{bitar2019private,yu2018lagrange}. We give a complete proof in Appendix~\ref{app:proof_privacy} for completeness and provide next a sketch of the proof.

Let $\cW_{i,t}$ be the set of random variables representing the tasks sent to worker $w_i$ at round $t$. For a set $\mathcal{A}\subseteq [n]$ we define $\mathcal{W}_{\mathcal{A},t}$ as the set of random variables representing the tasks sent to the workers indexed by $\mathcal{A}$ at round $t$, i.e., $\mathcal{W}_{\mathcal{A},t}\triangleq \{\cW_{i,t}| i\in \mathcal{A}\}$. We want to prove that at every round $t$ 
\begin{equation}\label{eq:privacy_per_round}
    {I}\left(\rvA,\rvB;\cW_{\mathcal{Z},t}\right) = 0, \forall \cZ \subset [n], \text{ s.t. } |\cZ| = z.
\end{equation}

To prove~\eqref{eq:privacy_per_round} it is enough to show that given the input matrices $\mA$ and $\mB$, any collection of $z$ workers $w_{i_1}, \dots,w_{i_z},$ can use the tasks given to them at round $t$ to obtain the random matrices $\mR_{t,1},\dots,\mR_{t,z}$ and $\mS_{t,1},\dots,\mS_{t,z}$. Decoding the random matrices holds due to the use of Lagrange polynomials and setting the random matrices as the first $z$ coefficients.

%% file: rate_analysis.tex

We assume that the workers in the same clusters have very similar response time. 
We compare RPM3 to the scheme in \cite{kakar2018rate} that has an improved rate over using the Lagrange polynomials but does not exist for all values of $m$ and $k$.

{\em Rate of RPM3: }Let $\tau_u$ be the number of rounds finished (tasks successfully computed) by all the workers in cluster $u$, $u=1,\dots,c$. There exist real numbers $\gamma_u \geq 1$ for $u \in [c]$ such that $\tau_{u} = \gamma_{u} \tau_{c}$. This means that the number of tasks computed by workers in cluster $u$ is $\gamma_{u}$ times more than the number of tasks computed by workers in the slowest cluster $c$. Given the values of $\tau_1,\dots,\tau_c$ we can compute the rate of RPM3 as follows.

\begin{lemma}\label{lemma:rate}
    Consider a private distributed matrix-matrix multiplication with $n$ workers out of which at most $z$ can collude. Let the input matrices $\mA$ and $\mB$ be split into $m$ and $k$ submatrices, $\mA=\bbm \mA_1^T, \dots, \mA_m^T\ebm^T$ and $\mB=\bbm \mB_1, \dots, \mB_k\ebm$, respectively. 
    
    Let $c$ be the number of clusters of workers and $\tau_u$ be the number of rounds in which the polynomials $\htr{t}{u}(x), t=1,\dots,\tau_u$ is interpolated at the master. Then, for an $\varepsilon$ overhead required by the Fountain code decoding process, the rate of RPM3 under this setting is
    \begin{align}\label{eq:rho_rpm3}
        \rho_{\text{RPM3}} 
        & = \dfrac{mk}{\displaystyle 2mk(1+\varepsilon)+ (z-1)\tau_c\sum_{u=1}^c\gamma_{u} + z\tau_c \gamma_{1,c}}.
    \end{align} 
 \end{lemma}
  \begin{IEEEproof}[Proof]
We count the number of results $N$ (that will be in the denominator of the rate) collected by the master at the end of the computation process. From each cluster of workers $u$, $u=1,\dots,c$, the master collects $n_u \tau_u$ results. Recall that $n_1= 2d_1 +2z -1$ and $n_u = 2d_u+z-1$ for $u=2,\dots,c$. We can write the following
 \begin{align}
     N & = \sum_{u=1}^{c} n_u \tau_u \nonumber\\
     & = \sum_{u=2}^{c} (2d_u + z -1)\tau_u + (2d_1 + 2z -1)\tau_1 \nonumber\\
     & = \sum_{u=1}^{c} 2d_u\tau_u + (z-1)\sum_{u=1}^{c} \tau_u + z\tau_1 \nonumber\\
     & = 2mk(1+\varepsilon)+ (z-1)\tau_c\sum_{u=1}^c\gamma_{u} + z\tau_c \gamma_{1}. \label{eq:sumoftasks}
 \end{align}

 Equation~\eqref{eq:sumoftasks} follows from the fact that $\sum_{u=1}^{c} d_u\tau_u = mk(1+\varepsilon)$. This is true because the master needs $mk(1+\varepsilon)$ different values of $\fcA{i}{t}{u}\fcB{j}{t}{u}$ in total to compute $\mA\mB$ and each interpolated polynomial $\htr{t}{u}(x)$ encodes $d_u$ such values.
 \end{IEEEproof}
 Lemma~\ref{lemma:rate} shows a tradeoff between the rate of the scheme and its adaptivity to heterogeneous systems. Dividing the workers into $c$ clusters and sending several polynomials to the workers affects the rate of the scheme. The loss in the rate appears in the term $(z-1)\tau_c\sum_{u=2}^c\gamma_{u}$. However, sending several polynomials to the workers allows the master a flexibility in assigning a number of tasks proportional to the resources of the workers; Hence, increasing the speed of the computing process.

{\begin{remark}[Properties of the optimal encoding polynomials]\label{rem:optimal}
A flexible-rate scheme that clusters the workers into $c$ clusters has the same rate as a fixed-rate scheme using the same encoding polynomials if the encoding polynomials $h_t^{(1)}(x)$ and $h_t^{(u)}(x)$, $u=2,\dots,c$, share $2z-1$ evaluations. In addition, $h_t^{(u)}(x)$ can be interpolated from the shared evaluations and an additional $n_u$ evaluations of the form $h_t^{(u)}(\beta_i)$, $i\in [n_u]$.
This constraint can be interpreted as finding \begin{align*}
\ftr{t}{u}(x) = r_t(x) + p_t^{(u)}(x), & & 
\gtr{t}{u}(x) = s_t(x) + q_t^{(u)}(x),
\end{align*}
such that 
\begin{align*}
    h_t^{(1)}(x) = \ftr{t}{1}(x)\gtr{t}{1}(x), & &
    h_t^{(u)}(x) & =\ftr{t}{u}(x)\gtr{t}{u}(x),
\end{align*}
share $2z-1$ evaluations for $u=2,\dots,c$. The polynomials $r_t(x)$ and $s_t(x)$ do not depend on the cluster number and are the polynomials that encode the randomness to guarantee privacy. The polynomials $q_t^{(u)}(x)$ and $p_t^{(u)}(x)$ encode the Fountain-coded data and change from a cluster to another to guarantee that the master obtains new coded packets. For $z=1$, RPM3 satisfies this property and thus has an optimal encoding polynomials. The problem of finding optimal encoding polynomials for $z>1$ is left open.
\end{remark}}
 
 The main property of RPM3 is that the rate of the scheme is independent of the degree of the encoding polynomials and from the number of available workers $n$. The rate only depends on the number of assigned tasks to the workers in different clusters. This property reflects the ability of RPM3 to flexibly assign the tasks to the workers based on their available resources. In addition, this property reflects the fact that RPM3 can design tasks to have arbitrarily small size to fit the computational power of the available workers. 
 
{\em Comparison to the scheme of \cite{kakar2018rate}: }We refer to this scheme as the {\em \fr\ scheme} for brevity. The \fr\ scheme has the highest rate amongst known schemes that tolerate stragglers and have a model similar to the one considered in this paper. In particular, it has a better rate than naively using Lagrange polynomials to send the tasks to the workers. The better rate is achieved by carefully choosing the coefficients and the degrees of $x$ in $\ftr{t}{u}(x)$ and $\gtr{t}{u}(x)$ to reduce the number of evaluations needed from the workers to interpolate $\htr{t}{u}(x)$.

\begin{figure*}[t]
\hspace*{-.5cm}
\begin{subfigure}[b]{0.33\textwidth}
\centering
 \setlength\figureheight{0.89\textwidth}
 \setlength\figurewidth{0.9\textwidth}
 \resizebox{.9\textwidth}{!}{
  \input{Figures/Setting1-rate-vs-z-m=2000-k=3000} }
  \captionsetup{width = 0.8\textwidth}
  \caption{Rate of RPM3 and the \fr\ scheme.}
  \label{subfig:set1-rate}
\end{subfigure}%
\begin{subfigure}[b]{0.33\textwidth}
\centering
 \setlength\figureheight{0.85\textwidth}
 \setlength\figurewidth{0.95\textwidth}
\resizebox{.9\textwidth}{!}{
 \input{Figures/scaled-set1-vs-z-m=2000-k=3000-s=100-mu=1000-exp=100} }
 \captionsetup{width = 0.8\textwidth}
 \caption{Empirical average waiting time under model~1.}
 \label{subfig:set1-scaled}
\end{subfigure}%
\begin{subfigure}[b]{0.33\textwidth}
\centering
 \setlength\figureheight{0.87\textwidth}
 \setlength\figurewidth{0.97\textwidth}
 \resizebox{.9\textwidth}{!}{
  \input{Figures/erlang-set1-vs-z-m=2000-k=3000-s=100-mu=1000-exp=100} }
  \captionsetup{width = 0.8\textwidth}
  \caption{Empirical average waiting time under model~2.}
  \label{subfig:set1-erlang}
\end{subfigure}       
\caption{Comparison between the rate of RPM3 and the \fr\ scheme for the first setting (see Table~\ref{tab:parameters}). The \fr\ scheme has higher rate than RPM3 for this particular setting. Note that RPM3 is restricted to $z \leq 109$ because of clustering (see~\eqref{eq:clustering}), while for the \fr\ scheme $z$ is restricted due to the rate calculation (see~\eqref{eq:rate_kakar}). Figures~\ref{subfig:set1-scaled} and~\ref{subfig:set1-erlang} are an extension of Figures~\ref{subfig:intro_model1} and~\ref{subfig:intro_model2}, that show the mean waiting for a largest range of $z$. The rate of RPM3 depends on the relative computation power $\gamma_{u}$ of the different clusters. The $\gamma_u$'s can be considered as a random variables. Thus, the plot in~\ref{subfig:set1-scaled} is the expected rate of RPM3 rather than the conventional rate. For the \fr\ scheme, the rate is fixed and depends only on the number of workers, number of stragglers and the privacy parameter $z$. Under model 1 the \fr\ scheme outperforms RPM3 for homogeneous clusters and RPM3 outperforms the \fr\ scheme for heterogeneous clusters for $z \leq 90$. However, under model~2 RPM3 outperforms the \fr\ scheme for both heterogeneous and homogeneous clusters for the values of $z$ for which it has a non-zero rate.} 
\label{fig:setting1}
\end{figure*}
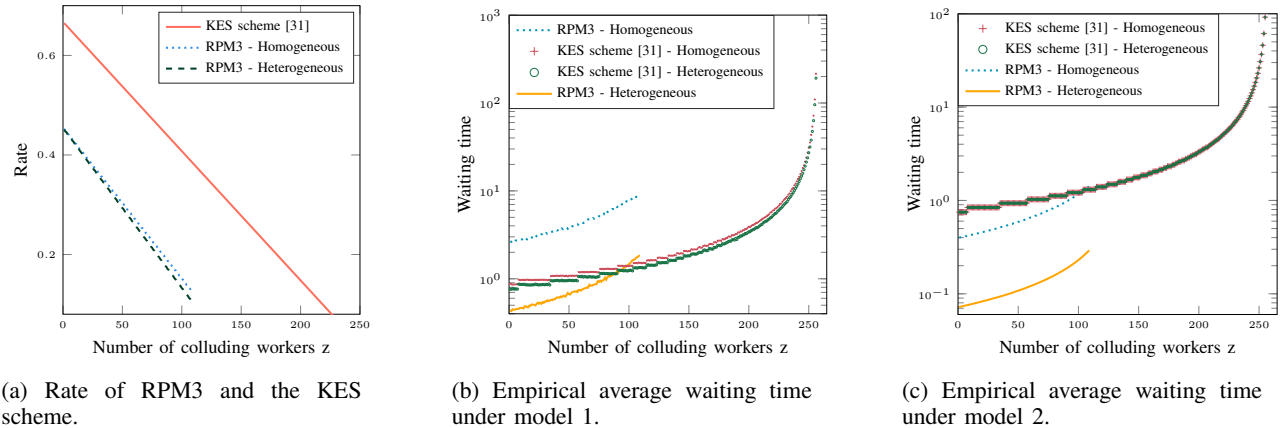

\begin{remark}
One could use the polynomials of the schemes in \cite{kakar2018rate,d2018gasp} instead of Lagrange polynomials to potentially improve the rate of RPM3. However, the polynomials $\htr{t}{u}$ constructed in \cite{kakar2018rate,d2018gasp} are not guaranteed to share any evaluations (thus may require a large number of workers per cluster) and do not exist for all values of $m$ and $k$. For values of $m_u$, $k_u$ and $n_u$, $u=1,\dots,c,$ for which the polynomials in \cite{kakar2018rate,d2018gasp} can be used per cluster, the rate of RPM3 is improved if the following holds
\begin{equation*}
    \sum_{u=1}^{c} \dfrac{m_uk_u}{mk} \geq \sum_{u=1}^c {d_u}.
\end{equation*}
\end{remark}

We assume that the master sends several tasks to the workers. Let $m_I$ be the number of rows in $\mA_i$ encoded for a given task. Let $k_I$ be the number of columns in $\mB_i$ encoded for a given task. Each task is of size $m_I k_I$ where $(m_I+z)(k_I+1) - 1 = n-n_s$ to tolerate $n_s$ stragglers. The master must send $\lceil m / m_I \rceil \lceil k / k_I\rceil$ tasks to the workers. The rate\footnote{In~\cite{BXWZ20} we defined a scaled version of this rate to reflect the size of the tasks of the \fr\ scheme in comparison to those sent using RPM3. However, we change the definition of the rate to keep the rate a number between $0$ and $1$. We shall explain the effect of the size of the tasks in the sequel.} of the \fr\ scheme is given by
\begin{equation}\label{eq:rate_kakar}
    \rho_{\text{\fr}} = 
    \dfrac{m_I k_I}{(m_I+z)(k_I+1) - 1}.
\end{equation}

{\em Numerical comparison:} We compare numerically the rate $\rho_{\text{\fr}}$ of the \fr\ scheme and the rate $\rho_{RPM3}$ of RPM3. We plot $\rho_{RPM3}$ and $\rho_{\text{\fr}}$ in Figure~\ref{subfig:set1-rate} and~\ref{subfig:set2-rate} for $n=1000$, $m=2000$, $k=3000$, $c=5$ and the system parameters {summarized} in Table~\ref{tab:parameters}.
\begin{table}[h]
\centering
\renewcommand{\arraystretch}{1.7}
\begin{tabular}{ |C{1.3cm}|c|c|c|C{1.3cm}| } 
\cline{1-2}\cline{4-5}
  \multicolumn{2}{|c|}{\small Clustering of the workers} & &\multicolumn{2}{c|}{\small Relative computation power}  \\
\cline{1-2}\cline{4-5}
\multirow{5}{1.3cm}{Setting 1} & $n_1=220$ & & $\gamma_{1} = 12$ & \multirow{5}{1.3cm}{{\em Homogeneous} environment} \\ 
& $n_2 = 240$ & & $\gamma_{2} = 9$ & \\ 
& $n_3 = 160$ & & $\gamma_{3} = 6$ & \\ 
& $n_4 = 150$ & & $\gamma_{4} = 3$ & \\ 
& $n_5 = 230$ & & $\gamma_{5} = 1$ & \\ 
\cline{1-2}\cline{4-5}
\multirow{5}{1.3cm}{Setting 2} & $n_1=220$ & & $\gamma_{1} = 100$ & \multirow{5}{1.3cm}{{\em Heterogeneous} environment} \\ 
& $n_2 = 300$ & & $\gamma_{2} = 60$ & \\ 
& $n_3 = 190$ & & $\gamma_{3} = 10$ & \\ 
& $n_4 = 160$ & & $\gamma_{4} = 3$ & \\ 
& $n_5 = 130$ & & $\gamma_{5} = 1$ & \\ 
\cline{1-2}\cline{4-5}
\end{tabular}
\caption{Parameters for the numerical simulations. The first two columns represent two different settings, where clusters consist of different size. The other columns show two scenarios of heterogeneity of the clusters, i.e., different $\gamma_{u,c}$ parameters}
\label{tab:parameters}
\end{table}
We {set} the maximum number of tolerated stragglers $n_s$ for the \fr\ scheme to be equal to the number of workers in the slowest clusters, i.e., $n_s=n_5$. For both settings we consider two scenarios; homogeneous and heterogeneous clusters. The respective expected service rates of workers in each cluster are shown in the last two columns of Table~\ref{tab:parameters}.

The \fr\ scheme tolerates $z\leq 256$ for setting~1 and $z \leq 65$ for setting~2. This restriction is dictated in a way such that the values of $m_I$ and $k_I$ satisfy $(m_I+z)(k_I+1) - 1=n-n_s$. RPM3 tolerates $z\leq 109$ for both setting since it is restricted by the number of workers in cluster $1$, see~\eqref{eq:clustering}. {The rate of the \fr\ scheme depends on $n_s$. For the two considered settings, the rate behaviour is different. For the first setting, the rate of RPM3 is always smaller than the rate of the \fr\ scheme (\fig{\ref{subfig:set1-rate}}). However, RPM3 can have higher rate than the \fr\ scheme in the second setting (\fig{\ref{subfig:set2-rate}}). More precisely, RPM3 has a better rate for $z \geq 45$. Despite the rate loss due to the Lagrange polynomials and the penalty overhead of clustering, the decrease in rate for RPM3 is slower in $z$. This allows RPM3 to have higher rate for large values of $z$.}
%
%
%
%
%

{\em Analytical comparison: }We compare the rates of RPM3 and the \fr\ scheme as follows
\begin{equation}\label{eq:ratio1}
\displaystyle
    \dfrac{\rho_{\text{\fr}}}{\rho_{RPM3}} =
     \dfrac{m_Ik_I}{mk}\dfrac{2mk(1+\varepsilon)+ (z-1)\tau_c\sum_{u=1}^c\gamma_{u} + z\tau_c \gamma_{1}} {(m_I+z)(k_I+1) - 1}.
\end{equation}

Let $\displaystyle D\triangleq(z-1)\tau_c\sum_{u=1}^c\gamma_{u} + z\tau_c \gamma_{1}$. From~\eqref{eq:ratio1} we deduce that $\rho_{RPM3}$ is smaller than $\rho_{\text{\fr}}$ when the following holds.%
\begin{align*}
     D \frac{m_Ik_I}{mk}& \leq m_I + z (k_I + 1) - m_I k_I(1+2\epsilon) - 1 .
\end{align*}
For small values of $m$ and $k$, the left hand side is larger than the right hand side and therefore the rate of the \fr\ scheme is better than the rate of RPM3. However, in the regime of interest where the tasks sent to the workers are small, i.e., $m$ and $k$ are large, the inequality depends mostly on $z$ and $k_I$. For small values of $z$ the left hand side is larger than the right hand side. However, for large values of $k_I$ when $z$ increases the left hand side has a smaller increase the right hand side which makes RPM3 better for larger $z$. For $z = 1$, the left hand side is equal to $\tau_c\gamma_{1}m_Ik_I/mk\geq 0$ and the right hand side is equal to $n-n_s - 2m_I k_I(1+\epsilon)$. When $z$ increases by $\Delta z$, $m_I$ and/or $k_I$ decrease so that $(m_I+z)(k_I+1) - 1= n-n_s$, and $\tau_c$ increases because $d_u$ decrease and $\sum_{u=1}^{c}2d_u\tau_c\gamma_{u} = 2mk(1+\epsilon)$. The values of $\gamma_{u}$ do not depend on $z$. For fixed $k_I$, $m_I+z$ remains a constant. Hence, $m_Ik_I$ decreases as $k_I\Delta z$ and the right hand side of the equation increases by a factor of $\Delta z+2k_I\Delta z(1+\epsilon)$. Whereas on the left hand side, $D$ increases with $z$ by a factor $\Delta \tau_c$ and $m_I k_I$ decreases as $k_I \Delta z$.


Despite the loss of rate for RPM3, the crucial advantage of RPM3 is the reduced time spent at the master to finish its computation. In RPM3, the master waits until each worker of the slowest cluster computes $\tau_c$ tasks. Whereas in the \fr\ scheme the master waits until every non-straggling worker computes $\lceil m/ m_I\rceil \lceil k/k_I \rceil$ tasks. In particular, assume that the slowest non-straggler in the \fr\ scheme belongs to the slowest cluster in RPM3. If $\tau_c<\lceil m/ m_I\rceil \lceil k/k_I \rceil$, then in RPM3 the master waits for the slowest workers to compute a smaller number of tasks which increases the speed of the computation with high probability. In Figure~\ref{subfig:set1-scaled} and~\ref{subfig:set1-erlang} we plot the average waiting time at the master for the same schemes and parameters used for Figure~\ref{subfig:set1-rate} when the time spent at the workers to compute a task is an exponential random variable. To understand the improvement brought by RPM3, we analyze next the waiting time at the master for different schemes and show for which parameter regimes RPM3 outperforms the \fr\ scheme.

%% file: Figures/Setting1-rate-vs-z-m=2000-k=3000.tex
%
%
\definecolor{bleudefrance}{rgb}{0.19, 0.55, 0.91}
\definecolor{bittersweet}{rgb}{1.0, 0.44, 0.37}
\definecolor{britishracinggreen}{rgb}{0.0, 0.26, 0.15}
\begin{tikzpicture}
%
\begin{axis}[%
width=0.951\figurewidth,
height=\figureheight,
at={(0\figurewidth,0\figureheight)},
scale only axis,
xmin=0,
xmax=250,
xlabel style={font=\color{white!15!black}, font = \small, yshift=0.2cm},
xlabel={Number of colluding workers z},
tick label style={font=\tiny},
ymin=0.08,
ymax=0.7,
yminorticks=true,
ylabel style={font=\color{white!15!black}, font = \small, yshift=-0.5cm},
ylabel={Rate}, 
axis background/.style={fill=white},
legend style={legend cell align=left, align=left, draw=white!15!black, font =\scriptsize}
]

\addplot [color=bittersweet, line width = 1pt]
  table[row sep=crcr]{%
1	0.664935064935065\\
2	0.662337662337662\\
3	0.65974025974026\\
4	0.657142857142857\\
5	0.654545454545455\\
6	0.651948051948052\\
7	0.649350649350649\\
8	0.646753246753247\\
9	0.644155844155844\\
10	0.641558441558442\\
11	0.638961038961039\\
12	0.636363636363636\\
13	0.633766233766234\\
14	0.631168831168831\\
15	0.628571428571429\\
16	0.625974025974026\\
17	0.623376623376623\\
18	0.620779220779221\\
19	0.618181818181818\\
20	0.615584415584416\\
21	0.612987012987013\\
22	0.61038961038961\\
23	0.607792207792208\\
24	0.605194805194805\\
25	0.602597402597403\\
26	0.6\\
27	0.597402597402597\\
28	0.594805194805195\\
29	0.592207792207792\\
30	0.58961038961039\\
31	0.587012987012987\\
32	0.584415584415584\\
33	0.581818181818182\\
34	0.579220779220779\\
35	0.576623376623377\\
36	0.574025974025974\\
37	0.571428571428571\\
38	0.568831168831169\\
39	0.566233766233766\\
40	0.563636363636364\\
41	0.561038961038961\\
42	0.558441558441558\\
43	0.555844155844156\\
44	0.553246753246753\\
45	0.550649350649351\\
46	0.548051948051948\\
47	0.545454545454545\\
48	0.542857142857143\\
49	0.54025974025974\\
50	0.537662337662338\\
51	0.535064935064935\\
52	0.532467532467532\\
53	0.52987012987013\\
54	0.527272727272727\\
55	0.524675324675325\\
56	0.522077922077922\\
57	0.519480519480519\\
58	0.516883116883117\\
59	0.514285714285714\\
60	0.511688311688312\\
61	0.509090909090909\\
62	0.506493506493506\\
63	0.503896103896104\\
64	0.501298701298701\\
65	0.498701298701299\\
66	0.496103896103896\\
67	0.493506493506494\\
68	0.490909090909091\\
69	0.488311688311688\\
70	0.485714285714286\\
71	0.483116883116883\\
72	0.480519480519481\\
73	0.477922077922078\\
74	0.475324675324675\\
75	0.472727272727273\\
76	0.47012987012987\\
77	0.467532467532468\\
78	0.464935064935065\\
79	0.462337662337662\\
80	0.45974025974026\\
81	0.457142857142857\\
82	0.454545454545455\\
83	0.451948051948052\\
84	0.449350649350649\\
85	0.446753246753247\\
86	0.444155844155844\\
87	0.441558441558442\\
88	0.438961038961039\\
89	0.436363636363636\\
90	0.433766233766234\\
91	0.431168831168831\\
92	0.428571428571429\\
93	0.425974025974026\\
94	0.423376623376623\\
95	0.420779220779221\\
96	0.418181818181818\\
97	0.415584415584416\\
98	0.412987012987013\\
99	0.41038961038961\\
100	0.407792207792208\\
101	0.405194805194805\\
102	0.402597402597403\\
103	0.4\\
104	0.397402597402597\\
105	0.394805194805195\\
106	0.392207792207792\\
107	0.38961038961039\\
108	0.387012987012987\\
109	0.384415584415584\\
110	0.381818181818182\\
111	0.379220779220779\\
112	0.376623376623377\\
113	0.374025974025974\\
114	0.371428571428571\\
115	0.368831168831169\\
116	0.366233766233766\\
117	0.363636363636364\\
118	0.361038961038961\\
119	0.358441558441558\\
120	0.355844155844156\\
121	0.353246753246753\\
122	0.350649350649351\\
123	0.348051948051948\\
124	0.345454545454545\\
125	0.342857142857143\\
126	0.34025974025974\\
127	0.337662337662338\\
128	0.335064935064935\\
129	0.332467532467532\\
130	0.32987012987013\\
131	0.327272727272727\\
132	0.324675324675325\\
133	0.322077922077922\\
134	0.319480519480519\\
135	0.316883116883117\\
136	0.314285714285714\\
137	0.311688311688312\\
138	0.309090909090909\\
139	0.306493506493506\\
140	0.303896103896104\\
141	0.301298701298701\\
142	0.298701298701299\\
143	0.296103896103896\\
144	0.293506493506493\\
145	0.290909090909091\\
146	0.288311688311688\\
147	0.285714285714286\\
148	0.283116883116883\\
149	0.280519480519481\\
150	0.277922077922078\\
151	0.275324675324675\\
152	0.272727272727273\\
153	0.27012987012987\\
154	0.267532467532468\\
155	0.264935064935065\\
156	0.262337662337662\\
157	0.25974025974026\\
158	0.257142857142857\\
159	0.254545454545455\\
160	0.251948051948052\\
161	0.249350649350649\\
162	0.246753246753247\\
163	0.244155844155844\\
164	0.241558441558442\\
165	0.238961038961039\\
166	0.236363636363636\\
167	0.233766233766234\\
168	0.231168831168831\\
169	0.228571428571429\\
170	0.225974025974026\\
171	0.223376623376623\\
172	0.220779220779221\\
173	0.218181818181818\\
174	0.215584415584416\\
175	0.212987012987013\\
176	0.21038961038961\\
177	0.207792207792208\\
178	0.205194805194805\\
179	0.202597402597403\\
180	0.2\\
181	0.197402597402597\\
182	0.194805194805195\\
183	0.192207792207792\\
184	0.18961038961039\\
185	0.187012987012987\\
186	0.184415584415584\\
187	0.181818181818182\\
188	0.179220779220779\\
189	0.176623376623377\\
190	0.174025974025974\\
191	0.171428571428571\\
192	0.168831168831169\\
193	0.166233766233766\\
194	0.163636363636364\\
195	0.161038961038961\\
196	0.158441558441558\\
197	0.155844155844156\\
198	0.153246753246753\\
199	0.150649350649351\\
200	0.148051948051948\\
201	0.145454545454545\\
202	0.142857142857143\\
203	0.14025974025974\\
204	0.137662337662338\\
205	0.135064935064935\\
206	0.132467532467532\\
207	0.12987012987013\\
208	0.127272727272727\\
209	0.124675324675325\\
210	0.122077922077922\\
211	0.119480519480519\\
212	0.116883116883117\\
213	0.114285714285714\\
214	0.111688311688312\\
215	0.109090909090909\\
216	0.106493506493506\\
217	0.103896103896104\\
218	0.101298701298701\\
219	0.0987012987012987\\
220	0.0961038961038961\\
221	0.0935064935064935\\
222	0.0909090909090909\\
223	0.0883116883116883\\
224	0.0857142857142857\\
225	0.0831168831168831\\
226	0.0805194805194805\\
227	0.0779220779220779\\
228	0.0753246753246753\\
229	0.0727272727272727\\
230	0.0701298701298701\\
231	0.0675324675324675\\
232	0.0649350649350649\\
233	0.0623376623376623\\
234	0.0597402597402597\\
235	0.0571428571428571\\
236	0.0545454545454545\\
237	0.051948051948052\\
238	0.0493506493506494\\
239	0.0467532467532468\\
240	0.0441558441558442\\
241	0.0415584415584416\\
242	0.038961038961039\\
243	0.0363636363636364\\
244	0.0337662337662338\\
245	0.0311688311688312\\
246	0.0285714285714286\\
247	0.025974025974026\\
248	0.0233766233766234\\
249	0.0207792207792208\\
250	0.0181818181818182\\
251	0.0155844155844156\\
252	0.012987012987013\\
253	0.0103896103896104\\
254	0.00779220779220779\\
255	0.00519480519480519\\
256	0.0025974025974026\\
};
\addlegendentry{\fr\ scheme~\cite{kakar2018rate}}

\addplot [color=bleudefrance, line width = 1pt, dotted]
  table[row sep=crcr]{%
1	0.452709921590642\\
2	0.448352751989192\\
3	0.446633278347814\\
4	0.442391746739573\\
5	0.440509228668341\\
6	0.436178379510084\\
7	0.434550848243256\\
8	0.430137127716316\\
9	0.428554286399973\\
10	0.424067828235567\\
11	0.422337723768604\\
12	0.417979819934294\\
13	0.41629893593992\\
14	0.411881971102361\\
15	0.410249677954003\\
16	0.405959485243373\\
17	0.404198272726048\\
18	0.399861381387786\\
19	0.398152572065616\\
20	0.393777269075884\\
21	0.392119957337349\\
22	0.387714115120075\\
23	0.385947397942643\\
24	0.381678418121074\\
25	0.379966157680889\\
26	0.375524795902273\\
27	0.373867182437215\\
28	0.369419912225829\\
29	0.367815640992318\\
30	0.363368572009961\\
31	0.361675861270784\\
32	0.357375150991001\\
33	0.355601915508985\\
34	0.351311092999073\\
35	0.349597380350297\\
36	0.34519341186867\\
37	0.343538723684934\\
38	0.339161637653273\\
39	0.337441875636922\\
40	0.333098683816067\\
41	0.331439707802754\\
42	0.327018960559333\\
43	0.325306221589923\\
44	0.320935934768701\\
45	0.31928616128634\\
46	0.31486212187683\\
47	0.313168745759173\\
48	0.308809088045588\\
49	0.307078780037423\\
50	0.302787461369366\\
51	0.301026500366249\\
52	0.296712426316414\\
53	0.295021221859892\\
54	0.290604185475215\\
55	0.288892098801098\\
56	0.284568423528876\\
57	0.282840624964645\\
58	0.27844501453483\\
59	0.276790558120481\\
60	0.272420180887\\
61	0.270678971131185\\
62	0.266345631398954\\
63	0.264605799100693\\
64	0.260244977271939\\
65	0.258583685438118\\
66	0.254209282028251\\
67	0.252486996919659\\
68	0.248116383124778\\
69	0.246410213867638\\
70	0.242056902736695\\
71	0.240370747841471\\
72	0.235987086786611\\
73	0.234325205893748\\
74	0.229930162545296\\
75	0.228240327935703\\
76	0.223853200056113\\
77	0.222145211882103\\
78	0.217783314313591\\
79	0.216116218657745\\
80	0.211696664083967\\
81	0.210026379313242\\
82	0.205668216033894\\
83	0.203957315812947\\
84	0.199588050264255\\
85	0.197892314917914\\
86	0.193494707919738\\
87	0.191821498586276\\
88	0.187460164714998\\
89	0.185741032422955\\
90	0.181365975984727\\
91	0.179686985271657\\
92	0.175292007268775\\
93	0.173625981854927\\
94	0.1692419652377\\
95	0.167537676431135\\
96	0.163165852650357\\
97	0.161469154009279\\
98	0.157112482062992\\
99	0.155409010614435\\
100	0.151025614951133\\
101	0.149331145797971\\
102	0.144962971624948\\
103	0.143268801881215\\
104	0.138890432115912\\
105	0.137192905114643\\
106	0.132831054837973\\
107	0.131129770391772\\
108	0.12675877804538\\
109	0.125057317937388\\
};
\addlegendentry{RPM3 - Homogeneous}

\addplot [color=britishracinggreen, line width = 1pt, dashed]
  table[row sep=crcr]{%
1	0.450927106130204\\
2	0.447028600889885\\
3	0.445744035944799\\
4	0.440678762809063\\
5	0.438189052284718\\
6	0.434506791341149\\
7	0.432086140693009\\
8	0.428505316322625\\
9	0.426150891507665\\
10	0.421518816599973\\
11	0.419240336510243\\
12	0.415868430318472\\
13	0.412550331140399\\
14	0.409284761236913\\
15	0.406070483007304\\
16	0.402906297425429\\
17	0.399791042548428\\
18	0.39571154211426\\
19	0.393702854083224\\
20	0.389746041479372\\
21	0.386830235682768\\
22	0.383009690145161\\
23	0.381127578645676\\
24	0.376502243953374\\
25	0.374683392533309\\
26	0.370212230331241\\
27	0.368453502396176\\
28	0.364128930771808\\
29	0.361582574612564\\
30	0.357416876748364\\
31	0.354963213978925\\
32	0.350947793006312\\
33	0.348581852828742\\
34	0.344708721130645\\
35	0.342425881917859\\
36	0.337949726598671\\
37	0.335755247854524\\
38	0.332160437920321\\
39	0.329339542481508\\
40	0.325196906727023\\
41	0.323164426059979\\
42	0.319174741787634\\
43	0.316569233691408\\
44	0.312739767154819\\
45	0.31023784901758\\
46	0.305954486210631\\
47	0.303559539156145\\
48	0.299457383221602\\
49	0.297162690629866\\
50	0.293230481113025\\
51	0.291029877127186\\
52	0.286726292992218\\
53	0.284621879832642\\
54	0.280504384283526\\
55	0.277990904137617\\
56	0.274061704079134\\
57	0.271661864288599\\
58	0.267446421566879\\
59	0.26516055471588\\
60	0.261142970553519\\
61	0.258963146091469\\
62	0.254710877682742\\
63	0.252225893510228\\
64	0.248190279214064\\
65	0.245830308254818\\
66	0.241995202041794\\
67	0.239381056340725\\
68	0.235385317919257\\
69	0.232911298061246\\
70	0.229126919510768\\
71	0.226782053375424\\
72	0.222552258979613\\
73	0.220339381404531\\
74	0.216043070346504\\
75	0.213662430452879\\
76	0.209620168255122\\
77	0.207378241321912\\
78	0.203301342737602\\
79	0.200931249363718\\
80	0.196851427041612\\
81	0.194384617178935\\
82	0.190329968722442\\
83	0.188022938798533\\
84	0.184008214126679\\
85	0.181638084904906\\
86	0.177481606989462\\
87	0.175275620913887\\
88	0.171024172556549\\
89	0.168790995112938\\
90	0.164669771240754\\
91	0.162258289235136\\
92	0.158284616845704\\
93	0.155898416591749\\
94	0.151779769578072\\
95	0.149440197021956\\
96	0.145378560926701\\
97	0.142966750699346\\
98	0.138995452068808\\
99	0.136548349039428\\
100	0.132467057650547\\
101	0.130133325930193\\
102	0.126112946755114\\
103	0.12369930184114\\
104	0.11959824557347\\
105	0.117247864330151\\
106	0.113225492342183\\
107	0.110878430671044\\
108	0.106757690646105\\
109	0.104386894016682\\
};
\addlegendentry{RPM3 - Heterogeneous}

\end{axis}
\end{tikzpicture}%

%% file: Figures/scaled-set1-vs-z-m=2000-k=3000-s=100-mu=1000-exp=100.tex
%
%
%
\definecolor{mycolor1}{rgb}{0.0, 0.58, 0.71}
\definecolor{mycolor2}{rgb}{1.0, 0.65, 0.0}
\definecolor{mycolor3}{rgb}{0.8, 0.31, 0.36}
\definecolor{mycolor4}{rgb}{0.09, 0.45, 0.27}
\begin{tikzpicture}

\begin{axis}[%
width=0.951\figurewidth,
height=\figureheight,
at={(0\figurewidth,0\figureheight)},
scale only axis,
xmin=0,
xmax=265,
xlabel style={font=\color{white!15!black}, font = \small, yshift=0.2cm},
xlabel={Number of colluding workers z},
tick label style={font=\tiny},
ymode=log,
ymin=0.4,
ymax=1000,
yminorticks=true,
ylabel style={font=\color{white!15!black}, font = \small, yshift=-0.5cm},
ylabel={Waiting time}, 
axis background/.style={fill=white},
legend style={legend cell align=left, align=left, draw=white!15!black, font =\scriptsize, at = {(0,1)}, anchor = north west}
]
\addplot [color=mycolor1, dotted, mark=none, mark options={solid, mycolor1}, line width = 1pt]
  table[row sep=crcr]{%
1	2.64981021526051\\
2	2.62880187917758\\
3	2.68175285590732\\
4	2.60199416388582\\
5	2.70631668735611\\
6	2.72820421269209\\
7	2.73595745139773\\
8	2.74780556468997\\
9	2.75175591809391\\
10	2.76550267201894\\
11	2.82064461339855\\
12	2.90731007625124\\
13	2.81653297545948\\
14	2.87028285747134\\
15	2.80969349851864\\
16	2.91006456888509\\
17	2.98798061827175\\
18	2.97746881443346\\
19	2.94953071519945\\
20	2.93906401572888\\
21	3.03377673148333\\
22	2.94856702857895\\
23	3.07790710873341\\
24	3.1365899121219\\
25	3.08744859761023\\
26	3.18417492059472\\
27	3.21851938557707\\
28	3.22581920783744\\
29	3.2992242565346\\
30	3.21671816056814\\
31	3.31322871360057\\
32	3.24849458645311\\
33	3.27648162554987\\
34	3.29020321500755\\
35	3.37321744480175\\
36	3.43674680994762\\
37	3.46594552800184\\
38	3.51044070270294\\
39	3.59897281128164\\
40	3.53877073287988\\
41	3.57778947657727\\
42	3.61738516239577\\
43	3.55429250804843\\
44	3.73575769998943\\
45	3.66990227066825\\
46	3.74825940578157\\
47	3.79737116373415\\
48	3.84083188703877\\
49	3.70815562217224\\
50	3.84256721127915\\
51	3.90227152355229\\
52	4.01013703334107\\
53	3.95340140324411\\
54	3.97889886040311\\
55	4.09734688697814\\
56	4.13256328912186\\
57	4.16175463136977\\
58	4.12597537184908\\
59	4.24812523054642\\
60	4.36322649468651\\
61	4.33515129805725\\
62	4.30797048000216\\
63	4.43565765389892\\
64	4.54226653957916\\
65	4.60089138224536\\
66	4.74699046016749\\
67	4.75801237745882\\
68	4.77285270835812\\
69	4.72844489507639\\
70	4.88943321824746\\
71	4.83055561217075\\
72	4.90632023775044\\
73	4.95814489678511\\
74	4.96924797551685\\
75	5.12606955123641\\
76	5.35569106761249\\
77	5.44812109906841\\
78	5.42965023012641\\
79	5.41641476252975\\
80	5.85130962106556\\
81	5.62531299147585\\
82	5.7472757134105\\
83	5.85792303534535\\
84	5.9155890500043\\
85	6.03550504038336\\
86	6.22198487478911\\
87	6.21881497716803\\
88	6.24417991186473\\
89	6.64369568094623\\
90	6.53092929070135\\
91	6.67191377736543\\
92	6.76185624108436\\
93	6.64478331909053\\
94	7.02651050489631\\
95	7.09703479795951\\
96	7.2264327538551\\
97	7.18417193115282\\
98	7.62561576583249\\
99	7.60262686501089\\
100	7.65631536206243\\
101	8.01841673949865\\
102	8.02878489331865\\
103	8.2074975892786\\
104	8.14316955555644\\
105	8.58106172560717\\
106	8.50679475619414\\
107	8.95189102619333\\
108	9.14306793953416\\
109	9.39065963854684\\
};
\addlegendentry{RPM3 - Homogeneous}

\addlegendimage{color=mycolor3, only marks, mark=+, mark size = 2pt};
\addlegendentry{\fr\ scheme~\cite{kakar2018rate} - Homogeneous};
\addlegendimage{color=mycolor4, only marks, mark=o, mark size = 2pt};
\addlegendentry{\fr\ scheme~\cite{kakar2018rate} - Heterogeneous};
\addlegendimage{color=mycolor2, solid, line width = 1pt};
\addlegendentry{RPM3 - Heterogeneous};
\addplot [color=mycolor3, only marks, mark=+, mark options={solid, mycolor3}, mark size = 0.5pt]
  table[row sep=crcr]{%
1	0.879310799492365\\
2	0.865950385076752\\
3	0.859563154650205\\
4	0.862585421407764\\
5	0.861084301484378\\
6	0.865197351494956\\
7	0.868997648115174\\
8	0.980053492508223\\
9	0.966498359398763\\
10	0.967830443879773\\
11	0.975112023951606\\
12	0.974367890227269\\
13	0.970891118198766\\
14	0.971973662635536\\
15	0.979667186671699\\
16	0.972117823421057\\
17	0.974724047856215\\
18	0.965692596651316\\
19	0.97212112334427\\
20	0.968987314068088\\
21	0.968645293532332\\
22	0.975760643570873\\
23	0.971735241436995\\
24	0.974831525929003\\
25	0.96914130364626\\
26	0.976944438162248\\
27	0.970186959885842\\
28	0.980026593864967\\
29	0.972198052882721\\
30	0.973138478529255\\
31	0.973275949834295\\
32	0.962927875629078\\
33	0.973657774692276\\
34	0.971164027479141\\
35	1.07013379280142\\
36	1.08234728097931\\
37	1.0751837345991\\
38	1.07677291881841\\
39	1.08195527817028\\
40	1.0844036957972\\
41	1.07776574680297\\
42	1.07870794327193\\
43	1.08554846146426\\
44	1.08451080896833\\
45	1.08463598972413\\
46	1.07037900956961\\
47	1.08045401834197\\
48	1.08193357195218\\
49	1.06389241899029\\
50	1.08204969658426\\
51	1.08247935733099\\
52	1.08141343704405\\
53	1.08958331846101\\
54	1.07976335486538\\
55	1.08169369253879\\
56	1.0879446553268\\
57	1.08145308492864\\
58	1.19401381241364\\
59	1.18282719081211\\
60	1.18691766571208\\
61	1.19590602352342\\
62	1.18864039440011\\
63	1.19083354541058\\
64	1.1858021707088\\
65	1.1977968114747\\
66	1.18251310419583\\
67	1.19661402806221\\
68	1.1972401057646\\
69	1.19670219097261\\
70	1.19697012939462\\
71	1.19562700157455\\
72	1.19169910289445\\
73	1.18923459923175\\
74	1.18227252847568\\
75	1.18540933034962\\
76	1.29563560049909\\
77	1.28598212911549\\
78	1.29753724164464\\
79	1.28783434475323\\
80	1.29428789607114\\
81	1.29738489655096\\
82	1.30343943329321\\
83	1.30284893514175\\
84	1.29396486421914\\
85	1.29591701240944\\
86	1.28864172542306\\
87	1.29605447355336\\
88	1.29380815466875\\
89	1.30377004477274\\
90	1.29498799717056\\
91	1.4075642665353\\
92	1.40893686430264\\
93	1.39750319497015\\
94	1.4101232030834\\
95	1.40135413803702\\
96	1.38831255460668\\
97	1.4093278158397\\
98	1.39149907805992\\
99	1.40506526604644\\
100	1.39119261981918\\
101	1.41150923867863\\
102	1.40789762363677\\
103	1.40323255683446\\
104	1.51277850527911\\
105	1.50631628739772\\
106	1.51192523359519\\
107	1.50001116432069\\
108	1.53072697013123\\
109	1.51995149540916\\
110	1.51934138131265\\
111	1.50031911949527\\
112	1.52110210608712\\
113	1.51668418178114\\
114	1.50544244062426\\
115	1.60727091192059\\
116	1.63327477121505\\
117	1.60799339317126\\
118	1.62952643542755\\
119	1.62387166082317\\
120	1.62171445696596\\
121	1.62483250199515\\
122	1.62693592315339\\
123	1.60970968898583\\
124	1.73264466233216\\
125	1.73924979150141\\
126	1.74319284167785\\
127	1.72973003132712\\
128	1.73519587247569\\
129	1.71849586800672\\
130	1.72823352823808\\
131	1.72892854696342\\
132	1.71135793226496\\
133	1.83426473380515\\
134	1.84081060164517\\
135	1.8319683362415\\
136	1.82972602671237\\
137	1.83957497749535\\
138	1.85155012584904\\
139	1.84637094238066\\
140	1.94965138107208\\
141	1.9468649004554\\
142	1.93063378824433\\
143	1.94486703959567\\
144	1.94565844290292\\
145	1.94266200039272\\
146	2.07643507044523\\
147	2.06230131258937\\
148	2.06880217609037\\
149	2.0458135959573\\
150	2.06715098790112\\
151	2.06690695602669\\
152	2.16292118449178\\
153	2.15084668519641\\
154	2.15362471956585\\
155	2.15912341227348\\
156	2.1564367563536\\
157	2.16946726011422\\
158	2.2676587206627\\
159	2.29162268884064\\
160	2.25583721391151\\
161	2.26206513753742\\
162	2.35940006382086\\
163	2.39058751314201\\
164	2.37931147769333\\
165	2.36889826288184\\
166	2.38607955952732\\
167	2.47580399567975\\
168	2.49457600395347\\
169	2.47458428284139\\
170	2.46745433358111\\
171	2.59927552769384\\
172	2.60742203506519\\
173	2.57994432514923\\
174	2.70799139297643\\
175	2.7147245071626\\
176	2.72723378958895\\
177	2.69970007427533\\
178	2.83049230113194\\
179	2.80945076298366\\
180	2.82409654555684\\
181	2.90321878323189\\
182	2.90639812462445\\
183	3.028722123821\\
184	3.01929253174602\\
185	3.01865667755717\\
186	3.13633071877338\\
187	3.12175474353605\\
188	3.14506769516198\\
189	3.22582336506195\\
190	3.23928546278525\\
191	3.37220293197054\\
192	3.35454924147435\\
193	3.46341538972957\\
194	3.44892364604957\\
195	3.56148973030689\\
196	3.57363177920106\\
197	3.68179081926872\\
198	3.66887688459925\\
199	3.80740503615691\\
200	3.91731663376451\\
201	3.88541834156997\\
202	3.97594771765232\\
203	4.09277054300325\\
204	4.10637108892888\\
205	4.19438575199192\\
206	4.30399047773743\\
207	4.34300867016463\\
208	4.43264476995002\\
209	4.54872774697073\\
210	4.63787515721101\\
211	4.76555574094998\\
212	4.86572104974515\\
213	4.96318991197106\\
214	5.05952130298364\\
215	5.15071310536721\\
216	5.29520831469565\\
217	5.42057077188134\\
218	5.6805329466969\\
219	5.72496576066828\\
220	5.9818848691356\\
221	6.03176487646831\\
222	6.28353673393007\\
223	6.36221102624498\\
224	6.54840758784422\\
225	6.79630318691604\\
226	7.07440154374865\\
227	7.17090801918308\\
228	7.52321225014604\\
229	7.80969041421808\\
230	8.06345401735089\\
231	8.30454730470528\\
232	8.62457733977346\\
233	9.08110957454905\\
234	9.34076812480192\\
235	9.9217859133018\\
236	10.4491107207634\\
237	10.806835947303\\
238	11.560614095069\\
239	12.0891242791514\\
240	12.8270522881517\\
241	13.6521128914738\\
242	14.4677871445783\\
243	15.5356971017637\\
244	16.8266202070907\\
245	18.2278102498266\\
246	19.5132127962011\\
247	21.5178706285991\\
248	23.8809879813617\\
249	27.2579345318869\\
250	31.0248423429328\\
251	35.5060826290653\\
252	43.3941005630055\\
253	53.8887035986303\\
254	71.8173913788893\\
255	109.48071270997\\
256	215.07164904932\\
};

\addplot [color=mycolor4, only marks, mark=o, mark options={solid, mycolor4}, mark size = 0.5pt]
  table[row sep=crcr]{%
1	0.763375912728432\\
2	0.759710514423163\\
3	0.772836775258338\\
4	0.764002022748855\\
5	0.765266540374356\\
6	0.767638086833723\\
7	0.763536077599703\\
8	0.867845697723851\\
9	0.866616042563247\\
10	0.860637823297394\\
11	0.863925481914477\\
12	0.862739315628761\\
13	0.865538157790476\\
14	0.852528490046849\\
15	0.854522111857067\\
16	0.849823777259183\\
17	0.852241900796817\\
18	0.848340973654875\\
19	0.853193044932312\\
20	0.854508002642754\\
21	0.854707332651197\\
22	0.871879076976538\\
23	0.861842986264457\\
24	0.852438843956144\\
25	0.852591082804575\\
26	0.855106532104441\\
27	0.854577384629191\\
28	0.863003450746603\\
29	0.857294955435274\\
30	0.854260136176278\\
31	0.864343133850721\\
32	0.857258653011654\\
33	0.875279951119076\\
34	0.848813748254946\\
35	0.951019509200043\\
36	0.947436062867624\\
37	0.946461025711445\\
38	0.951308217847093\\
39	0.955307009392025\\
40	0.959875284487428\\
41	0.957053245195442\\
42	0.946515108113832\\
43	0.95115936810187\\
44	0.965133069104929\\
45	0.946374214173125\\
46	0.952607632815452\\
47	0.947980227475836\\
48	0.951502220478408\\
49	0.952222006509388\\
50	0.962115864377146\\
51	0.959146096634362\\
52	0.951915912794546\\
53	0.960792711765281\\
54	0.947801037587487\\
55	0.957003175603927\\
56	0.961277673986182\\
57	0.945943877282371\\
58	1.06050974841385\\
59	1.06377692192996\\
60	1.05347930815732\\
61	1.07035573683182\\
62	1.04322485194029\\
63	1.05991907828321\\
64	1.05420700411856\\
65	1.04461797908527\\
66	1.04576599309763\\
67	1.05921190683229\\
68	1.05325193980057\\
69	1.04402746766981\\
70	1.05317835927116\\
71	1.04777029904069\\
72	1.0379792611226\\
73	1.0443884978413\\
74	1.05051299986778\\
75	1.06020121765895\\
76	1.15012069362737\\
77	1.15120602579702\\
78	1.14589747343256\\
79	1.15473518327684\\
80	1.14294536698395\\
81	1.1425647668089\\
82	1.14698247647102\\
83	1.14682469641177\\
84	1.13154187382727\\
85	1.14204181654017\\
86	1.14564531085159\\
87	1.14890154182329\\
88	1.13994829655719\\
89	1.13657113601503\\
90	1.14349724893679\\
91	1.25301442806729\\
92	1.25039938026637\\
93	1.24389696671633\\
94	1.24301312662152\\
95	1.23473473769425\\
96	1.23129861740062\\
97	1.24619587823634\\
98	1.22793948230619\\
99	1.22903267220411\\
100	1.23926058937003\\
101	1.25511520427971\\
102	1.24417262849722\\
103	1.22839079700477\\
104	1.32986388727522\\
105	1.33156416736873\\
106	1.33195442286975\\
107	1.32094524611208\\
108	1.3392402215088\\
109	1.34359622637916\\
110	1.34335666036199\\
111	1.33433463278195\\
112	1.33554158392638\\
113	1.32735907659997\\
114	1.33860940518125\\
115	1.41990536951073\\
116	1.44262042991234\\
117	1.4276403758472\\
118	1.42588551285798\\
119	1.42990270126787\\
120	1.42701199609788\\
121	1.42148802158891\\
122	1.43326596729534\\
123	1.43718693501454\\
124	1.52567610799041\\
125	1.52426069521736\\
126	1.51538788345122\\
127	1.51771805479426\\
128	1.53020280870045\\
129	1.52364005746599\\
130	1.52331575306666\\
131	1.53455289148738\\
132	1.51937555345502\\
133	1.6379355965443\\
134	1.60109145521983\\
135	1.62442848786636\\
136	1.61783805043913\\
137	1.62005100936013\\
138	1.62035492715111\\
139	1.61472605906497\\
140	1.71089432814304\\
141	1.71859974084756\\
142	1.71299122970713\\
143	1.72654531230403\\
144	1.70754304855809\\
145	1.71441173620863\\
146	1.83272210347763\\
147	1.81798453567062\\
148	1.81205651945255\\
149	1.82398142387919\\
150	1.82477043040602\\
151	1.82753960759941\\
152	1.89986956362326\\
153	1.91425944094854\\
154	1.92194694652004\\
155	1.90613196414739\\
156	1.90889549501867\\
157	1.92624985341788\\
158	1.98205188930666\\
159	2.00422814617294\\
160	1.99683812905607\\
161	2.00322314852879\\
162	2.09102367338731\\
163	2.13114051588005\\
164	2.10643480694831\\
165	2.08298617389203\\
166	2.10464649141565\\
167	2.21289327971634\\
168	2.19412240043835\\
169	2.18208292812091\\
170	2.17182083130255\\
171	2.27601702209054\\
172	2.27490744101496\\
173	2.29284035127136\\
174	2.38626793325672\\
175	2.40443947373077\\
176	2.37587358893907\\
177	2.39691275333682\\
178	2.50529592826001\\
179	2.48597490726347\\
180	2.50118414304857\\
181	2.57053980298048\\
182	2.5714045797779\\
183	2.65844555186785\\
184	2.68000476940808\\
185	2.67672567608001\\
186	2.77345179861314\\
187	2.79683445652376\\
188	2.75300388933102\\
189	2.88019528928524\\
190	2.79863149009639\\
191	2.96655309702827\\
192	2.9708095983042\\
193	3.07569718621403\\
194	3.06143113123328\\
195	3.15738627080871\\
196	3.16926769966176\\
197	3.27658653459868\\
198	3.25449236063714\\
199	3.3514172262007\\
200	3.42859802696068\\
201	3.42507406664027\\
202	3.53817811709883\\
203	3.59518370390477\\
204	3.64845482188182\\
205	3.70001930303728\\
206	3.81863918684923\\
207	3.8003459245813\\
208	3.8908236072925\\
209	3.99724035289179\\
210	4.10611230093566\\
211	4.14778336539574\\
212	4.32903217381811\\
213	4.42498075583549\\
214	4.51115622630925\\
215	4.54301094483467\\
216	4.69830357198442\\
217	4.78128979370449\\
218	4.94395170883727\\
219	5.00082425641603\\
220	5.22371526279043\\
221	5.37182074468053\\
222	5.53247564726795\\
223	5.59952260750684\\
224	5.85999713761211\\
225	6.00359647138191\\
226	6.22775892064466\\
227	6.38918220360292\\
228	6.61326781802256\\
229	6.8764424499927\\
230	7.12851607499402\\
231	7.33004668792487\\
232	7.55016777671517\\
233	7.99698000522026\\
234	8.35655830086399\\
235	8.63364838321096\\
236	9.11782288255568\\
237	9.50427901955212\\
238	10.0772448803329\\
239	10.6856830805068\\
240	11.2505830970629\\
241	11.9606013170954\\
242	12.8173433634176\\
243	13.736731291261\\
244	14.7141494509354\\
245	15.8373542316498\\
246	17.4351441812248\\
247	19.1261637157905\\
248	21.2745427337596\\
249	23.7618580629396\\
250	27.1980506477164\\
251	31.5843758345921\\
252	37.7972350646249\\
253	47.4842665613613\\
254	63.0181905186068\\
255	95.6025054840482\\
256	191.510553686552\\
};

\addplot [color=mycolor2, mark=none, mark options={solid, mycolor2}, line width = 1pt]
  table[row sep=crcr]{%
1	0.438362667605437\\
2	0.429542125907438\\
3	0.4485066970975\\
4	0.446872319837296\\
5	0.44577371114046\\
6	0.452081482166505\\
7	0.454058688677409\\
8	0.454571054890661\\
9	0.465365065076253\\
10	0.467489397178146\\
11	0.465226475235491\\
12	0.459768677588848\\
13	0.486371338519949\\
14	0.488312016931621\\
15	0.475147658015721\\
16	0.473327853789495\\
17	0.496012009586637\\
18	0.493442642189989\\
19	0.509234953121497\\
20	0.509675979181333\\
21	0.504411320616168\\
22	0.497662507571395\\
23	0.518389605697941\\
24	0.534854941370773\\
25	0.520439406932147\\
26	0.534373510576465\\
27	0.520559172077219\\
28	0.541425293193153\\
29	0.533027134988294\\
30	0.552861914513089\\
31	0.550269490221019\\
32	0.570071608046356\\
33	0.556380242034015\\
34	0.565406320542248\\
35	0.576429099121181\\
36	0.571132956678732\\
37	0.585643180921689\\
38	0.581974326786339\\
39	0.60189553798915\\
40	0.61385116348438\\
41	0.612545512275636\\
42	0.610923896052482\\
43	0.628112105733589\\
44	0.616098788298847\\
45	0.637088344644633\\
46	0.642160045104582\\
47	0.626460207473605\\
48	0.655824165011707\\
49	0.689333970511497\\
50	0.668459180754596\\
51	0.662446334429014\\
52	0.693779444031541\\
53	0.705876028814092\\
54	0.715238443273996\\
55	0.713948543785817\\
56	0.723407627060655\\
57	0.706773833496394\\
58	0.74459942731525\\
59	0.744779299077716\\
60	0.732133554829519\\
61	0.756866216348889\\
62	0.766175682630512\\
63	0.77959531914591\\
64	0.782760791703589\\
65	0.798228979754516\\
66	0.792348889876669\\
67	0.855662125186119\\
68	0.835805810506846\\
69	0.832430744272724\\
70	0.852225835240183\\
71	0.857984477184297\\
72	0.865444317480912\\
73	0.871869183731368\\
74	0.937554869165771\\
75	0.920315023836095\\
76	0.951960711689018\\
77	0.92668286931046\\
78	0.956424066119576\\
79	0.968829671420866\\
80	1.02771654050934\\
81	0.978416283278185\\
82	1.03788103416805\\
83	1.08581786558246\\
84	1.05717666888008\\
85	1.07328203104535\\
86	1.10001987820158\\
87	1.1161938738178\\
88	1.16311211466122\\
89	1.15111368320011\\
90	1.2172934986428\\
91	1.1736685050299\\
92	1.21969297032856\\
93	1.26800794787698\\
94	1.27313391595645\\
95	1.29157291904929\\
96	1.36451388390494\\
97	1.35380171721583\\
98	1.40001254403882\\
99	1.42817588033128\\
100	1.46744267079365\\
101	1.50867926865824\\
102	1.56379938667266\\
103	1.62872139294354\\
104	1.64103606466724\\
105	1.66722068334357\\
106	1.73210522392929\\
107	1.76826176500397\\
108	1.80329245008804\\
109	1.85833783285227\\
};

\end{axis}
\end{tikzpicture}%

%% file: Figures/erlang-set1-vs-z-m=2000-k=3000-s=100-mu=1000-exp=100.tex
%
%
\definecolor{mycolor1}{rgb}{0.0, 0.58, 0.71}
\definecolor{mycolor2}{rgb}{1.0, 0.65, 0.0}
\definecolor{mycolor3}{rgb}{0.8, 0.31, 0.36}
\definecolor{mycolor4}{rgb}{0.09, 0.45, 0.27}
\begin{tikzpicture}
%
\begin{axis}[%
width=0.951\figurewidth,
height=\figureheight,
at={(0\figurewidth,0\figureheight)},
scale only axis,
xmin=0,
xmax=265,
xlabel style={font=\color{white!15!black}, font = \small, yshift=0.2cm},
xlabel={Number of colluding workers z},
tick label style={font=\tiny},
ymode=log,
ymin=0.06,
ymax=100,
yminorticks=true,
ylabel style={font=\color{white!15!black}, font = \small, yshift=-0.5cm},
ylabel={Waiting time}, 
axis background/.style={fill=white},
legend style={legend cell align=left, align=left, draw=white!15!black, font =\scriptsize, at = {(0,1)}, anchor = north west}
]

\addlegendimage{color=mycolor3, only marks, mark=+, mark size = 2pt};
\addlegendentry{\fr\ scheme~\cite{kakar2018rate} - Homogeneous};

\addlegendimage{color=mycolor4, only marks, mark=o, mark size = 2pt};
\addlegendentry{\fr\ scheme~\cite{kakar2018rate} - Heterogeneous};

\addlegendimage{color=mycolor1, dotted, line width = 1 pt};
\addlegendentry{RPM3 - Homogeneous}

\addlegendimage{color=mycolor2, solid, line width = 1 pt};
\addlegendentry{RPM3 - Heterogeneous};

\addplot [color=mycolor1, dotted, mark=none, mark options={solid, mycolor1}, line width = 1pt]
  table[row sep=crcr]{%
1	0.398797505877992\\
2	0.402555576149379\\
3	0.404096888679591\\
4	0.407953668023388\\
5	0.409666224340223\\
6	0.413634058802602\\
7	0.415112886931128\\
8	0.419189788120906\\
9	0.420690177914351\\
10	0.425044082510061\\
11	0.426669296389991\\
12	0.43098464853228\\
13	0.432712902775327\\
14	0.4371483040391\\
15	0.438838273026781\\
16	0.443340602442795\\
17	0.445190933327267\\
18	0.450013351353087\\
19	0.451922011694236\\
20	0.456941271908463\\
21	0.458865548546028\\
22	0.463868345089385\\
23	0.465933337654212\\
24	0.470948646498484\\
25	0.472943069554714\\
26	0.478386308735325\\
27	0.48047137654007\\
28	0.486184500399961\\
29	0.488334459336137\\
30	0.494160781327209\\
31	0.496441851770145\\
32	0.502222817264471\\
33	0.50464972194514\\
34	0.510756062759285\\
35	0.513130015873959\\
36	0.519375965887215\\
37	0.521883694722551\\
38	0.528532808915061\\
39	0.531089696572224\\
40	0.537941668223232\\
41	0.540504105946518\\
42	0.547656547646788\\
43	0.550532670636309\\
44	0.557798999023488\\
45	0.560571416258833\\
46	0.568370281494652\\
47	0.571371371030846\\
48	0.579202358686338\\
49	0.582418781033966\\
50	0.590499838477102\\
51	0.593865144027299\\
52	0.602320460055623\\
53	0.605682485444778\\
54	0.614783382381929\\
55	0.618358113907632\\
56	0.627473509081678\\
57	0.631167119926951\\
58	0.641070243156672\\
59	0.644814617724145\\
60	0.655026631426341\\
61	0.659154369451873\\
62	0.669712406752134\\
63	0.673900140737516\\
64	0.684944846540484\\
65	0.689167016452883\\
66	0.700803605170963\\
67	0.705475019912782\\
68	0.717640127572539\\
69	0.722691866715767\\
70	0.735711346281806\\
71	0.740710912824958\\
72	0.75399753984974\\
73	0.759150618043039\\
74	0.773354082484491\\
75	0.778925323484388\\
76	0.793901522336433\\
77	0.79983953963735\\
78	0.815604294598659\\
79	0.821700248841216\\
80	0.838642970280036\\
81	0.845267121238903\\
82	0.862854581188277\\
83	0.869877063286925\\
84	0.8883866314647\\
85	0.895942232549266\\
86	0.916045702936263\\
87	0.923860108287978\\
88	0.944679938981879\\
89	0.953348157847422\\
90	0.976191712548126\\
91	0.985134187119625\\
92	1.00940157482201\\
93	1.01893485601343\\
94	1.04467906862935\\
95	1.0549596859627\\
96	1.08285855508889\\
97	1.09422832790629\\
98	1.12396367552123\\
99	1.13615131926717\\
100	1.16846625124633\\
101	1.18140418799808\\
102	1.21676597710386\\
103	1.23080071235025\\
104	1.26920454547392\\
105	1.28435471989672\\
106	1.32566565779357\\
107	1.34279497573684\\
108	1.38852122354887\\
109	1.40711207353609\\
};

\addplot [color=mycolor2, mark=none, mark options={solid, mycolor2}, line width = 1pt]
  table[row sep=crcr]{%
1	0.0722242876699268\\
2	0.0728455770383496\\
3	0.0730254972545038\\
4	0.0738194281871396\\
5	0.0742055896543366\\
6	0.0748277496668231\\
7	0.0752228158225371\\
8	0.0757974273778697\\
9	0.076170087206631\\
10	0.0769640445867082\\
11	0.0773722494264231\\
12	0.0779710700544566\\
13	0.0785308224107284\\
14	0.0791053557296572\\
15	0.0796794853436661\\
16	0.0802331482801014\\
17	0.0808348889577849\\
18	0.0816279403672203\\
19	0.0820583601976449\\
20	0.0828622008140717\\
21	0.0834569640028979\\
22	0.0842057908105111\\
23	0.084600295367134\\
24	0.0855348496413465\\
25	0.0859301967851714\\
26	0.0869639058809019\\
27	0.0873773387008448\\
28	0.0882937422815934\\
29	0.0888474119375729\\
30	0.0898098455922475\\
31	0.0903498803679849\\
32	0.0912870312565677\\
33	0.0918542923781244\\
34	0.092775428431474\\
35	0.0933448342116971\\
36	0.0945466294691716\\
37	0.095083813760873\\
38	0.0960166239345186\\
39	0.0968173304961924\\
40	0.0979477992800906\\
41	0.0985089355247137\\
42	0.0997015448602476\\
43	0.100466339686425\\
44	0.101616859546032\\
45	0.102404603289399\\
46	0.103800229025286\\
47	0.104612660415736\\
48	0.105995302299001\\
49	0.106794681277755\\
50	0.108176574840051\\
51	0.108891625912196\\
52	0.110442332878395\\
53	0.111194756440637\\
54	0.11278169184568\\
55	0.11377024677877\\
56	0.11528661354545\\
57	0.116313868526745\\
58	0.118104900497612\\
59	0.119068870670654\\
60	0.120744268016815\\
61	0.121676602769711\\
62	0.123676276169548\\
63	0.124795507545306\\
64	0.126683661020161\\
65	0.127803900999703\\
66	0.129783665405981\\
67	0.131072016622578\\
68	0.1332036581747\\
69	0.134526333918887\\
70	0.13663250480384\\
71	0.137945751525793\\
72	0.140473027539685\\
73	0.141835047434473\\
74	0.144446291639559\\
75	0.145935347713409\\
76	0.148546918936699\\
77	0.150045739316082\\
78	0.153039303073977\\
79	0.154703999973067\\
80	0.157700539287367\\
81	0.159687152231696\\
82	0.162941969263496\\
83	0.164864542621741\\
84	0.16836775080219\\
85	0.170477301428916\\
86	0.174361611381579\\
87	0.176531710748142\\
88	0.180579272069072\\
89	0.182749802012817\\
90	0.187071136475288\\
91	0.189700568967051\\
92	0.194369209445302\\
93	0.197185790347861\\
94	0.202421615458041\\
95	0.20544843385189\\
96	0.210977570747765\\
97	0.214351844650093\\
98	0.220350095882562\\
99	0.224176786316515\\
100	0.230892935798299\\
101	0.235013294176888\\
102	0.242220418144878\\
103	0.246706956557019\\
104	0.255117675724155\\
105	0.260060900609735\\
106	0.269060387257996\\
107	0.2744747873944\\
108	0.28451936300092\\
109	0.290809156417885\\
};

\addplot [color=mycolor3, only marks, mark=+, mark options={solid, mycolor3}, mark size = 1.5pt]
  table[row sep=crcr]{%
1	0.749731321362557\\
2	0.749731321362557\\
3	0.749731321362557\\
4	0.749731321362557\\
5	0.749731321362557\\
6	0.749731321362557\\
7	0.749731321362557\\
8	0.842405484128658\\
9	0.842405484128658\\
10	0.842405484128658\\
11	0.842405484128658\\
12	0.842405484128658\\
13	0.842405484128658\\
14	0.842405484128658\\
15	0.842405484128658\\
16	0.842405484128658\\
17	0.842405484128658\\
18	0.842405484128658\\
19	0.842405484128658\\
20	0.842405484128658\\
21	0.842405484128658\\
22	0.842405484128658\\
23	0.842405484128658\\
24	0.842405484128658\\
25	0.842405484128658\\
26	0.842405484128658\\
27	0.842405484128658\\
28	0.842405484128658\\
29	0.842405484128658\\
30	0.842405484128658\\
31	0.842405484128658\\
32	0.842405484128658\\
33	0.842405484128658\\
34	0.842405484128658\\
35	0.934890325127308\\
36	0.934890325127308\\
37	0.934890325127308\\
38	0.934890325127308\\
39	0.934890325127308\\
40	0.934890325127308\\
41	0.934890325127308\\
42	0.934890325127308\\
43	0.934890325127308\\
44	0.934890325127308\\
45	0.934890325127308\\
46	0.934890325127308\\
47	0.934890325127308\\
48	0.934890325127308\\
49	0.934890325127308\\
50	0.934890325127308\\
51	0.934890325127308\\
52	0.934890325127308\\
53	0.934890325127308\\
54	0.934890325127308\\
55	0.934890325127308\\
56	0.934890325127308\\
57	0.934890325127308\\
58	1.02742027727898\\
59	1.02742027727898\\
60	1.02742027727898\\
61	1.02742027727898\\
62	1.02742027727898\\
63	1.02742027727898\\
64	1.02742027727898\\
65	1.02742027727898\\
66	1.02742027727898\\
67	1.02742027727898\\
68	1.02742027727898\\
69	1.02742027727898\\
70	1.02742027727898\\
71	1.02742027727898\\
72	1.02742027727898\\
73	1.02742027727898\\
74	1.02742027727898\\
75	1.02742027727898\\
76	1.11985318588893\\
77	1.11985318588893\\
78	1.11985318588893\\
79	1.11985318588893\\
80	1.11985318588893\\
81	1.11985318588893\\
82	1.11985318588893\\
83	1.11985318588893\\
84	1.11985318588893\\
85	1.11985318588893\\
86	1.11985318588893\\
87	1.11985318588893\\
88	1.11985318588893\\
89	1.11985318588893\\
90	1.11985318588893\\
91	1.2127648161775\\
92	1.2127648161775\\
93	1.2127648161775\\
94	1.2127648161775\\
95	1.2127648161775\\
96	1.2127648161775\\
97	1.2127648161775\\
98	1.2127648161775\\
99	1.2127648161775\\
100	1.2127648161775\\
101	1.2127648161775\\
102	1.2127648161775\\
103	1.2127648161775\\
104	1.30514618967008\\
105	1.30514618967008\\
106	1.30514618967008\\
107	1.30514618967008\\
108	1.30514618967008\\
109	1.30514618967008\\
110	1.30514618967008\\
111	1.30514618967008\\
112	1.30514618967008\\
113	1.30514618967008\\
114	1.30514618967008\\
115	1.3972467872374\\
116	1.3972467872374\\
117	1.3972467872374\\
118	1.3972467872374\\
119	1.3972467872374\\
120	1.3972467872374\\
121	1.3972467872374\\
122	1.3972467872374\\
123	1.3972467872374\\
124	1.48976383543461\\
125	1.48976383543461\\
126	1.48976383543461\\
127	1.48976383543461\\
128	1.48976383543461\\
129	1.48976383543461\\
130	1.48976383543461\\
131	1.48976383543461\\
132	1.48976383543461\\
133	1.58269179185286\\
134	1.58269179185286\\
135	1.58269179185286\\
136	1.58269179185286\\
137	1.58269179185286\\
138	1.58269179185286\\
139	1.58269179185286\\
140	1.67485336121274\\
141	1.67485336121274\\
142	1.67485336121274\\
143	1.67485336121274\\
144	1.67485336121274\\
145	1.67485336121274\\
146	1.76703255341535\\
147	1.76703255341535\\
148	1.76703255341535\\
149	1.76703255341535\\
150	1.76703255341535\\
151	1.76703255341535\\
152	1.85955404519911\\
153	1.85955404519911\\
154	1.85955404519911\\
155	1.85955404519911\\
156	1.85955404519911\\
157	1.85955404519911\\
158	1.95205001996333\\
159	1.95205001996333\\
160	1.95205001996333\\
161	1.95205001996333\\
162	2.04462306371799\\
163	2.04462306371799\\
164	2.04462306371799\\
165	2.04462306371799\\
166	2.04462306371799\\
167	2.13715541049757\\
168	2.13715541049757\\
169	2.13715541049757\\
170	2.13715541049757\\
171	2.22920598329332\\
172	2.22920598329332\\
173	2.22920598329332\\
174	2.32126184793004\\
175	2.32126184793004\\
176	2.32126184793004\\
177	2.32126184793004\\
178	2.41337356991074\\
179	2.41337356991074\\
180	2.41337356991074\\
181	2.50562464426544\\
182	2.50562464426544\\
183	2.59769347227863\\
184	2.59769347227863\\
185	2.59769347227863\\
186	2.68959683076906\\
187	2.68959683076906\\
188	2.68959683076906\\
189	2.78219280031205\\
190	2.78219280031205\\
191	2.87438524705382\\
192	2.87438524705382\\
193	2.96640376554957\\
194	2.96640376554957\\
195	3.05888283530292\\
196	3.05888283530292\\
197	3.15066806779468\\
198	3.15066806779468\\
199	3.24284291540182\\
200	3.33504620899404\\
201	3.33504620899404\\
202	3.42736200152417\\
203	3.51935327743723\\
204	3.51935327743723\\
205	3.61165602326887\\
206	3.70393402096591\\
207	3.70393402096591\\
208	3.79616505775071\\
209	3.88835696046899\\
210	3.98072902464795\\
211	4.07276268206215\\
212	4.16474034947654\\
213	4.25687662193995\\
214	4.34889781007742\\
215	4.44075110547329\\
216	4.53280302991874\\
217	4.62489940131541\\
218	4.80949027936095\\
219	4.90155098372066\\
220	5.08583141619017\\
221	5.17777134100427\\
222	5.36227387118886\\
223	5.45423326226269\\
224	5.63845526591634\\
225	5.82264119866887\\
226	6.00602394583151\\
227	6.19016360554938\\
228	6.37411138267844\\
229	6.64973743566913\\
230	6.92548827552279\\
231	7.1094475316889\\
232	7.38503789243435\\
233	7.75277456844769\\
234	8.02919878182696\\
235	8.39711987597035\\
236	8.85635599013144\\
237	9.22423685732075\\
238	9.77565963532219\\
239	10.3272000238849\\
240	10.8794263675221\\
241	11.5229564621127\\
242	12.3500644261108\\
243	13.1772364427073\\
244	14.1885034636865\\
245	15.383151575779\\
246	16.7600013224273\\
247	18.4139242575986\\
248	20.5273362771407\\
249	23.005815823797\\
250	26.3141899250284\\
251	30.720149002085\\
252	36.7808941016256\\
253	45.9610631653932\\
254	61.2897987211074\\
255	91.8533074333356\\
256	183.587577846564\\
};

\addplot [color=mycolor4, only marks, mark=o, mark options={solid, mycolor4}, mark size = 0.5pt]
  table[row sep=crcr]{%
1	0.749256730113756\\
2	0.749256730113756\\
3	0.749256730113756\\
4	0.749256730113756\\
5	0.749256730113756\\
6	0.749256730113756\\
7	0.749256730113756\\
8	0.842129023080965\\
9	0.842129023080965\\
10	0.842129023080965\\
11	0.842129023080965\\
12	0.842129023080965\\
13	0.842129023080965\\
14	0.842129023080965\\
15	0.842129023080965\\
16	0.842129023080965\\
17	0.842129023080965\\
18	0.842129023080965\\
19	0.842129023080965\\
20	0.842129023080965\\
21	0.842129023080965\\
22	0.842129023080965\\
23	0.842129023080965\\
24	0.842129023080965\\
25	0.842129023080965\\
26	0.842129023080965\\
27	0.842129023080965\\
28	0.842129023080965\\
29	0.842129023080965\\
30	0.842129023080965\\
31	0.842129023080965\\
32	0.842129023080965\\
33	0.842129023080965\\
34	0.842129023080965\\
35	0.934857110416997\\
36	0.934857110416997\\
37	0.934857110416997\\
38	0.934857110416997\\
39	0.934857110416997\\
40	0.934857110416997\\
41	0.934857110416997\\
42	0.934857110416997\\
43	0.934857110416997\\
44	0.934857110416997\\
45	0.934857110416997\\
46	0.934857110416997\\
47	0.934857110416997\\
48	0.934857110416997\\
49	0.934857110416997\\
50	0.934857110416997\\
51	0.934857110416997\\
52	0.934857110416997\\
53	0.934857110416997\\
54	0.934857110416997\\
55	0.934857110416997\\
56	0.934857110416997\\
57	0.934857110416997\\
58	1.02719995224394\\
59	1.02719995224394\\
60	1.02719995224394\\
61	1.02719995224394\\
62	1.02719995224394\\
63	1.02719995224394\\
64	1.02719995224394\\
65	1.02719995224394\\
66	1.02719995224394\\
67	1.02719995224394\\
68	1.02719995224394\\
69	1.02719995224394\\
70	1.02719995224394\\
71	1.02719995224394\\
72	1.02719995224394\\
73	1.02719995224394\\
74	1.02719995224394\\
75	1.02719995224394\\
76	1.11979343956886\\
77	1.11979343956886\\
78	1.11979343956886\\
79	1.11979343956886\\
80	1.11979343956886\\
81	1.11979343956886\\
82	1.11979343956886\\
83	1.11979343956886\\
84	1.11979343956886\\
85	1.11979343956886\\
86	1.11979343956886\\
87	1.11979343956886\\
88	1.11979343956886\\
89	1.11979343956886\\
90	1.11979343956886\\
91	1.21208719228578\\
92	1.21208719228578\\
93	1.21208719228578\\
94	1.21208719228578\\
95	1.21208719228578\\
96	1.21208719228578\\
97	1.21208719228578\\
98	1.21208719228578\\
99	1.21208719228578\\
100	1.21208719228578\\
101	1.21208719228578\\
102	1.21208719228578\\
103	1.21208719228578\\
104	1.30440146938838\\
105	1.30440146938838\\
106	1.30440146938838\\
107	1.30440146938838\\
108	1.30440146938838\\
109	1.30440146938838\\
110	1.30440146938838\\
111	1.30440146938838\\
112	1.30440146938838\\
113	1.30440146938838\\
114	1.30440146938838\\
115	1.39682122379278\\
116	1.39682122379278\\
117	1.39682122379278\\
118	1.39682122379278\\
119	1.39682122379278\\
120	1.39682122379278\\
121	1.39682122379278\\
122	1.39682122379278\\
123	1.39682122379278\\
124	1.48880870364757\\
125	1.48880870364757\\
126	1.48880870364757\\
127	1.48880870364757\\
128	1.48880870364757\\
129	1.48880870364757\\
130	1.48880870364757\\
131	1.48880870364757\\
132	1.48880870364757\\
133	1.58111953106286\\
134	1.58111953106286\\
135	1.58111953106286\\
136	1.58111953106286\\
137	1.58111953106286\\
138	1.58111953106286\\
139	1.58111953106286\\
140	1.67368725328853\\
141	1.67368725328853\\
142	1.67368725328853\\
143	1.67368725328853\\
144	1.67368725328853\\
145	1.67368725328853\\
146	1.76613173058666\\
147	1.76613173058666\\
148	1.76613173058666\\
149	1.76613173058666\\
150	1.76613173058666\\
151	1.76613173058666\\
152	1.85853386037734\\
153	1.85853386037734\\
154	1.85853386037734\\
155	1.85853386037734\\
156	1.85853386037734\\
157	1.85853386037734\\
158	1.95104526910268\\
159	1.95104526910268\\
160	1.95104526910268\\
161	1.95104526910268\\
162	2.04334837414389\\
163	2.04334837414389\\
164	2.04334837414389\\
165	2.04334837414389\\
166	2.04334837414389\\
167	2.13572850521339\\
168	2.13572850521339\\
169	2.13572850521339\\
170	2.13572850521339\\
171	2.2280501169605\\
172	2.2280501169605\\
173	2.2280501169605\\
174	2.32027761568997\\
175	2.32027761568997\\
176	2.32027761568997\\
177	2.32027761568997\\
178	2.412708732359\\
179	2.412708732359\\
180	2.412708732359\\
181	2.50472168452462\\
182	2.50472168452462\\
183	2.596589883748\\
184	2.596589883748\\
185	2.596589883748\\
186	2.68874015934004\\
187	2.68874015934004\\
188	2.68874015934004\\
189	2.78123871937428\\
190	2.78123871937428\\
191	2.8735576934362\\
192	2.8735576934362\\
193	2.96578127011771\\
194	2.96578127011771\\
195	3.0578341696657\\
196	3.0578341696657\\
197	3.15019621316359\\
198	3.15019621316359\\
199	3.24236302449583\\
200	3.33417197413602\\
201	3.33417197413602\\
202	3.42630026189207\\
203	3.51855774645629\\
204	3.51855774645629\\
205	3.6107083059369\\
206	3.70273769324052\\
207	3.70273769324052\\
208	3.79491972535114\\
209	3.88703360868599\\
210	3.9793310924373\\
211	4.07139112059702\\
212	4.16347065475632\\
213	4.25565383897586\\
214	4.34751896916156\\
215	4.43919787290773\\
216	4.53145760379904\\
217	4.62351949681395\\
218	4.80773044599771\\
219	4.90008702097114\\
220	5.08411934873971\\
221	5.17599431195675\\
222	5.36013773136129\\
223	5.45249489272888\\
224	5.63675901702296\\
225	5.82076394556694\\
226	6.00487594246378\\
227	6.189221550735\\
228	6.37296269587981\\
229	6.64910181054127\\
230	6.92548025242477\\
231	7.10961813245816\\
232	7.38536434278325\\
233	7.75359465564018\\
234	8.02943870560922\\
235	8.39838259106334\\
236	8.85783156075978\\
237	9.22563472559387\\
238	9.77717344931324\\
239	10.3290059410463\\
240	10.879764350512\\
241	11.5229003937838\\
242	12.3505328584333\\
243	13.177410208462\\
244	14.1872748413636\\
245	15.3816271871347\\
246	16.7595824255571\\
247	18.4131190564683\\
248	20.5270449127638\\
249	23.0073713679926\\
250	26.3135427104456\\
251	30.7204297926149\\
252	36.7794789519169\\
253	45.9574144429915\\
254	61.2874924593564\\
255	91.8455710070936\\
256	183.593401209663\\
};

\end{axis}
\end{tikzpicture}%

%% file: time_analysis.tex

We analyse the performance of RPM3 by computing the expected waiting time at the master to compute $\mA\mB$, under the service model~1 and model~2 and some simplifying assumptions for tractability of the analysis. We compare the mean waiting at the master when using RPM3 to the one when using the \fr\ scheme.

\subsection{Clustering} In the encoding process of RPM3, at every round the master clusters the workers into $c$ different clusters of fixed size $n_u$, $u\in [c]$. We assume that $n_u$ is fixed during the computing process. The workers of each cluster are assumed to have similar expected speed in computing a new task, i.e., similar service rate $\lambda$. In the delay analysis, we assume that each worker of cluster $u$, $u\in [c]$, has a compute time following a shifted exponential distribution with service rate $\lambda_u$ and shift $s_u$. Therefore, at every completed round the master obtains $n_u$ tasks computed at rate $\lambda_u$ and shift $s_u$, for all $u\in [c]$.

\subsection{Decoding} Let $\tau_u$ be the number of tasks successfully computed by workers of cluster $u$ during the whole computation process, i.e., $\tau_u$ is the number of tasks computed by all $n_u$ workers of cluster $u$. Recall that $\sum_{u=1}^c \tau_u d_u \geq km(1+\varepsilon)$ so that the master receives enough packets to decode $\mA\mB$. The variable $\varepsilon$ is the required overhead of Fountain codes and $d_u$ is the number of Fountain-coded packets $\widetilde{\mA}\widetilde{\mB}$ encoded within each task sent to cluster $u$. 

\subsection{Waiting time} The waiting time, i.e., the time spent at the master to compute $\mA\mB$, can now be expressed as the time spent to receive the last packet from a given worker that makes $\displaystyle\sum_{u=1}^c \tau_u d_u \geq km(1+\varepsilon)$. Let $\rvT_u^{i}$ be the random variable representing the time spent by worker $w_i$ in cluster $u$ during $\tau_u$ different rounds, i.e., $\rvT_u^i$ is the time spent until worker $w_i$ receives, computes and sends the result of $\tau_u$ tasks to the master. Recall that the master needs $n_u$ responses from workers in cluster $u$ to decode the $\tau_u$ packets. Let $\rvT^\star_u \triangleq \displaystyle \max_{i\in [n_u]} \rvT_u^i$ be the time spent by all the workers of cluster $u$ to receive, compute and send the result of $\tau_u$ packets to the master. The waiting time at the master is given by
\begin{equation}\label{eq:waiting_time}
    \rvT_{RPM3} = \max_{u \in \{1,\dots,c\}} \rvT^\star_u.
\end{equation}

\subsection{Probability distribution of the waiting time}
{In the following analysis we ignore the real identity of the workers and only assign identities that depend on the speed of the designated worker at a given round. More precisely, instead of referring to a worker by $w_j$, $j\in [n]$, we refer to a worker $w_i$ of cluster $u$ as a worker who falls into cluster $u \in [c]$ at position $i \in [n_u]$. This worker $w_i$ of cluster $u$ can be any of the $w_j$'s and $j$ could be different at different rounds of the algorithm. This abstraction will help us in simplifying the notations and the concepts explained in the sequel.}

\begin{theorem}[Waiting time under worker-dependent fixed service time]\label{thm:waiting_exp}
Consider a master running a private distributed matrix-matrix multiplication of two matrices $\mA$ and $\mB$ using RPM3. The multiplication is divided into $km$ different smaller multiplications. 

Consider the worker-dependent fixed service time where $\lambda_u s_u \triangleq t_m$, i.e., the handshake time $s_u$ is a constant factor of the mean service time $1/\lambda_u$ for all $u\in [c]$.
Let $u^\star$ be the value of $u$ that minimizes the ratio $\lambda_u / \tau_u$. Let $H_n$ be the $n^{\text{th}}$ harmonic sum defined as $H_n \triangleq \sum_{i=1}^n \frac{1}{i}$.
The probability distribution on the waiting time of the master $\rvT_{RPM3}$ is bounded by
\begin{equation}\label{eq:thm_cdfbound}
    \Pr(\rvT_{RPM3}>x) \leq  1-\left(1 - e^{\left( t_m - \frac{\lambda_{u^\star} mk}{\tau_{u^\star}}x\right)}\right)^n,
\end{equation}
for $x\geq \frac{s_m}{km}$ and $0$ otherwise.

The mean waiting time at the master is upper bounded by
\begin{align}\label{eq:avg_waiting_model1}
        \mathbb{E}[{\rvT_{RPM3}}] & \leq \frac{(t_m+H_n) \tau_{u^\star}}{\lambda_{u^\star} m k}.
\end{align}
\end{theorem}
The proof of Theorem~\ref{thm:waiting_exp} is given in Appendix~\ref{app:proof_waiting}. {Numerical simulations indicate that this bound is a good representation of the empirical mean waiting time. This is illustrated in~\fig{\ref{subfig:intro_bound}} for setting 1 described in Table~\ref{tab:parameters}. }

While the bound in~\eqref{eq:thm_cdfbound} is not surprising, it has important practical implications: it allows the master to tune the parameters and set a probabilistic deadline on the computation time. Assume the master wants a probabilistic guarantee on the maximum computing time, i.e., a probabilistic deadline $t_D$ given by $\Pr(\rvT_{RPM3}>t_D)\leq g_D$ where $g_D<1$ is the probabilistic guarantee. Given the number of clusters and their respective service rates $\lambda_u$, the master finds the minimum ratio $\lambda_u/\tau_u$ (maximum number of tasks per cluster) that satisfies
\begin{equation*}
    \left(1 - e^{\left( t_m -\frac{\lambda_{u^\star}mk}{\tau_{u^\star}}t_D\right)}\right)^n = 1 - g_D.
\end{equation*}
Assume that for the given $\lambda_u$'s the allowed values of $\tau_u$'s, $u=1\dots,c$, such that $\lambda_u/\tau_u \leq \lambda_{u^\star}/\tau_{u^{\star}}$, satisfy
\begin{equation*}
    \sum_{{u=}1}^c d_u \tau_u \geq km(1+\epsilon).
\end{equation*}
Then, there exists at least one possible task assignment strategy that satisfies the deadline guarantees.
\begin{figure*}[t]
\hspace*{-.5cm}
\begin{subfigure}[b]{0.33\textwidth}
\centering
 \setlength\figureheight{0.7\textwidth}
 \setlength\figurewidth{0.75\textwidth}
 \resizebox{.9\textwidth}{!}{
  \input{Figures/Setting2-rate-vs-z-m=2000-k=3000} }
  \captionsetup{width=0.8\textwidth}
  \caption{Rate of RPM3 and the \fr\ scheme.}
  \label{subfig:set2-rate}
\end{subfigure}   
\begin{subfigure}[b]{0.33\textwidth}
\centering
 \setlength\figureheight{0.9\textwidth}
 \setlength\figurewidth{\textwidth}
\resizebox{.9\textwidth}{!}{
 \input{Figures/scaled-set2-vs-z-m=2000-k=3000-s=100-mu=1000-exp=100} }
 \captionsetup{width = 0.8\textwidth}
 \caption{Empirical average waiting time under model~1.}
 \label{subfig:set2-scaled}
\end{subfigure}     
\begin{subfigure}[b]{0.33\textwidth}
\centering
 \setlength\figureheight{0.9\textwidth}
 \setlength\figurewidth{\textwidth}
 \resizebox{.9\textwidth}{!}{
  \input{Figures/erlang-set2-vs-z-m=2000-k=3000-s=100-mu=1000-exp=100} }
 \captionsetup{width = 0.8\textwidth}
 \caption{Empirical average waiting time under model~2.}
  \label{subfig:set2-erlang}
\end{subfigure}       
\caption{Comparison of the performance of RPM3 and the \fr\ scheme for setting 2, c.f., Table~\ref{tab:parameters}. In contrast to setting~1 (Figure~\ref{subfig:set1-rate}), RPM3 enjoys a higher rate for large values of $z$, regardless of the {clustering environment}. In addition, for $z \geq 50$, the waiting time of RPM3 is smaller for both settings under both service time models.} 
\label{fig:setting2}
\end{figure*}
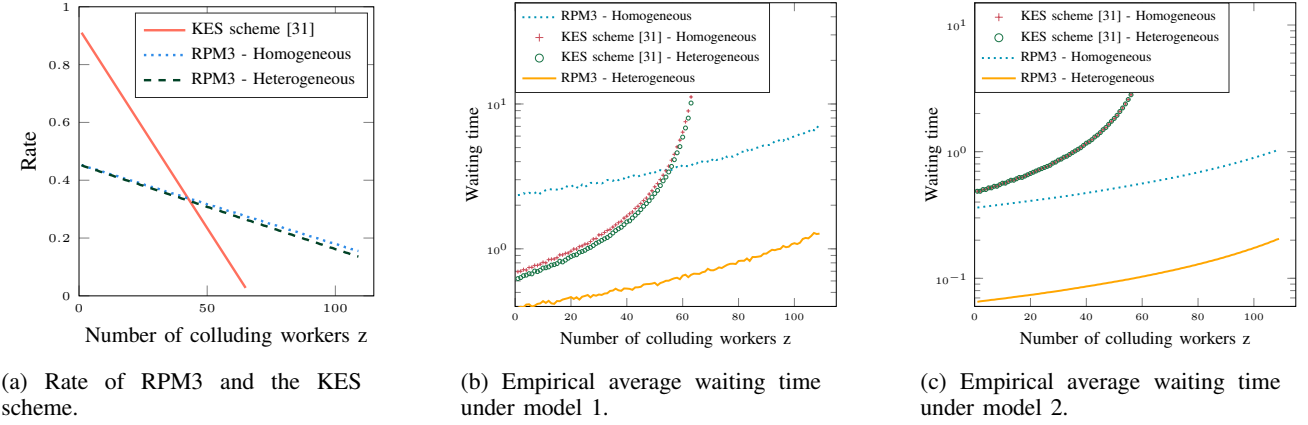

Next we find a lower bound on the mean waiting time of the master when using the \fr\ scheme so that we can compare the waiting time of the master under both schemes.
\begin{corollary}\label{cor:kakar_exp}
Consider the worker-dependent fixed service time where $\lambda_u s_u \triangleq t_m$, i.e., the handshake time $s_u$ is a constant factor of the mean service time $1/\lambda_u$ for all $u\in [c]$. Let $n_s$ be the number of stragglers and let $m_I\leq m$ and $k_I\leq k$ be the number of matrices in which $\mA$ and $\mB$ must be divided to use the \fr\ scheme.
The mean waiting time $\mathbb{E}[\rvT_{\text{\fr}}]$ of the master is bounded by
\begin{align*}
        \mathbb{E}[{T_\text{\fr}}] & \geq \frac{t_m + H_n- H_{n-n_s}}{\lambda_{1} m_I k_I}.
\end{align*}
\end{corollary}

The proof of Corollary~\ref{cor:kakar_exp} follows the same steps as the proof of Theorem~\ref{thm:waiting_exp}. A sketch of the proof is given in Appendix~\ref{app:proof_waiting}.

We can now compare the mean waiting time of the master when using RPM3 and the \fr\ scheme as follows.
\begin{align}\label{eq:comparison}
        \dfrac{\mathbb{E}[{T_{RPM3}}]}{\mathbb{E}[{T_{\text{\fr}}}]} & \leq \dfrac{\lambda_{u^\star}}{\lambda_{1}} \dfrac{t_m + H_n}{t_m + H_n - H_{n-n_s}} \dfrac{m_I k_I \tau_{u^\star}}{mk}.
\end{align}
The inequality in~\eqref{eq:comparison} can be understood as follows. The first ratio $\lambda_{u^\star}/\lambda_{1}$ indicates how fast are the workers of cluster $1$ compared to cluster $u^\star$. This is an artifact of our bounding technique. To understand the remaining ratios, recall that $(m_I+z)(k_I+1) - 1 = n-n_s$. For a fixed $n_s$, the ratio $$\frac{t_m + H_n}{t_m + H_n - H_{n-n_s}} \approx \frac{t_m + \log n}{t_m + \log\frac{n}{n-n_s}}$$ reflects the speed brought by mitigating the stragglers. Here we approximate $H_n$ by $H_n\approx \log(n)+\gamma,$ where $\gamma \approx 0.577218$ is called the Euler-Mascheroni constant. 

The ratio ${m_I k_I \tau_{u^\star}}/{mk}$ is the most important one. It reflects the respective number of tasks assigned to the workers under the different schemes. RPM3 assigns $\tau_{u^\star}$ tasks of size $mk$ each to the workers of cluster $u^\star$, i.e., the slowest cluster. Whereas, the \fr\ scheme assigns one task of size $m_I k_I$, or equivalently ${m_Ik_I}/{mk}$ tasks of size $mk$, to all the workers. For fixed system parameters, when $z$ increases, $m_Ik_I$ increases and RPM3 is expected to outperform the \fr\ scheme. Similarly for large values of $n_s$, $m_Ik_I$ increases and RPM3 is expected to outperform the \fr\ scheme.


Under the second model of the service time, we have the following upper bound on the master's mean waiting time.

\begin{theorem}[Mean waiting time under cluster-dependent additive service time]\label{thm:waiting_erlang}
Consider a master running a private distributed matrix-matrix multiplication of two matrices $\mA$ and $\mB$ using RPM3. The multiplication is divided into $km$ different smaller multiplications. 

Consider the cluster-dependent additive service time model where $\lambda_u s_u \triangleq t_m$, i.e., the shift $s_u$ is a constant factor of the mean service time $1/\lambda_u$ for all $u\in [c]$. Let $\displaystyle \tau_{\text{max}} \triangleq \max_{u\in\{1,\dots,c\}} \tau_u$ and let $s_m = \displaystyle \max_{u} s_u\tau_u$.
The probability distribution $P_1(x)\triangleq \Pr(\rvT_{RPM3}>x)$ is bounded from above by 
\begin{align*}
    P_1(x) <1 \!- \!\Big(1 - \!\!\!\sum_{j=0}^{\tau_{\max}-1}\!\!\frac{ e^{{\lambda_c} s_u - {\lambda_c} km x} }{j!} \!\left({\lambda_c} k mx - {\lambda_c} s_u\right)^j\!\Big)^n\!\!,
\end{align*}
for $x\geq \frac{s_m}{km}$ and is equal to $0$ otherwise. The mean waiting time at the master is bounded from above by
\begin{align}
    \mathbb{E}[{T_{RPM3}}] & \leq \frac{s_{m}}{km} + \Phi (n),\label{eq:avg_waiting_model2_1}
\end{align}
where $\Phi(n)$ is given by
\begin{align}
    \Phi(n) = \dfrac{{\lambda_c} n}{(\tau_{\text{max}} -1 )!} &\sum_{j=0}^{n - 1} (-1)^j
 \binom{n - 1}{j} \nonumber \\
 &\cdot \sum_{\ell = 0}^{(\tau_{\text{max}} - 1)j} a_\ell(\tau_{\text{max}},j) \frac{(\tau_{\text{max}}+\ell)!}{(j+1)^{\tau_{\text{max}}+\ell+1}}\label{eq:avg_waiting_model2}
\end{align}
and $a_\ell(\tau_{\text{max}},j)$ is the coefficient of $x^\ell$ in the expansion $$\displaystyle \left(\sum_{i=0}^{\tau_{\text{max}}-1} \frac{x^i}{i!}\right)^j.$$ 
\end{theorem}
{The bound in~\eqref{eq:avg_waiting_model2} can be interpreted as the mean of the $n^\text{th}$ ordered statistic of $n$ \emph{iid} Erlang random variables with shift $\max_u s_u\tau_u$, service rate $\lambda_u$ and shape $\tau_{max}$ chosen such that cluster $u$ has computed the highest number of tasks in respect to its service rate. The shape of an Erlang random variable is the number of \emph{iid} exponential random variables being summed. Recall that the service rate $\lambda_u$ reflects the expected number of tasks that a worker can process per time unit. Thus, the ratio $\lambda_u/\tau_u$ represents how large is the number of tasks computed at cluster $u$ in respect to its rate. In implementation we expect $\lambda_u/\tau_u$ to be almost the same for all values of $u$ and therefore we expect the upper bound to be a good estimate of the actual mean waiting time at the master.}

For the \fr\ scheme, we prove that the mean waiting time of the master under this setting is bounded from below by the mean of the $(n-n_s)^\text{th}$ ordered statistic of $n$ \emph{iid} Erlang random variables with rate ${\lambda_1}/{mk}$ and shift $\frac{s_1}{mk} \left\lceil\frac{m}{m_I}\right\rceil \left\lceil\frac{k}{k_I}\right\rceil$ and shape $\left\lceil\frac{m}{m_I}\right\rceil \left\lceil\frac{k}{k_I}\right\rceil$, i.e., sum of $\left\lceil\frac{m}{m_I}\right\rceil \cdot\left\lceil\frac{k}{k_I}\right\rceil$ \emph{iid} exponential random variables.

\begin{corollary}\label{cor:erlang_kakr}
Consider the cluster-dependent additive service time model where $\lambda_u s_u \triangleq t_m$, i.e., the shift $s_u$ is a constant factor of the mean service time $1/\lambda_u$ for all $u\in [c]$.
The mean waiting time at the master when using the \fr\ scheme that tolerates $n_s$ stragglers is bounded from below by
\begin{align*}
    \mathbb{E}[{\rvT_{\text{\fr}}}] & \geq \frac{s_{1}}{k_Im_I} + \Phi_I (n),
\end{align*}
where $\Phi_I(n)$ is given by
\begin{align}
    \Phi_I(n) &= \dfrac{{\lambda_1} (n-n_s)}{(\tau_{1} -1 )!} \sum_{j=0}^{n-n_s - 1} (-1)^j
 \binom{n -n_s - 1}{j} \nonumber \\
 &\cdot \sum_{\ell = 0}^{(\tau_{1} - 1)(n-n_s+j)}  \frac{a_\ell(\tau_{1},n-n_s+j)(\tau_{1}+\ell)!}{(n-n_s+j+1)^{\tau_{\text{max}}+\ell+1}},
\end{align}
where again $a_\ell(o,p)$ is the coefficient of $x^\ell$ in the expansion $$\displaystyle \left(\sum_{i=0}^{o-1} \frac{x^i}{i!}\right)^p.$$ 
\end{corollary}
\subsection{Numerical results} 
We provide numerical simulations showing the empirical average waiting time at the master for both considered models under the two settings and two clustering environments summarized in Table~\ref{tab:parameters}.

{{\em Model 1:} The empirical average waiting time under the worker-dependent fixed service time model is simulated numerically and shown in~\fig{\ref{subfig:set1-scaled}} and~\fig{\ref{subfig:set2-scaled}} for settings 1 and~2, respectively. Recall that in this model we assume that the time spent to compute $\tau$ tasks at a worker is a scaled shifted exponential random variable. For the first setting, the \fr\ scheme enjoys a smaller waiting time in the homogeneous scenario. When the workers compute powers are different, i.e., in the heterogeneous scenario, RPM3 has a slightly better waiting time for small values of $z$. However, for large values of $z$ the overhead of clustering becomes significant. Thus, the \fr\ scheme outperforms RPM3 in this parameter regime.}

{{\em Model 2:} The empirical average waiting time under the cluster-dependent additive service time model is simulated numerically and shown in~\fig{\ref{subfig:set1-erlang}} and~\fig{\ref{subfig:set2-erlang}} for settings 1 and 2, respectively. Recall that in this model we assume that the time spent to compute $\tau$ tasks at a worker is the sum of $\tau$ \emph{iid} shifted exponential random variables, i.e., it is a random variable that follows a shifted Erlang distribution. We make a small change to the \fr\ scheme. We assign multiple smaller tasks to the workers, rather than one large task as is done in the original scheme of~\cite{kakar2018rate}. This change allows a fair comparison to RPM3 under model~2. Interestingly, regardless of the heterogeneity of the clusters, RPM3 outperforms the \fr\ scheme for the values of $z$ it can tolerate. As expected, in the heterogeneous scenario RPM3 has much smaller average waiting time compared to the \fr\ scheme. This is an important observation, since model~2 is a better approximation of the real waiting time. The reason is that in RPM3 the master sends multiple independent tasks to the workers and it is more realistic to assign them independent random variables.}

%% file: Figures/Setting2-rate-vs-z-m=2000-k=3000.tex
%
%
\definecolor{mycolor1}{rgb}{1.00000,0.00000,1.00000}%
\definecolor{mycolor2}{rgb}{0.85000,0.32500,0.09800}%
\definecolor{mycolor3}{rgb}{0.92900,0.69400,0.12500}%
\definecolor{bleudefrance}{rgb}{0.19, 0.55, 0.91}
\definecolor{bittersweet}{rgb}{1.0, 0.44, 0.37}
\definecolor{britishracinggreen}{rgb}{0.0, 0.26, 0.15}
\begin{tikzpicture}

\begin{axis}[%
width=0.951\figurewidth,
height=\figureheight,
at={(0\figurewidth,0\figureheight)},
scale only axis,
xmin=0,
xmax=115,
xlabel style={font=\color{white!15!black}, font = \small, yshift=0.2cm},
xlabel={Number of colluding workers z},
tick label style={font=\tiny},
ymin=0,
ymax=1,
yminorticks=true,
ylabel style={font=\color{white!15!black}, font = \small, yshift=-0.5cm},
ylabel={Rate}, 
axis background/.style={fill=white},
legend style={legend cell align=left, align=left, draw=white!15!black, font =\scriptsize}
]

\addplot [color=bittersweet, line width = 1pt]
  table[row sep=crcr]{%
1	0.910344827586207\\
2	0.896551724137931\\
3	0.882758620689655\\
4	0.868965517241379\\
5	0.855172413793103\\
6	0.841379310344828\\
7	0.827586206896552\\
8	0.813793103448276\\
9	0.8\\
10	0.786206896551724\\
11	0.772413793103448\\
12	0.758620689655172\\
13	0.744827586206897\\
14	0.731034482758621\\
15	0.717241379310345\\
16	0.703448275862069\\
17	0.689655172413793\\
18	0.675862068965517\\
19	0.662068965517241\\
20	0.648275862068965\\
21	0.63448275862069\\
22	0.620689655172414\\
23	0.606896551724138\\
24	0.593103448275862\\
25	0.579310344827586\\
26	0.56551724137931\\
27	0.551724137931034\\
28	0.537931034482759\\
29	0.524137931034483\\
30	0.510344827586207\\
31	0.496551724137931\\
32	0.482758620689655\\
33	0.468965517241379\\
34	0.455172413793103\\
35	0.441379310344828\\
36	0.427586206896552\\
37	0.413793103448276\\
38	0.4\\
39	0.386206896551724\\
40	0.372413793103448\\
41	0.358620689655172\\
42	0.344827586206897\\
43	0.331034482758621\\
44	0.317241379310345\\
45	0.303448275862069\\
46	0.289655172413793\\
47	0.275862068965517\\
48	0.262068965517241\\
49	0.248275862068966\\
50	0.23448275862069\\
51	0.220689655172414\\
52	0.206896551724138\\
53	0.193103448275862\\
54	0.179310344827586\\
55	0.16551724137931\\
56	0.151724137931034\\
57	0.137931034482759\\
58	0.124137931034483\\
59	0.110344827586207\\
60	0.096551724137931\\
61	0.0827586206896552\\
62	0.0689655172413793\\
63	0.0551724137931034\\
64	0.0413793103448276\\
65	0.0275862068965517\\
};
\addlegendentry{\fr\ scheme~\cite{kakar2018rate}}

\addplot [color=bleudefrance, line width = 1pt, dotted]
  table[row sep=crcr]{%
1	0.45278883966068\\
2	0.448945539164886\\
3	0.447284535584467\\
4	0.443301383617502\\
5	0.441912449778491\\
6	0.437797382993177\\
7	0.436442672163905\\
8	0.432428380851206\\
9	0.430887139167928\\
10	0.426973936087695\\
11	0.425257457952669\\
12	0.42144538910648\\
13	0.419772986768755\\
14	0.415853730381234\\
15	0.414225326150666\\
16	0.410408506947532\\
17	0.408822387113918\\
18	0.404910211160675\\
19	0.403366225608108\\
20	0.399368731158948\\
21	0.397866639081246\\
22	0.393793551105543\\
23	0.39233302796746\\
24	0.388371886794773\\
25	0.386774379033734\\
26	0.382750945873275\\
27	0.381199252849464\\
28	0.377290388464472\\
29	0.375782567196186\\
30	0.371820009369864\\
31	0.370193500142525\\
32	0.36618881061264\\
33	0.36476825057147\\
34	0.360725635688751\\
35	0.359194542158663\\
36	0.355273862857183\\
37	0.353640761105646\\
38	0.349695182366038\\
39	0.34811285122411\\
40	0.344148979799028\\
41	0.342616332520571\\
42	0.338640392687399\\
43	0.337156311285184\\
44	0.333174150128272\\
45	0.331607500519518\\
46	0.327627696990248\\
47	0.326112655616883\\
48	0.322140213675604\\
49	0.320675385117776\\
50	0.316596461401351\\
51	0.315064162816758\\
52	0.311125860781548\\
53	0.309645935355218\\
54	0.30562020271788\\
55	0.304082767274815\\
56	0.30009302883894\\
57	0.298610565040865\\
58	0.294557027958862\\
59	0.293027126497857\\
60	0.289122767791531\\
61	0.287550914483796\\
62	0.283599980337068\\
63	0.281993449292173\\
64	0.278101328072892\\
65	0.276556320694709\\
66	0.272548258075946\\
67	0.270977374744096\\
68	0.267043970124901\\
69	0.265535720528965\\
70	0.26151493860501\\
71	0.259988430514842\\
72	0.256055171354254\\
73	0.254514989872424\\
74	0.250521711464124\\
75	0.248973916247664\\
76	0.245009363441173\\
77	0.243458671267495\\
78	0.239463028105776\\
79	0.237914630679171\\
80	0.233968023590216\\
81	0.232425800968363\\
82	0.228472554353621\\
83	0.226940826314243\\
84	0.222935284859276\\
85	0.221418718295607\\
86	0.217436367247125\\
87	0.215883250338217\\
88	0.211936474161235\\
89	0.210408339464399\\
90	0.206405449103856\\
91	0.204856533847933\\
92	0.200916985120088\\
93	0.199355086295833\\
94	0.195396523049137\\
95	0.193830128567524\\
96	0.189872636599913\\
97	0.18835128896635\\
98	0.18437088046314\\
99	0.182817528910307\\
100	0.178838195546452\\
101	0.177301978069518\\
102	0.1733433718406\\
103	0.171795034779905\\
104	0.167809308214517\\
105	0.16629246243384\\
106	0.162305780844996\\
107	0.160764122592289\\
108	0.156802360816344\\
109	0.155249007764779\\
};
\addlegendentry{RPM3 - Homogeneous}

\addplot [color=britishracinggreen, line width = 1pt, dashed]
  table[row sep=crcr]{%
1	0.452382244901652\\
2	0.448073842569255\\
3	0.445246878262825\\
4	0.442455361784688\\
5	0.439698630558615\\
6	0.435627346942332\\
7	0.434286955105586\\
8	0.430314818321084\\
9	0.427706849725198\\
10	0.425130302437697\\
11	0.422584612003938\\
12	0.418822731184912\\
13	0.41635180061745\\
14	0.412699591840104\\
15	0.410300175608475\\
16	0.406752911842408\\
17	0.404421949596892\\
18	0.400975171617373\\
19	0.398709775167558\\
20	0.395359272855225\\
21	0.393156714232076\\
22	0.389898509417998\\
23	0.387756209915702\\
24	0.383541468503575\\
25	0.381468271376528\\
26	0.377388396816352\\
27	0.375381011726903\\
28	0.372409658072072\\
29	0.369484974893496\\
30	0.366605871193027\\
31	0.363771289714731\\
32	0.360059337778866\\
33	0.358231625404354\\
34	0.354631307561094\\
35	0.351978205509515\\
36	0.348501877553865\\
37	0.346789337615026\\
38	0.342580729148824\\
39	0.340925749780955\\
40	0.336857423411254\\
41	0.334462702391743\\
42	0.331322207533604\\
43	0.329005269019383\\
44	0.32521488573575\\
45	0.322982289266168\\
46	0.31932864345999\\
47	0.317175866088349\\
48	0.313651689798479\\
49	0.31157452628988\\
50	0.307501656665175\\
51	0.305504892660856\\
52	0.301588163267768\\
53	0.299667219552687\\
54	0.295897820564603\\
55	0.294048459186074\\
56	0.290418231294888\\
57	0.288047470223093\\
58	0.284563025018781\\
59	0.282286520818631\\
60	0.278389073785632\\
61	0.276209902953651\\
62	0.27247733669752\\
63	0.270389387757309\\
64	0.266811456350313\\
65	0.264809118966821\\
66	0.260893272475629\\
67	0.258978459466634\\
68	0.255231935640715\\
69	0.252944911127805\\
70	0.24936971803766\\
71	0.247186095287768\\
72	0.243350448981578\\
73	0.241270530614214\\
74	0.237614916513999\\
75	0.235238767348859\\
76	0.231762332363408\\
77	0.229501236437911\\
78	0.225829216654905\\
79	0.223681870696221\\
80	0.219849315279308\\
81	0.217813673471166\\
82	0.214177936888187\\
83	0.211926817431405\\
84	0.208175900308725\\
85	0.206048555342066\\
86	0.202501090974628\\
87	0.200203206254348\\
88	0.196578357116038\\
89	0.194412204420545\\
90	0.190734135688264\\
91	0.188694198408176\\
92	0.184984613904738\\
93	0.18282805752502\\
94	0.179115812702177\\
95	0.176871253645759\\
96	0.173181914612657\\
97	0.171082739890079\\
98	0.167231351195871\\
99	0.165273138652594\\
100	0.161491144633084\\
101	0.159303905653855\\
102	0.155615502105089\\
103	0.15341658740143\\
104	0.149833609776343\\
105	0.147639393733594\\
106	0.143876921925908\\
107	0.141710100812566\\
108	0.138104951476825\\
109	0.135976166097606\\
};
\addlegendentry{RPM3 - Heterogeneous}

\end{axis}
\end{tikzpicture}%

%% file: Figures/scaled-set2-vs-z-m=2000-k=3000-s=100-mu=1000-exp=100.tex
%
%
\definecolor{mycolor1}{rgb}{0.0, 0.58, 0.71}
\definecolor{mycolor2}{rgb}{1.0, 0.65, 0.0}
\definecolor{mycolor3}{rgb}{0.8, 0.31, 0.36}
\definecolor{mycolor4}{rgb}{0.09, 0.45, 0.27}
\begin{tikzpicture}

\begin{axis}[%
width=0.951\figurewidth,
height=\figureheight,
at={(0\figurewidth,0\figureheight)},
scale only axis,
xmin=0,
xmax=115,
xlabel style={font=\color{white!15!black}, font = \small, yshift=0.2cm},
xlabel={Number of colluding workers z},
tick label style={font=\tiny},
ymode=log,
ymin=0.4,
ymax=50,
yminorticks=true,
ylabel style={font=\color{white!15!black}, font = \small, yshift=-0.5cm},
ylabel={Waiting time}, 
axis background/.style={fill=white},
legend style={legend cell align=left, align=left, draw=white!15!black, font =\scriptsize, at = {(0,1)}, anchor = north west}
]

\addlegendimage{color=mycolor1, dotted, line width = 1 pt};
\addlegendentry{RPM3 - Homogeneous}

\addlegendimage{color=mycolor3, only marks, mark=+, mark size = 2pt};
\addlegendentry{\fr\ scheme~\cite{kakar2018rate} - Homogeneous};

\addlegendimage{color=mycolor4, only marks, mark=o, mark size = 2pt};
\addlegendentry{\fr\ scheme~\cite{kakar2018rate} - Heterogeneous};

\addlegendimage{color=mycolor2, solid, line width = 1 pt};
\addlegendentry{RPM3 - Heterogeneous};

\addplot [color=mycolor1, dotted, mark=none, mark options={solid, mycolor1}, line width = 1pt]
  table[row sep=crcr]{%
1	2.34857124326515\\
2	2.3596229263528\\
3	2.42173107613722\\
4	2.38066087255643\\
5	2.43727748756688\\
6	2.39641232471001\\
7	2.42036062415351\\
8	2.49338797095063\\
9	2.40817656341537\\
10	2.48204624383507\\
11	2.54094708866519\\
12	2.58492814561774\\
13	2.57194691214397\\
14	2.57437178888237\\
15	2.56259547784786\\
16	2.53122638654419\\
17	2.62110603368669\\
18	2.7184682888403\\
19	2.67088085957848\\
20	2.6734501850759\\
21	2.75732498105092\\
22	2.66567605285491\\
23	2.6211662180872\\
24	2.77326108273035\\
25	2.67239806005689\\
26	2.75663419989907\\
27	2.85007505497427\\
28	2.85531350886534\\
29	2.83519386233012\\
30	2.87189142516866\\
31	2.82561848354863\\
32	2.89190727221851\\
33	2.98601163509\\
34	2.99441496673804\\
35	2.94472544333718\\
36	3.03784370055971\\
37	2.97770152975399\\
38	3.05378296972089\\
39	3.0563319535604\\
40	3.18234347955414\\
41	3.09394728997573\\
42	3.17979108198829\\
43	3.23474017687168\\
44	3.15508511021726\\
45	3.2229063863624\\
46	3.24944179023622\\
47	3.26659627935854\\
48	3.35771618302405\\
49	3.27172783366146\\
50	3.35910696134664\\
51	3.50133905826787\\
52	3.48968273980568\\
53	3.45028858399959\\
54	3.50074928423455\\
55	3.44730419472932\\
56	3.50781976792735\\
57	3.61902263731511\\
58	3.67424062581985\\
59	3.62669927944299\\
60	3.80396970701565\\
61	3.73357519283798\\
62	3.71490509601375\\
63	3.73424223311886\\
64	3.91718149413625\\
65	3.8812916487463\\
66	4.05016737696501\\
67	4.00142441194675\\
68	3.93203679875372\\
69	4.01333084714419\\
70	3.96751957733759\\
71	4.21909343109685\\
72	4.10141114084699\\
73	4.14650159872054\\
74	4.15569685058024\\
75	4.40183964424829\\
76	4.29767018681212\\
77	4.37819407750139\\
78	4.32242297414125\\
79	4.42410951963208\\
80	4.52740403544833\\
81	4.68856506258933\\
82	4.66328801501056\\
83	4.82199501102459\\
84	4.64670350605838\\
85	4.87343949522223\\
86	4.95856924008227\\
87	5.00807936926576\\
88	5.07089727699329\\
89	5.1746790960555\\
90	5.20403489322951\\
91	5.21287339686255\\
92	5.28406071629489\\
93	5.54590060111519\\
94	5.51855420339617\\
95	5.70294462377622\\
96	5.56855426672414\\
97	5.8721551416624\\
98	5.70367013340378\\
99	5.88723864746908\\
100	5.97273753974878\\
101	6.02176757869378\\
102	6.22335260577311\\
103	6.20087822496069\\
104	6.37682353042299\\
105	6.41227685664543\\
106	6.73316905939361\\
107	6.58461647687813\\
108	7.01023406967342\\
109	6.78470475532187\\
};

\addplot [color=mycolor2, mark=none, mark options={solid, mycolor2}, line width = 1pt]
  table[row sep=crcr]{%
1	0.392281988666876\\
2	0.397913154848278\\
3	0.395682611013824\\
4	0.412028682776042\\
5	0.413925511525537\\
6	0.398984248939371\\
7	0.404810571522628\\
8	0.416537032508856\\
9	0.419826915000119\\
10	0.435454662553511\\
11	0.42298797768535\\
12	0.437208883083554\\
13	0.419959150623081\\
14	0.41683919963287\\
15	0.43598066874126\\
16	0.434665945234339\\
17	0.446321733746661\\
18	0.451540892831732\\
19	0.45691476804749\\
20	0.464407855219075\\
21	0.450785894344335\\
22	0.462776814679002\\
23	0.446203216880208\\
24	0.466820500371075\\
25	0.472424416150414\\
26	0.47965351738519\\
27	0.462289599712825\\
28	0.479738739209664\\
29	0.474466749387114\\
30	0.483617790592619\\
31	0.480272471425516\\
32	0.491066651604417\\
33	0.49328068466785\\
34	0.51188876205779\\
35	0.514887376554522\\
36	0.530385043660349\\
37	0.517207396309792\\
38	0.512752550263191\\
39	0.532210492218066\\
40	0.531355263362369\\
41	0.529100970133332\\
42	0.520951373103551\\
43	0.54165874030801\\
44	0.542105351080282\\
45	0.546580613948066\\
46	0.566538932153567\\
47	0.570535099562846\\
48	0.571323400625476\\
49	0.574952194618931\\
50	0.582268483635203\\
51	0.560288279086037\\
52	0.580454461736771\\
53	0.589141889331684\\
54	0.601592884180018\\
55	0.6023832539332\\
56	0.615577292338228\\
57	0.625661018585411\\
58	0.631252755277389\\
59	0.616933790816841\\
60	0.653042187783881\\
61	0.662335940244304\\
62	0.637112316309138\\
63	0.661360030671241\\
64	0.673303712251406\\
65	0.671976316042416\\
66	0.671257543819649\\
67	0.68857112746444\\
68	0.695740915746023\\
69	0.727549707770271\\
70	0.711075062106092\\
71	0.710518423867004\\
72	0.724002638530184\\
73	0.7438450407895\\
74	0.737156926216777\\
75	0.757163679413394\\
76	0.776121364919106\\
77	0.787500892502617\\
78	0.789174711244038\\
79	0.785539355270206\\
80	0.823815529334195\\
81	0.82734093669386\\
82	0.82677333150198\\
83	0.857442988428602\\
84	0.824454851868311\\
85	0.877530881898973\\
86	0.877243177494333\\
87	0.882611472213636\\
88	0.889245098309075\\
89	0.925167161953531\\
90	0.926734341502693\\
91	0.957067669981963\\
92	0.984717942012919\\
93	0.978400734591939\\
94	0.992716424451537\\
95	0.998931890345318\\
96	1.04664681177599\\
97	1.02599724240056\\
98	1.03405147945289\\
99	1.07642551601208\\
100	1.09313519292385\\
101	1.07903640126559\\
102	1.10146810670566\\
103	1.16961073361618\\
104	1.18888676873815\\
105	1.18080575293522\\
106	1.22526583242717\\
107	1.28635976935969\\
108	1.26494697416346\\
109	1.27264183300473\\
};

\addplot [color=mycolor3, only marks, mark=+, mark options={solid, mycolor3}, mark size = 1pt]
  table[row sep=crcr]{%
1	0.694902302520987\\
2	0.699825289613405\\
3	0.718002657676405\\
4	0.713795598889502\\
5	0.742945882401074\\
6	0.745262572686064\\
7	0.768003997842614\\
8	0.768678700854252\\
9	0.783117939080538\\
10	0.806543148330551\\
11	0.800582105963634\\
12	0.841067676310256\\
13	0.846643329477146\\
14	0.850782841099247\\
15	0.868084640916416\\
16	0.907842225557298\\
17	0.908558919482473\\
18	0.928928490785875\\
19	0.936351844531305\\
20	0.962672154454278\\
21	0.995727715559612\\
22	0.9965908071849\\
23	1.03445979068201\\
24	1.04783869460481\\
25	1.07593710366751\\
26	1.08356836132455\\
27	1.12241083950395\\
28	1.16719964316479\\
29	1.17727331752827\\
30	1.24720097056637\\
31	1.25505275787718\\
32	1.2969121542391\\
33	1.33805851852997\\
34	1.36097417223002\\
35	1.4127192084862\\
36	1.46349122546124\\
37	1.51377630812644\\
38	1.54833074619332\\
39	1.62863450836526\\
40	1.67919777387066\\
41	1.74637461903411\\
42	1.79715483658214\\
43	1.89975640749122\\
44	1.96733274700603\\
45	2.05712772711975\\
46	2.15198087381019\\
47	2.24028304202222\\
48	2.3865030215227\\
49	2.5157459528304\\
50	2.67966643812989\\
51	2.82047735715599\\
52	3.00381973452385\\
53	3.220712050495\\
54	3.45797819511205\\
55	3.72167111750858\\
56	4.08972054611019\\
57	4.49168240317614\\
58	5.05470326421679\\
59	5.62626786700551\\
60	6.38680142622947\\
61	7.5456813994813\\
62	8.94866156085426\\
63	11.1806656574529\\
64	15.1840311468261\\
65	22.5878512889426\\
};

\addplot [color=mycolor4, only marks, mark=o, mark options={solid, mycolor4}, mark size = 1pt]
  table[row sep=crcr]{%
1	0.622913424545574\\
2	0.633017139820043\\
3	0.652330598270772\\
4	0.658154267521768\\
5	0.673468730175949\\
6	0.671573075316975\\
7	0.702606859926779\\
8	0.696143732772881\\
9	0.711256818900571\\
10	0.736468659021001\\
11	0.740226650663074\\
12	0.750097264180047\\
13	0.771564904576183\\
14	0.777665702412202\\
15	0.790321322006277\\
16	0.820561014856283\\
17	0.81162181246825\\
18	0.835540715918399\\
19	0.861361271113136\\
20	0.883408225887894\\
21	0.902218934303336\\
22	0.910421085032358\\
23	0.945524298051963\\
24	0.957925301779888\\
25	0.976602760305425\\
26	1.00534723681044\\
27	1.02349251896167\\
28	1.05404437165419\\
29	1.08960576536363\\
30	1.11387937148166\\
31	1.14311513400524\\
32	1.17221003064305\\
33	1.1965003203045\\
34	1.23564342730183\\
35	1.28570991267067\\
36	1.33594076594736\\
37	1.37508129666574\\
38	1.40979679478472\\
39	1.46785612232931\\
40	1.53511827932054\\
41	1.55912730147232\\
42	1.63324431841508\\
43	1.70484916508956\\
44	1.77457428496077\\
45	1.86704397411812\\
46	1.96456891567822\\
47	2.05293914944216\\
48	2.14833087387534\\
49	2.27199706955858\\
50	2.40888203632748\\
51	2.53065649710883\\
52	2.73213777148634\\
53	2.94572222332602\\
54	3.12891825353925\\
55	3.39405248954875\\
56	3.70520476362927\\
57	4.12408050303249\\
58	4.6020506240736\\
59	5.07716616764169\\
60	5.90442340396945\\
61	6.82536038242938\\
62	7.98795033655392\\
63	10.1501122510593\\
64	13.4928377499615\\
65	20.443480293436\\
};

\end{axis}
\end{tikzpicture}%

%% file: Figures/erlang-set2-vs-z-m=2000-k=3000-s=100-mu=1000-exp=100.tex
%
%
\definecolor{mycolor1}{rgb}{0.0, 0.58, 0.71}
\definecolor{mycolor2}{rgb}{1.0, 0.65, 0.0}
\definecolor{mycolor3}{rgb}{0.8, 0.31, 0.36}
\definecolor{mycolor4}{rgb}{0.09, 0.45, 0.27}
\begin{tikzpicture}

\begin{axis}[%
width=0.951\figurewidth,
height=\figureheight,
at={(0\figurewidth,0\figureheight)},
scale only axis,
xmin=0,
xmax=115,
xlabel style={font=\color{white!15!black}, font = \small, yshift=0.2cm},
xlabel={Number of colluding workers z},
tick label style={font=\tiny},
ymode=log,
ymin=0.06,
ymax=15,
yminorticks=true,
ylabel style={font=\color{white!15!black}, font = \small, yshift=-0.5cm},
ylabel={Waiting time}, 
axis background/.style={fill=white},
legend style={legend cell align=left, align=left, draw=white!15!black, font =\scriptsize, at = {(0,1)}, anchor = north west}
]

\addlegendimage{color=mycolor3, only marks, mark=+, mark size = 2pt};
\addlegendentry{\fr\ scheme~\cite{kakar2018rate} - Homogeneous};

\addlegendimage{color=mycolor4, only marks, mark=o, mark size = 2pt};
\addlegendentry{\fr\ scheme~\cite{kakar2018rate} - Heterogeneous};

\addlegendimage{color=mycolor1, dotted, line width = 1 pt};
\addlegendentry{RPM3 - Homogeneous}

\addlegendimage{color=mycolor2, solid, line width = 1 pt};
\addlegendentry{RPM3 - Heterogeneous};

\addplot [color=mycolor1, dotted, mark=none, mark options={solid, mycolor1}, line width=1pt]
  table[row sep=crcr]{%
1	0.361803927158678\\
2	0.364798119341049\\
3	0.36616942814156\\
4	0.369356796570814\\
5	0.37049171673823\\
6	0.373938266586446\\
7	0.375131871950876\\
8	0.378519940304956\\
9	0.379861526779844\\
10	0.383209351512054\\
11	0.384755533401417\\
12	0.38807918664544\\
13	0.389638460681232\\
14	0.39321271897874\\
15	0.394729815629665\\
16	0.39842670058793\\
17	0.399942491614209\\
18	0.403748880907172\\
19	0.405166658394928\\
20	0.409129808721713\\
21	0.410667580697207\\
22	0.414902973494467\\
23	0.41635193322411\\
24	0.420373190945603\\
25	0.422151446969544\\
26	0.426316206697116\\
27	0.42794689335645\\
28	0.432317369390165\\
29	0.434046737845584\\
30	0.438532968449154\\
31	0.440303185579546\\
32	0.445020559206747\\
33	0.446646577589789\\
34	0.451516559940715\\
35	0.453386851938291\\
36	0.458205427568678\\
37	0.460292135607327\\
38	0.465470570912665\\
39	0.46754165232952\\
40	0.47271426012252\\
41	0.474858869701895\\
42	0.480177678982879\\
43	0.482344608181698\\
44	0.487968286274829\\
45	0.490322674221309\\
46	0.496032746646143\\
47	0.498387755859402\\
48	0.504382002243175\\
49	0.506640277644008\\
50	0.513084351817287\\
51	0.515601376202445\\
52	0.521891783250066\\
53	0.524290655180859\\
54	0.531063539620229\\
55	0.533656953181653\\
56	0.540447926166516\\
57	0.543088756215379\\
58	0.550655194327439\\
59	0.553499872191475\\
60	0.560793649426116\\
61	0.563761893941106\\
62	0.57145205373133\\
63	0.574614009940388\\
64	0.582554227152075\\
65	0.585751041861594\\
66	0.594021290147559\\
67	0.597514510161442\\
68	0.60599413720841\\
69	0.609328695361743\\
70	0.618391771974121\\
71	0.621852186128098\\
72	0.631153598333499\\
73	0.635005463833772\\
74	0.644825716483054\\
75	0.648775153728027\\
76	0.659053641081238\\
77	0.663192054467596\\
78	0.674005771401955\\
79	0.678206123956473\\
80	0.689533925644092\\
81	0.694047315156182\\
82	0.70561358020444\\
83	0.710244147365502\\
84	0.722722033276713\\
85	0.727644738997386\\
86	0.740790337886249\\
87	0.74607500918709\\
88	0.759670465728587\\
89	0.765027742240487\\
90	0.779563825727644\\
91	0.785470049255916\\
92	0.800686543890571\\
93	0.806635824579815\\
94	0.822633185281135\\
95	0.829175397689964\\
96	0.846197143373618\\
97	0.853004128933861\\
98	0.871231953760449\\
99	0.878600556479064\\
100	0.89784529101147\\
101	0.905612841923257\\
102	0.925875976232871\\
103	0.93410652034419\\
104	0.955759137664279\\
105	0.964317495728149\\
106	0.987360644740253\\
107	0.996650134228362\\
108	1.021729817391\\
109	1.031648406713\\
};

\addplot [color=mycolor2, mark=none, mark options={solid, mycolor2}, line width = 1pt]
  table[row sep=crcr]{%
1	0.0652141909319091\\
2	0.0658266775195535\\
3	0.0662274934760203\\
4	0.0665937036432384\\
5	0.066999729614423\\
6	0.0675648946740663\\
7	0.0677728214728257\\
8	0.0683660521094456\\
9	0.0687484598877521\\
10	0.069138405760942\\
11	0.069523933058348\\
12	0.0701297782863158\\
13	0.0705306185793037\\
14	0.0711528007816188\\
15	0.0715262567669538\\
16	0.0720855875049078\\
17	0.0724485599379789\\
18	0.0730267793673593\\
19	0.0733912830415001\\
20	0.0739726957020858\\
21	0.0743723103689328\\
22	0.0749859818361576\\
23	0.0754206775080116\\
24	0.0761830314363785\\
25	0.0765393021396094\\
26	0.0773649541097595\\
27	0.077729946753584\\
28	0.0782950872321874\\
29	0.0788769511407939\\
30	0.079471935920553\\
31	0.080029102318825\\
32	0.0808469143011476\\
33	0.0812883370332693\\
34	0.0820814940554193\\
35	0.0826885407586287\\
36	0.0834864647597864\\
37	0.0839066950250698\\
38	0.0849035253000238\\
39	0.0852972407744186\\
40	0.0862838546097119\\
41	0.0868587339508763\\
42	0.0876135972864133\\
43	0.0882018129374756\\
44	0.089194220594553\\
45	0.0897513742668793\\
46	0.0907533374596872\\
47	0.0913572364514449\\
48	0.0922908368652873\\
49	0.0928806990258095\\
50	0.0939778388701949\\
51	0.0945815742631886\\
52	0.095627863463902\\
53	0.0962046339611159\\
54	0.0974253876684697\\
55	0.0979847194934323\\
56	0.0991005114444865\\
57	0.0999522417893897\\
58	0.10111367065535\\
59	0.101904067130988\\
60	0.103183102272975\\
61	0.104002434340947\\
62	0.105324885364286\\
63	0.1061484739706\\
64	0.107537324938401\\
65	0.10827120179794\\
66	0.10985958707344\\
67	0.110629072214299\\
68	0.112304685485246\\
69	0.113232550356167\\
70	0.114680387438031\\
71	0.115626347618968\\
72	0.117274075767428\\
73	0.118273612002653\\
74	0.120107153311596\\
75	0.121220487332242\\
76	0.122911239278382\\
77	0.124018726283947\\
78	0.125953356167321\\
79	0.127102633042383\\
80	0.129255829542786\\
81	0.130437824397948\\
82	0.132520863164573\\
83	0.133888095935159\\
84	0.136223661008784\\
85	0.137513805684662\\
86	0.139825645163103\\
87	0.141300783494958\\
88	0.143808992522298\\
89	0.145400802192445\\
90	0.148179969642102\\
91	0.149761388182534\\
92	0.152677744539419\\
93	0.154407516941612\\
94	0.157487193426353\\
95	0.159393578592873\\
96	0.162612865035903\\
97	0.164568828331948\\
98	0.168197475036611\\
99	0.170081100110308\\
100	0.173809521230247\\
101	0.17615768884122\\
102	0.180086278839704\\
103	0.182578870443858\\
104	0.18672143325627\\
105	0.189393510006053\\
106	0.194164609154971\\
107	0.196997211862704\\
108	0.20189347493268\\
109	0.204900130053282\\
};

\addplot [color=mycolor3, only marks, mark=+, mark options={solid, mycolor3}, mark size = 1pt]
  table[row sep=crcr]{%
1	0.486670417635986\\
2	0.486670417635986\\
3	0.501990835119817\\
4	0.501990835119817\\
5	0.517552818593454\\
6	0.517552818593454\\
7	0.53295389399307\\
8	0.53295389399307\\
9	0.548532025768059\\
10	0.56407950834901\\
11	0.56407950834901\\
12	0.579524260595296\\
13	0.595096919447319\\
14	0.595096919447319\\
15	0.610622967017164\\
16	0.626080708181753\\
17	0.626080708181753\\
18	0.641842172317421\\
19	0.657232562014869\\
20	0.672572785161217\\
21	0.687924715315992\\
22	0.703287516820451\\
23	0.718682750114173\\
24	0.734108636913964\\
25	0.749590466400083\\
26	0.76497204285783\\
27	0.780657422427719\\
28	0.811416852123686\\
29	0.826744677910419\\
30	0.857567292257625\\
31	0.87298330383341\\
32	0.903894233207408\\
33	0.919339752783533\\
34	0.950101521273241\\
35	0.981066494718299\\
36	1.01195917891367\\
37	1.04280939332715\\
38	1.0738383061213\\
39	1.12019185963012\\
40	1.16638483971222\\
41	1.19714566313916\\
42	1.24321648404613\\
43	1.30487544586296\\
44	1.35091099283624\\
45	1.41253984667363\\
46	1.48935137854671\\
47	1.5512379368431\\
48	1.64377238773665\\
49	1.73589181790568\\
50	1.82827740068247\\
51	1.93634298568279\\
52	2.07525928708837\\
53	2.21365572493403\\
54	2.38272591702251\\
55	2.58245867891334\\
56	2.81319322936983\\
57	3.08967300141123\\
58	3.44270894192375\\
59	3.8573370040093\\
60	4.4095842427853\\
61	5.14555148484291\\
62	6.15873288330841\\
63	7.69116532049955\\
64	10.2517022640635\\
65	15.3504289413167\\
};

\addplot [color=mycolor4, only marks, mark=o, mark options={solid, mycolor4}, mark size = 1pt]
  table[row sep=crcr]{%
1	0.486786765804684\\
2	0.486786765804684\\
3	0.502247493291464\\
4	0.502247493291464\\
5	0.51769079324723\\
6	0.51769079324723\\
7	0.533087308739831\\
8	0.533087308739831\\
9	0.548553919959967\\
10	0.56393346207219\\
11	0.56393346207219\\
12	0.579419823316042\\
13	0.594947724157121\\
14	0.594947724157121\\
15	0.610592637723699\\
16	0.625969890499668\\
17	0.625969890499668\\
18	0.641565735836572\\
19	0.656999876868341\\
20	0.672374907165265\\
21	0.687902462249404\\
22	0.703428003081861\\
23	0.719020737240986\\
24	0.734470199220564\\
25	0.749775765450823\\
26	0.765256106069532\\
27	0.780687213972859\\
28	0.811666901552619\\
29	0.827230216907795\\
30	0.85810903493234\\
31	0.873597546474486\\
32	0.90448688967171\\
33	0.92009582488567\\
34	0.950846347982933\\
35	0.981823764290576\\
36	1.01272487465277\\
37	1.04360173524028\\
38	1.07424991536504\\
39	1.1206051107164\\
40	1.16686554294514\\
41	1.19767826140528\\
42	1.24378097258781\\
43	1.30521505342602\\
44	1.35150460532906\\
45	1.41333977343136\\
46	1.49023198188635\\
47	1.55200212301741\\
48	1.64450779939554\\
49	1.73685366289176\\
50	1.82878901286215\\
51	1.93600504240122\\
52	2.07480555377031\\
53	2.21321763374008\\
54	2.38218264867528\\
55	2.58186839981626\\
56	2.81246809117104\\
57	3.08974655708818\\
58	3.44249674178609\\
59	3.85671427433452\\
60	4.40909985318404\\
61	5.14612352273606\\
62	6.15950446034428\\
63	7.69185246639813\\
64	10.2511407684551\\
65	15.3531064005799\\
};

\end{axis}
\end{tikzpicture}%

%% file: perfect_load_balancing.tex
\begin{table*}
\centering
\renewcommand{\arraystretch}{1.7}
\caption{The value of $\tau_c$ under perfect load balancing when using different schemes. The acronym LB refers to load balancing. The values of $m_i^u$ and $k_i^u$ are the number of matrices in which $\mA$ and $\mB$ are divided according to the rate of each scheme and the value of the number of workers per cluster $n_u$. The ratio of $\tau_c$ of two schemes is proportional to the ratio of the mean waiting time of those schemes under service model~1, see~\eqref{eq:lb_bound}.}
\begin{tabular}{c|c|c|c}
\rowcolor{lightgray} Setting  &  RPM3  &  {LB: GASP low $z$} & {LB: GASP large $z$} \\ 
 \multirow{2}{*}{$\tau_{c}$}    &%
 \multirow{2}{*}{\small $\frac{mk(1+\epsilon)}{\gamma_1 \left\lfloor \frac{n_1-2z+1}{2}\right\rfloor + \sum_{u=2}^{c}\left\lfloor \frac{n_u-z+1}{2}\right\rfloor\gamma_u}$}&%
\multirow{2}{*}{\small $\frac{mk}{\sum_{u=1}^c \gamma_u n_u - \sum_{u=1}^c (\gamma_{u}-\gamma_{u+1})(k_2^{u}+m_2^{u}) - \gamma_1(z^2+z-3)}$} &%
  \multirow{2}{*}{\small $\frac{2mk}{\sum_{u=1}^c \gamma_u n_u - \gamma_1(2z-1)}$}\\
  ~&~&~&~\\\hline
\rowcolor{lightgray} Setting & {LB: GASP $z=1$} & {LB: GASP medium $z$} & {LB: Ideal scheme} \\ 
  $\tau_{c}$ & \multirow{2}{*}{\small $\frac{mk}{\sum_{u=1}^c \gamma_u n_u - \sum_{u=1}^c (\gamma_{u}-\gamma_{u+1})(k_1^{u}+m_1^{u})}$} &%
 \multirow{2}{*}{\small $\frac{mk}{\sum_{u=1}^c \gamma_u n_u - \sum_{u=1}^c (\gamma_{u}-\gamma_{u+1})(k_3^{u}+zm_3^{u}) - \gamma_1(z-1)}$} &%
 \multirow{2}{*}{\small $\frac{mk}{\sum_{u=1}^c \gamma_u n_u - \gamma_1 z}$}\\
   ~&~&~&~\\
\end{tabular}
\label{tab:LB_tauc}
\end{table*}
In this section we study the waiting time of the master under perfect load balancing in the worker-dependent fixed service time model (model~1). To make load balancing possible, we assume that the master has previous knowledge of the behavior of the workers during the run time of the algorithm. More precisely, the master knows the overall computing power of each worker, i.e., the master knows $s_i$ and $\lambda_i$, for all $i=1,\dots,n$. Given this knowledge, the master knows the number of clusters $c$, number of workers per cluster $n_u$, $u=1,\dots,c$, and can assign tasks that are proportional to the workers computing power without the need of a rateless code. This knowledge alleviates the need of assigning one polynomial per cluster and paying the extra penalty of privacy per cluster.

The master groups workers with similar $s_i$ and $\lambda_i$ in the same cluster. Let $\tau^{(\text{LB})}_u$ be the number of tasks assigned to workers in cluster $u$, $u=1,\dots,c$. The master chooses the values of $\taulb{u}$ such that $\taulb{u}(s_u + \lambda_u)$ is a constant so that the average compute time is the same at all the workers. We assume that $s_u \lambda_u$ is a constant, hence $\taulb{u}$ depends on $\lambda_u$. We write $\taulb{u}$ as a function of $\taulb{c}$, i.e., $\taulb{u} = \gamma_{u}\taulb{c}$. Recall that the master divides $\mA$ into $m$ sub-matrices $\mA_i$, $i=1\dots, m,$ and $\mB$ into $k$ sub-matrices $\mB_j$, $j=1,\dots,k$. Thus, the master needs $mk$ computations of the form $\mA_i\mB_j$. Given a double-sided $z$-private task assignment to $n$ workers, the number of computations of the form $\mA_i\mB_j$ that can be computed depends on the rate of the scheme used to encode the tasks. Therefore, the total number of tasks that the master needs to assign to the workers, and hence the value of $\taulb{u}$, depends on the rate of the scheme. 

Given the value of $\taulb{c}$, the mean waiting time of the master under perfect load balancing is characterized as follows.

\begin{theorem}\label{thm:load_bal}
Given a scheme that assigns $\taulb{c}$ to the workers of cluster $c$. The mean waiting time at the master using load balancing under the worker-dependent fixed service time model is given by
\begin{equation}
    \mathbb{E}[\rvT_{\text{LB}}] = \dfrac{s_c\taulb{c}}{km} + \dfrac{H_n \taulb{c}}{\lambda_c mk}.
\end{equation}
Therefore, when $\lambda_us_u = t_m$, the mean waiting time achievable using RPM3 is bounded away from the mean waiting time with load balancing as follows
\begin{align}
    {\mathbb{E}[\rvT_{RPM3}]} & \leq \mathbb{E}[\rvT_{\text{LB}}] \dfrac{\tau_{u^\star}}{\taulb{c}} \dfrac{\lambda_c}{\lambda_{u^\star}}. \label{eq:lb_bound}
\end{align}
\end{theorem}
\begin{IEEEproof}[Sketch of proof]
The proof of Theorem~\ref{thm:load_bal} follows the same steps of the proofs of Theorem~\ref{thm:waiting_exp} and Theorem~\ref{thm:waiting_erlang}. The main difference is that $\taulb{u}$ is designed after assuming that the master has a full knowledge of the future. Thus, the mean waiting time at all the workers is the same. Therefore, the mean waiting time $\mathbb{E}[\rvT_\text{LB}]$ is exactly the $n^\text{th}$ ordered statistic of $n$ \emph{iid} random variables following a shifted exponential distribution with shift ${s_c\taulb{c}}/{km}$ and rate ${\lambda_c mk}/{\taulb{c}}$. 
Under the assumption that $\lambda_u s_u$ is a fixed constant $t_m$, i.e., $s_u=t_m/\lambda_u$, the value $u^\star$ of $u$ that maximizes $s_u \tau_u = \tau_u t_m / \lambda_u$ is the same as the value of $u$ that minimizes the ratio $\lambda_u\tau_u$. Thus, $s_m = s_{u^\star}\tau_{u^\star}$. We can now write

\begin{align}
    \dfrac{\mathbb{E}[\rvT_{\text{LB}}]}{\mathbb{E}[\rvT_{RPM3}]}  &\geq  \dfrac{\dfrac{s_c\taulb{c}}{km}+ \dfrac{H_n \taulb{c}}{\lambda_{c} mk}}%
    {\dfrac{s_m}{km}+ \dfrac{H_n \tau_{u^\star}}{\lambda_{u^\star} mk}} \label{eq:UB}\\ 
     &= \dfrac{\taulb{c}}{\tau_{u^\star}} \dfrac{\lambda_{u^\star}}{\lambda_c}. \label{eq:lb_bound}
\end{align}
In~\eqref{eq:UB} we replace $\mathbb{E}[\rvT_{RPM3}]$ by the upper bound obtained in Theorem~\ref{thm:waiting_exp}. The remaining follows from simple calculations.
\end{IEEEproof}

Thus, we only need to calculate the value of $\taulb{c}$ depending on the scheme used by the master for perfect load balancing. 
The values of $\taulb{c}$ of the different considered schemes are summarized in Table~\ref{tab:LB_tauc}.

We are particularly interested in comparing RPM3 to two extreme settings: load balancing using a scheme with the best theoretical rate possible, referred to as the ideal scheme; and load balancing using GASP when $z$ takes large values. Load balancing using the ideal scheme provides a theoretical lower bound on the mean waiting time for the master using any possible rateless scheme. Load balancing using GASP provides a lower bound on the mean waiting time that could be achievable if one finds encoding polynomials with the same rate of GASP and that satisfy the properties of Remark~\ref{rem:optimal}.
After explaining the schemes used for load balancing and finding their respective values of $\taulb{c}$, we can prove the following.
\begin{corollary}\label{cor:LB_ideal}
For a fixed overhead of Fountain codes $\epsilon$ and a given $\tauideal{c}$ for load balancing using the ideal scheme. The mean waiting time of RPM3 is bounded away from the mean waiting time of the ideal scheme as follows.
\begin{equation}\label{eq:sum1}
    {\mathbb{E}[\rvT_{RPM3}]} \leq \mathbb{E}[\rvT_{\text{LB}}]\dfrac{2\gamma_{u^\star}\lambda_c(1+\epsilon)}{\lambda_{u^\star}\left(1
    - \dfrac { \sum_{u=1}^{c} \left({z+1}\right) \gamma_u}%
    {\sum_{u=1}^c \gamma_u n_u - \gamma_1 z}\right)}.
\end{equation}
The mean waiting time of RPM3 is bounded away from the mean waiting time of load balancing using GASP scheme with large values of $z$ as follows.

\begin{align}\label{eq:sum2}
     {\mathbb{E}[\rvT_{RPM3}]} \leq \mathbb{E}[\rvT_{\text{LB}}] \dfrac{\gamma_{u^\star}(1+\epsilon)}{1- \dfrac{\sum_{u=2}^c\gamma_u(z+1) +2\gamma_1 }{\sum_{u=1}^c \gamma_u n_u - \gamma_1 (2z-1)}}\dfrac{\lambda_c}{\lambda_{u^\star}}.
\end{align}

\end{corollary}

The ratio of the sums in both denominators of~\eqref{eq:sum1} and~\eqref{eq:sum2} is less than one. The equality is attained when $z=n_1+1/2 = n_u+1$ for all $u=2,\dots,c$.

\begin{remark}\label{rem:open}
Finding coding strategies that achieve the mean waiting time under load balancing is equivalent to finding encoding polynomials with the desired rates (as GASP or the ideal scheme) that share $2z-1$ evaluations as explained in Remark~\ref{rem:optimal}. The problem of finding such polynomials, if they exist, is left open. 
\end{remark}

\subsection{Theoretical lower bound}
Given $n$ workers and a double-sided $z$-private task assignment with no straggler tolerance, the best rate the master can hope for is $(n-z)/n$. In other words, out of every $n$ tasks the master obtains $n-z$ computations of the form $\mA_i \mB_j$. This is clear from the privacy requirements. Any collection of $z$ workers should learn nothing about $\mA$ and $\mB$ and therefore any collection of $z$ tasks cannot give information about $\mA$ and $\mB$. This is standard in the literature of secret sharing. %
This rate is indeed achievable in the private matrix-vector multiplication setting where the input vector is not private. However, in the setting of matrix-matrix multiplication no known scheme achieves this rate. When restricting the encoding to polynomials, better bounds can be obtained by replacing the ideal rate with the lower bounds provided in \cite{d2019degree}. We use the ideal scheme to keep our lower bound theoretical and independent from the encoding strategy. 

Now we turn our attention to the task assignment. The master wants to maximize the rate to reduce the number of assigned tasks. Since the rate $(n-z)/n$ is a linear function of $n$, the master wants to maximize $n$ when possible. Let $\tauideal{u}$ be the number of tasks assigned to workers in cluster $u$ under this setting. Given the knowledge of the workers, the master assigns $\tauideal{c}$ tasks with maximal rate $n-z$, and $\tauideal{c-1}- \tauideal{c}$ tasks with rate $n-n_c-z$, and $\tauideal{c-2}- \tauideal{c-1}$ tasks with rate $n-n_c-n_{c-1}-z$ and so on. The only constraint is the following
\begin{equation*}
    \sum_{u=1}^{c} (\tauideal{u}-\tauideal{u+1}) (n-z-\sum_{i=u}^{c-1}n_{i+1}) = mk,
\end{equation*}
where $\tauideal{c+1}$ is defined as $0$ and $\sum_{i}^j$ is defined as $0$ if $j<i$. Next we compute $\tauideal{c}$ given $mk$ and $\gamma_1,\dots,\gamma_{c-1}$. We first use the telescopic expansion to write
\begin{align*}
    mk & = (\tauideal{c}) (n-z)\\
    & ~~ + (\tauideal{c-1} - \tauideal{c}) (n-n_c-z)\\
    & ~~ + (\tauideal{c-2} - \tauideal{c-1}) (n-n_c-n_{c-1}-z) \\
    & ~~ \qquad \qquad \vdots\\
    & ~~ + (\tauideal{1}-\tauideal{2})(n_1 - z)\\
    & = \sum_{u=2}^c \tauideal{u}n_u + \tauideal{1}(n_1-z).
\end{align*}

Using the notation $\tauideal{u} = \gamma_u \tauideal{c}$ with $\gamma_c = 1$, we can compute $\tauideal{c}$ as in~\eqref{eq:taucideal}. Given the value of $\tauideal{c}$, we characterize the average waiting time at the master as in the first part of Corollary~\ref{cor:LB_ideal}. We provide the proof here.
\begin{align}\label{eq:taucideal}
    \tauideal{c} = \frac{mk}{\sum_{u=1}^c \gamma_u n_u - \gamma_1 z}.
\end{align}

\begin{IEEEproof}[Proof of Corollary~\ref{cor:LB_ideal}]
To compare the waiting time of RPM3 to that of the ideal scheme, we express $\tau_c$ as follows
\begin{align*}
    \tau_c & = \dfrac{mk(1+\epsilon)}{\gamma_1 \left\lfloor \dfrac{n_1-2z+1}{2}\right\rfloor + \sum_{u=2}^{c}\left\lfloor \dfrac{n_u-z+1}{2}\right\rfloor\gamma_u}.
\end{align*}
Now we can write
    \begin{align*}
    \dfrac{\tauideal{c}}{\tau_{u^\star}}
    & = \dfrac { \gamma_1 \left\lfloor \dfrac{n_1-2z+1}{2}\right\rfloor + \sum_{u=2}^{c}\left\lfloor \dfrac{n_u-z+1}{2}\right\rfloor\gamma_u }%
    {\gamma_{u^\star}(1+\epsilon)  \left(\sum_{u=1}^c \gamma_u n_u - \gamma_1 z\right)}\\
    & \geq \dfrac {\gamma_1 \left(\dfrac{n_1-2z-1}{2}\right) + \sum_{u=2}^{c} \left(\dfrac{n_u-z-1}{2}\right) \gamma_u }%
    {\gamma_{u^\star}(1+\epsilon)  \left(\sum_{u=1}^c \gamma_u n_u - \gamma_1 z\right)}\\
    & = \dfrac { \left( \sum_{u=1}^{c} \left({n_u-z-1}\right) \gamma_u -\gamma_1 z\right)}%
    {2\gamma_{u^\star}(1+\epsilon) \left(\sum_{u=1}^c \gamma_u n_u - \gamma_1 z\right)}\\
    & = \dfrac{1}{2\gamma_{u^\star}(1+\epsilon)}\left(1
    - \dfrac { \sum_{u=1}^{c} \left({z+1}\right) \gamma_u}%
    {\sum_{u=1}^c \gamma_u n_u - \gamma_1 z}\right).
\end{align*}
We conclude the proof of the first part by combining the above inequality together with~\eqref{eq:lb_bound}.

\end{IEEEproof}

\subsection{Load balancing using GASP}
Given $n$ workers and a $z$-private task assignment with no straggler tolerance, the best achievable rate under our model is obtained by using GASP codes~\cite{d2018gasp}. The rate of these codes depends on the privacy parameter $z$ and on the number of sub-matrices $\mA_i$ and $\mB_j$ used in one encoding of GASP. We divide our analysis into four parts accordingly.

\paragraph{No collusion}
For $z=1$, let $m_1$ and $k_1$ be the number of sub-matrices that $\mA$ and $\mB$ are divided into, respectively. Then, the rate achieved by GASP is equal to $(k_1m_1)/(k_1m_1+k_1+m_1)$. Therefore, given $n$ workers the master obtains $n-k_1-m_1$ computations of the form $\mA_i\mB_j$. Notice that the rate of the scheme depends on $n$ since $n$ must satisfy $n=m_1k_1+m_1+k_1$. Since the rate is a linear function of $n$, the master wants to maximize $n$ when possible. Let $\taunc{u}$ be the number of tasks that could be assigned to workers in cluster $u$ when using this scheme. When using GASP, the value of $m_1$ and $k_1$ depends on the number of workers. We denote by $m_1^{u}$ and $k_1^{u}$ the number of divisions when the tasks are sent to all workers in clusters $1$ to $u$. Thus, the master assigns $\taunc{c}$ tasks with maximal rate $n-k_1^c-m_1^c$ to all $n$ workers, and $\taunc{c-1}- \taunc{c}$ tasks with rate $n-n_c-k_1^{c-1}-m_1^{c-1}$ to the $n-n_c$ workers not in the slowest cluster $c$ and so on. The only constraint on $\taunc{u}$ is the following
\begin{equation*}
    \sum_{u=1}^{c} \left(\taunc{u}-\taunc{u+1}\right) \left(n-k_1^{u}-m_1^u-\sum_{i=u}^{c-1}n_i\right) = mk,
\end{equation*}
where $\taunc{c+1}$, $m_1^{c+1}$ and $k_1^{c+1}$ are defined as $0$ and $\sum_{i}^j$ is defined as $0$ if $j<i$. Next we compute $\taunc{c}$ given $mk$ and $\gamma_1,\dots,\gamma_{c-1}$.

Using the telescopic expansion we write
\begin{align*}
    mk 
    & = \sum_{u=1}^c \taunc{u}n_u - \sum_{u=1}^c (\taunc{u}-\taunc{u+1})(k_1^{u}+m_1^{u}).
\end{align*}

Using the notation $\taunc{u} = \gamma_u \taunc{c}$ with $\gamma_c = 1$ and $\gamma_{c+1} = 0$, we can compute $\taunc{c}$ as
\begin{align}
    \taunc{c} = \frac{mk}{\sum_{u=1}^c \gamma_u n_u - \sum_{u=1}^c (\gamma_{u}-\gamma_{u+1})(k_1^{u}+m_1^{u})}.
\end{align}

\paragraph{Small number of collusion}
Let $m_2$ and $k_2$ be the number of sub-matrices that $\mA$ and $\mB$ are divided into, respectively. For $2\leq z < \min\{m_2,k_2\}$ the rate achieved by GASP is equal to $(k_2m_2)/(k_2m_2+k_2+m_2+z^2+z-3)$. Therefore, given $n$ workers the master obtains $n-k_2-m_2-z^2-z+3$ computations of the form $\mA_i\mB_j$. Notice that also here the rate of the scheme depends on $n$ since $n$ must satisfy $n=k_2m_2+k_2+m_2+z^2+z-3$. Since the rate is a linear function of $n$, the master wants to maximize $n$ when possible. Let $\taugasp{u}$ be the number of tasks that could be assigned to workers in cluster $u$ when using this scheme. Thus, the master assigns $\taugasp{c}$ tasks with maximal rate $n-k_2^c-m_2^c-z^2-z+3$ to all $n$ workers, and $\taugasp{c-1}- \taugasp{c}$ tasks with rate $n-n_c-k_2^{c-1}-m_2^{c-1}-z^2-z+3$ to the $n-n_c$ workers not in cluster $C$ and so on. Using the notation $\taugasp{u} = \gamma_u \taugasp{c}$ with $\gamma_c = 1$ and $\gamma_{c+1} = 0$, we can compute $\taugasp{c}$ by following the same steps as above. We express $\taugasp{c}$ as
\begin{align}
\frac{mk}{\sum_{u=1}^c \gamma_u n_u - \sum_{u=1}^c (\gamma_{u}-\gamma_{u+1})(k_2^{u}+m_2^{u}) - \gamma_1(z^2+z-3)}.
\end{align}

\paragraph{Medium number of collusion}
Let $m_3$ and $k_3$ be the number of sub-matrices that $\mA$ and $\mB$ are divided into, respectively. Let $m_3\leq k_3$. For $m_3 \leq z < k_3$ the rate achieved by GASP is equal to $(k_3m_3)/((k_3+z)(m_3+1)-1)$. This rate coincides with the rate of~\cite{kakar2018rate}. For $m_3\geq k_3$, the values of $m_3$ and $k_3$ are interchanged in the rate. Therefore, given $n$ workers the master obtains $n-k_3-zm_3-z+1$ computations of the form $\mA_i\mB_j$. Since the rate is a linear function of $n$, the master wants to maximize $n$ when possible. Let $\taukk{u}$ be the number of tasks that could be assigned to workers in cluster $u$ when using this scheme. Thus, the master assigns $\taukk{c}$ tasks with maximal rate $n-k_3^c-zm_3^c-z+1$ to all workers, and $\taukk{c-1}- \taukk{c}$ tasks with rate $n-n_c-k_3^{c-1}-zm_3^{c-1}-z+1$ to the $n-n_c$ workers not in cluster $c$ and so on. 

Using the notation $\taukk{u} = \gamma_u \taukk{c}$ with $\gamma_c = 1$ and $\gamma_{c+1} = 0$, we can compute $\taukk{c}$ by following the same steps as above. We express $\taukk{c}$ as
\begin{align}
\frac{mk}{\sum_{u=1}^c \gamma_u n_u - \sum_{u=1}^c (\gamma_{u}-\gamma_{u+1})(k_3^{u}+zm_3^{u}) - \gamma_1(z-1)}.
\end{align}

\paragraph{Large number of collusion}
Let $m_4$ and $k_4$ be the number of sub-matrices that $\mA$ and $\mB$ are divided into, respectively. For $\max\{m_4,k_4\} \leq z$ the rate achieved by GASP is equal to $(k_4m_4)/(2m_4k_4+2z-1)$. Notice that this rate coincides with the rate of regular Lagrange polynomials that we use in our scheme. Given $n$ workers the master obtains $\left\lfloor(n-2z+1)/2\right\rfloor$ computations of the form $\mA_i\mB_j$. Since the rate is a linear function of $n$, the master wants to maximize $n$ when possible. Let $\taulag{u}$ be the number of tasks that could be assigned to workers in cluster $u$ when using this scheme. Thus, the master assigns $\taulag{c}$ tasks with maximal rate $\left\lfloor(n-2z+1)/2\right\rfloor$ to all $n$ workers, and $\taulag{c-1}- \taulag{c}$ tasks with rate $\left\lfloor(n-n_c-2z+1)/2\right\rfloor$ to the $n-n_c$ workers not in cluster $c$ and so on. We can now prove the second part of Corollary~\ref{cor:LB_ideal}.

\begin{IEEEproof}[Proof of Corollary~\ref{cor:LB_ideal} (Continued)]
Assuming that $n_u$ is even for all $u=1,\dots,c$ and using the notation $\taulag{u} = \gamma_u \taulag{c}$ with $\gamma_c = 1$, we can compute $\taulag{c}$ by following the same steps as above. We express $\taulag{c}$ as
\begin{align}
\taulag{c} = \frac{2mk}{\sum_{u=1}^c \gamma_u n_u - \gamma_1(2z-1)}.
\end{align}

We compare $\taulag{c}$ to $\tau_c$ obtained for RPM3. Recall that $\tau_c$ is expressed as
\begin{align*}
    \tau_c & = \dfrac{mk(1+\epsilon)}{\gamma_1 \left\lfloor \dfrac{n_1-2z+1}{2}\right\rfloor + \sum_{u=2}^{c}\left\lfloor \dfrac{n_u-z+1}{2}\right\rfloor\gamma_u}.
\end{align*}

Following the steps of the proof of the first part of this corollary we can write
    \begin{align*}
    \dfrac{\taulag{c}}{\tau_{u^\star}}
   & \geq \dfrac {\sum_{u=1}^{c} {n_u} \gamma_u - \sum_{u=1}^c \gamma_u(z+1) -\gamma_1 z}%
    {\gamma_{u^\star}(1+\epsilon) \left(\sum_{u=1}^c \gamma_u n_u - \gamma_1 (2z-1)\right)}\\
    & = \dfrac{1}{\gamma_{u^\star}(1+\epsilon)}\left(1- \dfrac{\sum_{u=2}^c\gamma_u(z+1) +2\gamma_1 }{\sum_{u=1}^c \gamma_u n_u - \gamma_1 (2z-1)} \right).
\end{align*}
Thus, we prove the second part of Corollary~\ref{cor:LB_ideal}.
\end{IEEEproof}

%% file: conclusion.tex
We considered the heterogeneous and time-varying setting of private distributed matrix-matrix multiplication. The workers have different computing and communication resources that can change over time. We designed a scheme called RPM3 that allows the master to group the workers into clusters with similar resources. The workers are assigned a number of tasks proportional to their overall available resources, i.e., faster workers compute more tasks and slower workers compute less tasks. This flexibility increases the speed of the computation.

We analyzed the rate of RPM3 and the mean waiting time of the master under two models of the workers service times.
Using RPM3 results in a smaller mean waiting time than known fixed-rate straggler-tolerant schemes. The reduction of the mean waiting time is possible by leveraging the heterogeneity of the workers. We provide lower bounds on the mean waiting time of the master under the worker-dependent fixed service time model. The lower bounds are obtained by assuming perfect load balancing, i.e., the master has full knowledge of the future behavior of the workers.
In terms of rate, RPM3 has a worse rate than rates of known fixed-rate straggler-tolerant schemes. We show that there exists a tradeoff between the flexibility of RPM3 and its rate. Dividing the workers into several clusters provides a good granularity for the master to refine the task assignment. However, increasing the number of clusters negatively affects the rate of RPM3. 

Finding lower bounds on the rate of flexible schemes and finding codes that achieve this rate are left as open problems. The rate of a flexible scheme using polynomials is affected by two factors. The first factor is the rate of the polynomial $h_t^{(u)}(x)$ assigned to each cluster $u$ at round $t$. The rate of $h_t^{(u)}(x)$ is the ratio of the number of computations of the form $\mA_i\mB_i$ to the number of evaluations needed to interpolate $h_t^{(u)}(x)$. Increasing the rate of $h_t^{(u)}(x)$ increases the rate of the scheme. The second factor is the number of evaluations that are shared between $h_t^{(1)}(x)$ and all other $h_t^{(u)}(x)$ for $u=2,\dots,c$. Allowing a larger number of shared evaluations increases the rate of the scheme, see Remark~\ref{rem:loss} and Remark~\ref{rem:optimal}. Finding lower bounds on the rate of a flexible scheme or lower bounds on the number of evaluations that any two polynomials $h_t^{(u_1)}(x)$ and $h_t^{(u_2)}(x)$ can share, implies finding better lower bounds on the mean waiting time of the master.

%% file: proof_privacy.tex
We want to prove that RPM3 maintains information theoretic privacy of the master's data. We prove that at any given round $t$, the tasks sent to the workers do not reveal any information about the input matrices $\mA$ and $\mB$. This is sufficient since the random matrices generated at every round are drawn independently and uniformly at random. In other words, if the workers do not obtain any information about $\mA$ and $\mB$ at any given round, then the workers obtain no information about $\mA$ and $\mB$ throughout the whole process. 

Recall that we define $\cW_{i,t}$ as the set of random variables representing the tasks sent to worker $w_i$ at round $t$. In addition, for a set $\mathcal{A}\subseteq [n]$ we define $\mathcal{W}_{\mathcal{A},t}$ as the set of random variables representing the tasks sent to the workers indexed by $\mathcal{A}$ at round $t$, i.e., $\mathcal{W}_{\mathcal{A},t}\triangleq \{\cW_{i,t}| i\in \mathcal{A}\}$. The privacy constraint is then expressed as
\begin{equation*}
    {I}\left(\rvA,\rvB;\cW_{\mathcal{Z},t}\right) = 0, \forall \cZ \subset [n], \text{ s.t. } |\cZ| = z.
\end{equation*}

We start by proving the privacy constraint for $\mA$. For a set $\cA\subseteq[n]$, let $\cF_{\cA,t}$ be the set of random variables representing the evaluations of $\ftr{t}{u}(x)$ sent to workers indexed by the set $\cA$ at round $t$. We want to prove 
\begin{equation*}
    {I}\left(\rvA;\cF_{\mathcal{Z},t}\right) = 0, \forall \cZ \subset [n], \text{ s.t. } |\cZ| = z.
\end{equation*}
Proving the satisfaction of the privacy constraint for $\mB$ follows the same steps and is omitted.

Let $\cK$ be the set of random variable presenting the random matrices $\mR_{t,1},\dots,\mR_{t,z}$ generated by the master at round $t$. We start by showing that proving the privacy constraint is equivalent to proving that $H(\cK \mid \cF_{\cZ}, \rvA) = 0$ for all $\cZ\subseteq [n], |\cZ| = z$. To that end we write,
\begin{align}
H(\rvA \mid \cF_{\cZ})&=H(\rvA)-H(\cF_{\cZ})+H(\cF_{\cZ} \mid \rvA)\\
~&=H(\rvA)-H(\cF_{\cZ})+H(\cF_{\cZ} \mid \rvA) \nonumber\\
& \qquad ~~~  -H(\cF_{\cZ} \mid \rvA, \cK) \label{eq:sk}\\
~&=H(\rvA)-H(\cF_{\cZ})+I(\cF_{\cZ}; \cK \mid \rvA) \nonumber \\
~&=H(\rvA)-H(\cF_{\cZ})+H(\cK \mid \rvA) - H(\cK \mid \cF_{\cZ},\rvA) \nonumber\\
~&=H(\rvA)-H(\cF_{\cZ})+H(\cK) - H(\cK \mid \cF_{\cZ},\rvA) \label{eq:sk2} \\
~&=H(\rvA) - H(\cK \mid \cF_{\cZ},\rvA). \label{eq:keys}
\end{align}
\vspace{0.2cm}

\noindent Equation~\eqref{eq:sk} follows because $H(\cF_{\cZ} \mid A, \cK)=0$, i.e., the tasks sent to the workers are a function of the matrix $\mA$ and the random matrices $\mR_{t,1},\dots,\mR_{t,z}$ which is true by construction. In~\eqref{eq:sk2} we use the fact that the random matrices $\mR_{t,1},\dots,\mR_{t,z}$ are chosen independently from $\mA$, i.e., $H(\cK \mid \rvA)=H(\cK)$. Equation~\eqref{eq:keys} follows because for any collection of $z$ workers, the master assigns $z$ tasks each of which has the same dimension as $\mR_{t,\delta}$, $\delta\in \{1,\dots,z\}$. In addition, all matrices $\mR_{t,1},\dots,\mR_{t,z}$ are chosen independently and uniformly at random; hence, $H(\cF_{\cZ}) = H(\cK)$. 

Therefore, since the entropy $H(.)$ is positive, proving that $H(\rvA\mid \cF_{\cZ}) = H(\rvA)$ is equivalent to proving that $H(\cK \mid \cF_{\cZ},A) = 0$.

The explanation of $H(\cK \mid \cF_{\cZ}, \rvA) = 0$ is that given the matrix $\mA$ and all the tasks received at round $t$, any collection of $z$ workers can obtain the value of the matrices $\mR_{t,1},\dots,\mR_{t,z}$.

This follows immediately from the use of Lagrange polynomials and setting the random matrices as the first $z$ coefficients. More precisely, given the data matrix as side information, the tasks sent to any collection of $z$ workers become the evaluations of a Lagrange polynomial of degree $z-1$ whose coefficients are the random matrices $\mR_{t,1},\dots,\mR_{t,z}$. Thus, the workers can interpolate that polynomial and obtain the random matrices. Therefore, by repeating the same calculations for $\mB$, we show that information theoretic privacy of the input matrices is guaranteed.

%% file: app_proof_waiting_time.tex
\begin{IEEEproof}[Proof of Theorem~\ref{thm:waiting_exp}]
In this model we assume that for a given worker $w_i$ in cluster $u$, the random variable $\rvT_u^i$ follows a shifted exponential distribution with shift $s_u \tau_u /mk$ and rate $\lambda_u m k / \tau_u$. In the following we shall focus on $\rvT_u$ since $\rvT_u^i$ depends only on the identity of the cluster $u$ and is the same for all workers in this cluster.

We can write the pdf of $T_u$ as 

\begin{equation*}
    \Pr_{\rvT_u}(x) = 
    \begin{cases}
    0 & \text{if } x < \frac{s_u \tau_u}{mk}\\
    \dfrac{\lambda_u mk}{\tau_u} \exp\left(\lambda_u s_u - \frac{\lambda_u mk}{\tau_u}x\right) & \hfill \text{otherwise}.
    \end{cases}
\end{equation*}

\begin{remark}
It is worth noting that by coupling RPM3 with a mechanism like the one proposed in \cite{KS18}, we can reduce the shift of $\rvT_u$ to $s_u/mk$ by allowing the master to send computational tasks to the workers while they are busy computing other tasks and also allowing the workers to send results of the previous task to the master while computing a new task. Thus absorbing the delays of communicating every task to and from the workers except for the delays of sending the first task and receiving the last task. In our analysis we do not assume the use of such mechanism.
\end{remark}

Let $\mathds{1}_{\{x\geq \frac{s_u \tau_u}{mk}\}}$ be the indicator function that is equal to $1$ when $x\geq \nicefrac{s_u \tau_u}{mk}$ and is equal to $0$ otherwise. The cumulative density function ${F}_{\rvT_u}(x) \triangleq \Pr(\rvT_u<x)$ of $\rvT_u$ can then be expressed as

\begin{equation*}
    {F}_{\rvT_u}(x)= \left(1 - e^{\left({\lambda_u s_u} - \frac{\lambda_u mk}{\tau_u}x\right)}\right)\mathds{1}_{\{x\geq \frac{s_u \tau_u}{mk}\}}.
\end{equation*}
The random variable $\rvT_u^\star$ is the maximum of $n_u$ \emph{iid} copies of $\rvT_u$. Hence, we can write
\begin{equation*}
    {F}_{\rvT^\star_u}(x) = \left(\left(1 - e^{\left({\lambda_u s_u} - \frac{\lambda_u mk}{\tau_u}x\right)}\right)\mathds{1}_{\{x\geq \frac{s_u \tau_u}{mk}\}} \right)^{n_u}.
\end{equation*}

Recall that $\rvT_{RPM3} = \max_{u\in\{1,\dots,c\}} \rvT^\star_u$ is the maximum of $c$ independent random variables. Therefore, we can write

\begin{align*}
    {F}_{\rvT_{RPM3}}(x) & = \Pr(\rvT_{RPM3}<x)\\
   & = \Pr(\rvT^\star_1<x)\Pr(\rvT^\star_2<x) \cdots \Pr(\rvT^\star_c<x)\\
    & = \prod_{u=1}^{c} \left( \left(1 - e^{\left({\lambda_u s_u} - \frac{\lambda_u mk}{\tau_u}x\right)}\right)\mathds{1}_{\{x\geq \frac{s_u \tau_u}{mk}\}} \right)^{n_u}.
\end{align*}

Define $s_m \triangleq \max_u s_u\tau_u$ and $t_m\triangleq \lambda_u s_u$. 
We consider $\bar{F}_{\rvT_{RPM3}}(x) \triangleq 1 - {F}_{\rvT_{RPM3}}(x)$ and find a bound on $\bar{F}_{\rvT_{RPM3}}(x)$ as follows
\begin{align*}
    \bar{F}_{\rvT_{RPM3}}(x) &= \Pr(\rvT_{RPM3}>x)\\
    &= 1-\Pr(\rvT_{RPM3}<x)\\
    & = 1 -  \Pr(\rvT^\star_1<x)\Pr(\rvT^\star_2<x) \cdots \Pr(\rvT^\star_c<x).
\end{align*}

To obtain a non-trivial lower bound on $\bar{F}_{\rvT_{RPM3}}(x)$, we consider $x\geq s_m/mk$ and bound from below each term in the product $\prod_{u=1}^c\Pr(\rvT^\star_u<x)$ by 
$$\Pr(\rvT^\star_u<x)\geq \left(1 - e^{\left( t_m - \frac{\lambda_{u^\star} mk}{\tau_{u^\star}}x\right)} \right).$$

The bound is obtained by maximizing the exponent of $e$ in $F_{\rvT_u}(x)$ because $t_m - \nicefrac{\lambda_{u^\star}}{\tau_{u^\star}}mkx \leq 0$. To see that, recall that $s_m = \max_u s_u\tau_u$. Let $s_m = s_{u_1}\tau_{u_1}$ we can write

\begin{align*}
    x & \geq \frac{s_{u_1}\tau_{u_1}}{mk} = \frac{\lambda_{u^\star}s_{u_1}\tau_{u_1}}{\lambda_{u^\star}mk}\\
    & = \frac{\lambda_{u}s_{u}}{mk} \frac{\tau_{u_1}}{\lambda_{u^\star}} \geq \frac{\lambda_{u}s_{u}}{mk} \frac{\tau_{u^\star}}{\lambda_{u^\star}}.
\end{align*}

Let $F_\rvX(x)\triangleq 1 - e^{\left( t_m - \frac{\lambda_{u^\star} mk}{\tau_{u^\star}}x\right)}$, we can write 
$${F}_{\rvT_{RPM3}}(x) \geq F_\rvX(x)^n.$$

Notice that $F_\rvX(x)$ is the cumulative distribution function (CDF) of a random variable following a shifted exponential distribution with shift $s_m/km$ and rate $\lambda_{u^\star}km/\tau_{u^\star}$. Given $n$ \emph{iid} random variables $\rvX_1,\dots,\rvX_n$, we let $\rvX_{(1)}\leq \rvX_{(2)}\leq \dots \leq \rvX_{(n)}$ be the ordered values of the $\rvX_i$'s (known as ordered statistics). With this notation, ${F}_{\rvT_{RPM3}}(x)$ is bounded by the distribution of the $n^\text{th}$ ordered statistic of $n$ random variables following a shifted exponential distribution, i.e., we have $F_\rvX(x)^n = F_{\rvX_{(n)}}(x)$.

It follows that
\begin{align}
    \mathbb{E}[\rvT_{RPM3}] &= \int_0^\infty \Pr(\rvT_{RPM3}>x)dx \nonumber\\
     &= \int_0^\infty (1-F_{\rvT_{RPM3}}(x))dx \nonumber\\
     & \leq \int_0^\infty (1 -  F_\rvX(x)^n) dx \nonumber\\
     & = \int_0^\infty (1 -  F_{\rvX_{(n)}}(x)) dx \nonumber\\
     & = \mathbb{E}[\rvX_{(n)}]. \label{eq:bound_mean}
\end{align}

Next we obtain a bound on $\mathbb{E}[\rvX_{(n)}]$ the mean of $\rvX_{(n)}$. We express $\rvX$ as $$\rvX=\frac{sm}{km} + \rvX',$$ where $\rvX'$ is a random variable following an exponential distribution with rate $\lambda_{u^\star}km/\tau_{u^\star}$. The following equations hold
\begin{align}
    F_\rvX(x) & = F_{\rvX'}\left(x-\frac{sm}{km}\right), \quad \forall x\geq \frac{sm}{km}, \label{eq:distribution}\\
    \mathbb{E}[\rvX] & = \frac{sm}{km} + \mathbb{E}[\rvX'].\label{eq:mean_x}
\end{align}

We use the following Theorem from Renyi~\cite{Renyi53} to compute $\mathbb{E}[\rvX']$.

\begin{theorem*}[Renyi~\cite{Renyi53}]
\label{thm:Renyi}
The $d^{\text{th}}$ order statistic $\rvX'_{(d)}$ of $n$ \emph{iid} exponential random variables $\rvX'_i$ 
is equal to the following random variable in the distribution 
\begin{align*}
\rvX'_{(d)} &\triangleq \sum_{j=1}^{d}\frac{\rvX'_j}{n-j+1}. 
\end{align*}
\end{theorem*}

Using Renyi's theorem, the mean of the $d^{\text{th}}$ order statistic $\mathbb{E}[ \rvX'_{(d)}]$ can be written as
\begin{align*}
\mathbb{E}[\rvX'_{(d)}]&=\mathbb{E}[\rvX'_j]\sum_{j=0}^{d-1}\dfrac{1}{n-j}
=\dfrac{(H_n-H_{n-d})\tau_{u^\star}}{\lambda_{u^\star}mk},
\end{align*}
where $H_n$ is the $n^{\text{th}}$ harmonic sum defined as $H_n \triangleq \sum_{i=1}^n \frac{1}{i}$, with the notation $H_0 \triangleq 0$.
In particular,%
\begin{equation}\label{eq:mean_xn}
    \mathbb{E}[\rvX_{(n)}] = \dfrac{H_n\tau_{u^\star}}{\lambda_{u^\star}mk}.
\end{equation}

Combining the results of~\eqref{eq:bound_mean},~\eqref{eq:mean_x} and~\eqref{eq:mean_xn} we have
\begin{equation*}
    \mathbb{E}[\rvT_{RPM3}] \leq \frac{sm}{km} + \dfrac{H_n\tau_{u^\star}}{\lambda_{u^\star}mk}.
\end{equation*}

Under the assumption that $\lambda_u s_u$ is a fixed constant $t_m$, i.e., $s_u=t_m/\lambda_u$, the value $u^\star$ of $u$ that maximizes $s_u \tau_u = \tau_u t_m / \lambda_u$ is the same as the value of that minimizes the ratio $\lambda_u\tau_u$. Thus, we can write $s_m = s_{u^\star}\tau_{u^\star} = t_m \tau_{u^\star} /\lambda_{u^\star}$. This concludes the proof.

\end{IEEEproof}

\begin{IEEEproof}[Proof of Corollary~\ref{cor:kakar_exp}]
We only provide a sketch of the proof because the detailed steps are similar to steps of the proof of Theorem~\ref{thm:waiting_exp}. We first bound $F_{\rvT_I}(x)$ from above by $F_\rvX(x)^{n-n_s}$ where $\rvX$ is a shifted exponential random variable with rate $\frac{\lambda_1}{mk} \left\lceil\frac{m}{m_I}\right\rceil \left\lceil\frac{k}{k_I}\right\rceil$ and shift $\frac{s_1}{mk} \left\lceil\frac{m}{m_I}\right\rceil \left\lceil\frac{k}{k_I}\right\rceil$.
This is the $n-n_s$ ordered statistic of $n$ \emph{iid} random variable following the shifted exponential distribution. We use Renyi's theorem and the inequality $ \left\lceil\frac{m}{m_I}\right\rceil \left\lceil\frac{k}{k_I}\right\rceil\geq\frac{mk}{m_Ik_I}$ and the rest follows.
\end{IEEEproof}

\begin{IEEEproof}[Proof of Theorem~\ref{thm:waiting_erlang}]
In this model the random variable $\rvT_u$ (the time spent by a worker in cluster $u$ to compute $\tau_u$ tasks) is the sum of $\tau_u$ \emph{iid} random variables following the shifted exponential distribution with rate $\lambda_u mk$ and shift $s_u/mk$. Thus, $\rvT_u$ is a random variable following a shifted Erlang distribution.

The CDF $F_{\rvT_u}(x)$ of $\rvT_u$ is equal to $0$ if $x<s_u/mk$ and is expressed as follows otherwise.
\begin{align*}
    F_{\rvT_u}(x) & = 1 - \sum_{j=0}^{\tau_u-1}\frac{ e^{-\lambda_u km (x-s_u/mk)} }{j!} \left(\lambda_u k m (x-\frac{s_u}{km})^j\right)\\
    & = 1 - \sum_{j=0}^{\tau_u-1}\frac{ e^{\lambda_u s_u - \lambda_u km x} }{j!} \left(\lambda_u k mx - \lambda_u s_u\right)^j.
\end{align*}

Again, the random variable $\rvT_u^\star$ (the time spent by all the workers of cluster $u$ to compute $\tau_u$ tasks) is the maximum of $n_u$ \emph{iid} copies of $\rvT_u$. Hence, we have $F_{\rvT_u^\star}(x) = 0$ for $x< s_u/km$ and for $x\geq s_u/km$ we have
\begin{align*}
    F_{\rvT_u^\star}(x) & =
    \left(1 - \sum_{j=0}^{\tau_u-1}\frac{ e^{\lambda_u s_u - \lambda_u km x} }{j!} \left(\lambda_u k mx - \lambda_u s_u\right)^j\right)^{n_u}.
\end{align*}

Similarly to the flow of the proof of Theorem~\ref{thm:waiting_exp}, we want to bound the mean waiting time of the master. We first have the following set of inequalities.

\begin{align}
    \mathbb{E}[{\rvT_{RPM3}}] &= \int_0^\infty \Pr(\rvT_{RPM3}>x)dx\\
    &= \int_0^\infty (1-\Pr(\rvT_{RPM3}<x))dx \nonumber\\
    %
    %
    %
    & = \frac{s_{m}}{km} + \int_{\frac{s_{m}}{km}}^\infty\left(1- \prod_{u=1}^{c}  F_{\rvT^\star_u}(x)\right) dx \nonumber\\
    & =\frac{s_{m}}{km} + \int_{\frac{s_{m}}{km}}^\infty\left(1- \prod_{u=1}^{c}  F_{\rvT_u}(x)^{n_u}\right) dx. \label{eq:avg_waiting1}
\end{align}

Next, we bound $F_{\rvT_u}(x)$ from below for all values of $u \in [c]$. Since all the terms in the summation in $F_{\rvT_u}$ are positive, we can write 
\begin{align*}
    F_{\rvT_u}(x) & \geq 1 - \sum_{j=0}^{\tau_{\max}-1}\frac{ e^{\lambda_u s_u - \lambda_u km x} }{j!} \left(\lambda_u k mx - \lambda_u s_u\right)^j.
\end{align*}
Taking the derivative of $\frac{ e^{\lambda_u s_u - \lambda_u km x} }{j!} \left(\lambda_u k mx - \lambda_u s_u\right)^j$ with respect to $\lambda_u$, we see that this function is decreasing in $\lambda_u$. We can now bound $F_{\rvT_u}(x)$ as
\begin{align}\label{eq:avg_waiting2}
    F_{\rvT_u}(x) & \geq 1 - \sum_{j=0}^{\tau_{\max}-1}\frac{ e^{{\lambda_c} s_u - {\lambda_c} km x} }{j!} \left({\lambda_c} k mx - {\lambda_c} s_u\right)^j.
\end{align}
Let $F_{max}(x)\triangleq 1 - \sum_{j=0}^{\tau_{\max}-1}\frac{ e^{{\lambda_c} s_u - {\lambda_c} km x} }{j!} \left({\lambda_c} k mx - {\lambda_c} s_u\right)^j$. Combining~\eqref{eq:avg_waiting1} and~\eqref{eq:avg_waiting2} we can bound the mean waiting time from above by
\begin{align*}
    \mathbb{E}[{\rvT_{RPM3}}] &\leq \frac{s_{m}}{km} + \int_{\frac{s_{m}}{km}}^\infty\left(1- \prod_{u=1}^{c} \left(F_{max}(x) \right)^{n_u}\right) dx\\
    & = \frac{s_{m}}{km} + \int_{\frac{s_{m}}{km}}^\infty\left(1- \left(F_{max}(x) \right)^{n}\right) dx.
\end{align*}
Notice that $F_{max}(x)$ is the cumulative density function of an Erlang distribution with shape $\tau_{max}$ (i.e. sum of $\tau_{max}$ \emph{iid} exponential random variables) and rate ${\lambda_c}$. Hence, the mean waiting time of the master when using RPM3 is bounded by the mean of the $n^\text{th}$ order statistic of $n$ Erlang random variables. Using the derivation from~\cite{gupta1960order} we can bound the mean waiting time as in the statement of the theorem.
\end{IEEEproof}

%% file: algorithms.tex
%
We summarize the encoding process of RPM3 in the following four algorithms. We consider the clustering of the workers (result of Algorithm~\ref{alg:clustering}) as global knowledge for all the provided algorithms. The coordinator (Algorithm~\ref{alg:sch}) takes as input the numbers of clusters and keeps track of the workers. More precisely, when any worker is idle, the coordinator calls the Encode function (Algorithm~\ref{alg:encode}) to generate a new task. The encoder in Algorithm~\ref{alg:encode} takes as input the round, the cluster in which the idle worker is located. The encoder needs to know if the idle worker is the first worker of this cluster starting this round. If so, the master creates a fresh polynomial pair and sends an evaluation to the workers. 
Only for the first round, the clustering in the encoding does not play a role. This holds because at the first round one polynomial pair is generated for all the workers.
{The coordinator checks at real-time if the idle worker is the last worker of its cluster to respond. In this case, the coordinator calls the interpolation algorithm.} Algorithm~\ref{alg:interpolation} provides the Fountain-coded matrices. To do so, it requires the index of the cluster at hand and the considered round. If it is the first cluster of the considered round, the algorithm interpolates the polynomial $\htr{t}{1}(x)$ and saves the $z$ shared evaluations that are used in the interpolation of the other clusters. Otherwise, the interpolation algorithm extracts the $z$ shared evaluations from the memory and interpolates the polynomial. The coordinator collects all the Fountain-coded matrices obtained from the interpolation and saves them in a list. When enough Fountain-coded matrices are collected, the coordinator runs the peeling decoder to obtain all $mk$ components of the multiplication $C$ successfully.

\subsection{Scheduler}
\begin{algorithm}[H]
\SetAlgoLined
    \SetKwInOut{Input}{Input}
    \SetKwInOut{Result}{Result}
    \Input{$n$ workers}
    \Result{$\mA_1\mB_1,\dots,\mA_m\mB_k$}
$\text{Inter} \leftarrow []$ \tcp*[l]{storage for Fountain-coded matrices}
counter $\leftarrow 0$ \tcp*[l]{nb of coded matrices collected}

$t_1, t_2, \dots, t_n \leftarrow 1$ \tcp*[l]{put all workers to round $1$} 

Encode ($t = 1$, $n_t = n$, $u = c$) \tcp*[l]{encode for all the workers}

Clustering($n$ tasks, $\Delta,z$) \tcp*[l]{Cluster the workers and make it global}

%
$i \leftarrow 1$ \tcp*[l]{auxiliary variable representing worker $w_i$}

$n_{t}(1),\dots,n_t(c) \leftarrow 0$ \tcp*[l]{nb of workers in cluster $u$ at round $t$}

%
\While{ $\text{counter} \leq mk(1 + \varepsilon)$}
{ \eIf{worker $i$ is ready}
{ $t_i \leftarrow$ extract number of packets computed by this worker so far\;

  %
  $n_{t_i}({u'}) = n_{t_i}({u'}) +1 $ \;
  
  \If{$n_{t_i} = n_{u'}$}
  {
  $\text{Inter} \leftarrow$ $[\text{Inter}; \text{Interpolate}(n_{t_i}({u'}),u')]$ \; 
  
  counter $=$ counter $+\text{ }d_{u'}$ \;
  }
  $t_{i} = t_{i} +1 $\;
  
  
  Encode($i, t_{i}$) \; 
  %
}
{$i = i + 1 \mod n$ \tcp*[l]{Check the next worker}}  
}
$\mA_1\mB_1,\dots,\mA_m\mB_k \leftarrow$ Peeling\textunderscore decoder(Inter) as in~\cite{factored_lt} \;
\caption{Coordinator}
\label{alg:sch}
\end{algorithm}


\subsection{Interpolation}
\begin{algorithm}[H]
\SetAlgoLined
\SetKwInOut{Input}{Input}
\SetKwInOut{Result}{Result}
\Input{$u$}
\Result{Fountain-coded matrices $\widetilde{\mA}\widetilde{\mB}$ }

%
%
%


\eIf{$u = 1$}
{\tcp{first cluster finished computing} 
interpolate $\htr{t}{1}(x)$ using $2d_1+2z-1$ evaluations \; 

save $\htr{t}{1}(\alpha_{1}), \dots, \htr{t}{1}(\alpha_{z})$ in the memory as $z$ common evaluations\;
}
{ 
\tcp{cluster $u \neq 1$ finished computing}
extract $z$ common evaluations for round $t$ from the memory \;

interpolate $\htr{t}{u}$ using $2d_{u} + z - 1$ and $z$ common evaluations \;
}

%
\textbf{return} $\htr{t}{u}(\alpha_{z+1}), \dots, \htr{t}{u}(\alpha_{z+d_{u}})$;
\caption{Interpolate}
\label{alg:interpolation}
\end{algorithm}

\subsection{Encoding}
\begin{algorithm}[H]
\SetAlgoLined
    \SetKwInOut{Input}{Input}
    \SetKwInOut{Result}{Result}
    \Input{$t, n_t, u$}
    \Result{Tasks for the idle workers}
  \eIf{$t=1$}{
    
    generate $z$ random matrices $\mR_1, \mS_1$ \;
    
    encode $\sum_{u=1}^{c}d_u$ Fountain-coded matrices $\widetilde{\mA}, \widetilde{\mB}$ \;
    
    construct $c$ polynomial-pairs $\ftr{1}{u}(x), \gtr{1}{u}(x)$ as in \eqref{eq:fx}, \eqref{eq:gx} \;
    
    pick carefully distinct $\beta_1, \dots, \beta_n$ elements from $\F_q$ \;
    
    \textbf{return} $n$ evaluations $\ftr{1}{1}(\beta_i), \gtr{1}{1}(\beta_i)$
    }
    {
    %
    %
    \eIf{$n_t(u) = 1$}
    {
    \eIf{$u = 1$}
    {
        \tcp{the first worker at round $t$}
    generate $z$ random matrices $\mR_{t}, \mS_t$ \;
    
    encode $d_1$ Fountain-coded matrices $\widetilde{\mA}, \widetilde{\mB}$ \; 
    
    construct a polynomial-pair $\ftr{t}{1}(x), \gtr{t}{1}(x)$ \;
    
    send an evaluation to worker $i$, $\ftr{t}{1}(\beta_{i}), \gtr{t}{1}(\beta_{i})$ \;
    }
    {
    \tcp{the first worker of non-first cluster at round $t$}
    encode $d_{u}$ Fountain-coded matrices $\widetilde{\mA}, \widetilde{\mB}$ \; 
    
    extract $\mR_t, \mS_t$ from the memory \; 
    
    construct a polynomial-pair $\ftr{t}{u}(x), \gtr{t}{u}(x)$ \;
    
    send to worker $i$ an evaluation $\ftr{t}{u}(\beta_{i}), \gtr{t}{u}(\beta_{i})$ \;
    }
    
    
    
    }
    {
     \tcp{non-first worker of any cluster}
     extract the polynomial-pair $\ftr{t}{u}(x), \gtr{t}{u}(x)$ from the memory \;
    
    send $\ftr{t}{u}(\beta_{i}), \gtr{t}{u}(\beta_{i})$ to worker $i$ \;
    }
    
    
    
    
    
    
    }
    
    
    
    
    
 \caption{Encode}
 \label{alg:encode}
\end{algorithm}

\subsection{Clustering}
\begin{algorithm}[H]
\SetAlgoLined
\SetKwInOut{Input}{Input}
\SetKwInOut{Result}{Result}
\Input{$n$ idle workers and $n$ tasks, $\Delta, z$}
\Result{$c$, $n_1,\dots,n_c$}
Send $n$ tasks to $n$ workers \;

non\textunderscore assigned $\leftarrow n$ \tcp*[l]{nb of workers not assigned}
$u \leftarrow 1$ \tcp*[l]{indexing of the clusters}
$n_1 \leftarrow 0$ \tcp*[l]{nb of workers in the first cluster}
\While{non\textunderscore assigned $\geq z+1$}
{ 
    stop $\leftarrow$ \textit{False} \tcp*[l]{stopping criterion}
    \While{\textit{not} stop}
    {
    \If{\textit{any worker $w_i$ completed the task}}
    { 
    $n_u = n_u + 1$ \;
    
    \If{$n_u = 1$}
    {
        start\textunderscore time $\leftarrow$ get\textunderscore current\textunderscore time() \tcp*[l]{time when first worker of cluster $u$ finished computing}
    }
    non\textunderscore assigned $=$ non\textunderscore assigned $- 1$ \;
    
    \eIf{$u = 1$}
    {stop $= (n_1 \geq 2z-1)$ \textit{and} get\textunderscore current\textunderscore time() $-$ start\textunderscore time $\geq \Delta$ }
    {
    stop $= (n_u \geq z+1)$ \textit{and} get\textunderscore current\textunderscore time() $-$ start\textunderscore time $\geq \Delta$
    }
    }
    }
    $u = u + 1$ \;
    
    $n_u \leftarrow 0$ \; 
    
}
$c = u-1$ \tcp*[l]{nb of clusters}

$n_{c} = n_{c}\text{ }+ $ non\textunderscore assigned \; 
    
non\textunderscore assigned $= 0$ \;

\textbf{return} $c, n_1,\dots,n_c$ \;
\caption{Clustering}
\label{alg:clustering}
\end{algorithm}



 
 
  




